\documentclass[10pt, final, journal, letterpaper, oneside, twocolumn]{IEEEtran}
\usepackage{graphicx}
\usepackage{amsmath}
\usepackage{amssymb}
\usepackage{mathrsfs}
\usepackage{stfloats}
\usepackage{dsfont}
\usepackage{setspace}
\usepackage{array}
\usepackage{bbm}
\usepackage{accents}
\newcommand*{\dt}[1]{%
  \accentset{\mbox{\large .}}{#1}}
\newcommand*{\ddt}[1]{%
  \accentset{\mbox{\large .\hspace{-0.1ex}.}}{#1}}
\newtheorem{definition}{Definition}
\newtheorem{theorem}{Theorem}
\newtheorem{corollary}{Corollary}
\newtheorem{proposition}{Proposition}
\newtheorem{lemma}{Lemma}
\newtheorem{remark}{Remark}

\begin{document}
\title{System-Level Modeling and Optimization of the Energy Efficiency in Cellular Networks -- A Stochastic Geometry Framework}
\author{Marco~Di~Renzo,~\IEEEmembership{Senior~Member,~IEEE},
Alessio~Zappone,~\IEEEmembership{Senior~Member,~IEEE},
Thanh~Tu~Lam,~\IEEEmembership{Student~Member,~IEEE}, and
M\'erouane~Debbah,~\IEEEmembership{Fellow,~IEEE}
\thanks{Manuscript received August 2, 2017; revised November 19, 2017; accepted January 9, 2018. Date of publication January XY, 2018; date of current version January XY, 2018. This work was supported in part by the European Commission through the H2020-MSCA ETN-5Gwireless project under Grant Agreement 641985, the H2020-MSCA IF-BESMART project under Grant Agreement 749336, and the H2020-ERC PoC-CacheMire project under Grant Agreement 727682. The associate editor coordinating the review of this paper and approving it for publication was S. Mukherjee. (Corresponding author: Marco Di Renzo)}
\thanks{M. Di Renzo and T. Tu-Lam are with the Laboratoire des Signaux et Syst\`emes, CNRS, CentraleSup\'elec, Univ Paris Sud, Universit\'e Paris-Saclay, 3 rue Joliot Curie, Plateau du Moulon, 91192 Gif-sur-Yvette, France. (e-mail: marco.direnzo@l2s.centralesupelec.fr, lamthanh.tu@l2s.centralesupelec.fr).}
\thanks{A. Zappone and M. Debbah are with the LANEAS group of the Laboratoire des Signaux et Syst\`emes, CentraleSup\'elec, CNRS, Univ Paris Sud, Universit\'e Paris-Saclay, 3 rue Joliot Curie, Plateau du Moulon, 91192 Gif-sur-Yvette, France. (e-mail: alessio.zappone@l2s.centralesupelec.fr, merouane.debbah@l2s.centralesupelec.fr). M. Debbah is also with the Mathematical and Algorithmic Sciences Laboratory, France Research Center, Huawei Technologies, 20 Quai du Point du Jour, 92100 Boulogne-Billancourt, France.}
%
%
}
%
%
%
%
%
%
\maketitle
\begin{abstract}
In this paper, we analyze and optimize the energy efficiency of downlink cellular networks. With the aid of tools from stochastic geometry, we introduce a new closed-form analytical expression of the potential spectral efficiency (bit/sec/m$^2$). In the interference-limited regime for data transmission, unlike currently available mathematical frameworks, the proposed analytical formulation depends on the transmit power and deployment density of the base stations. This is obtained by generalizing the definition of coverage probability and by accounting for the sensitivity of the receiver not only during the decoding of information data, but during the cell association phase as well. Based on the new formulation of the potential spectral efficiency, the energy efficiency (bit/Joule) is given in a tractable closed-form formula. An optimization problem is formulated and is comprehensively studied. It is mathematically proved, in particular, that the energy efficiency is a unimodal and strictly pseudo-concave function in the transmit power, given the density of the base stations, and in the density of the base stations, given the transmit power. Under these assumptions, therefore, a unique transmit power and density of the base stations exist, which maximize the energy efficiency. Numerical results are illustrated in order to confirm the obtained findings and to prove the usefulness of the proposed framework for optimizing the network planning and deployment of cellular networks from the energy efficiency standpoint.
\end{abstract}
\section{Introduction} \label{Introduction}
The Energy Efficiency (EE) is regarded as a key performance metric towards the optimization of operational cellular networks, and the network planning and deployment of emerging communication systems \cite{AlessioSurvey2016}. The EE is defined as a benefit-cost ratio where the benefit is given by the amount of information data per unit time and area that can be reliably transmitted in the network, i.e., the network spectral efficiency, and the cost is represented by the amount of power per unit area that is consumed to operate the network, i.e., the network power consumption. Analyzing and designing a communication network from the EE standpoint necessitate appropriate mathematical tools, which are usually different from those used for optimizing the network spectral efficiency and the network power consumption individually \cite{AlessioMonograph2015}. The optimization problem, in addition, needs to be formulated in a sufficiently simple but realistic manner, so that all relevant system parameters appear explicitly and the utility function is physically meaningful.

Optimizing the EE of a cellular network can be tackled in different ways, which include \cite{AlessioSurvey2016}: the design of medium access and scheduling protocols for optimally using the available resources, e.g., the transmit power; the use of renewable energy sources; the development of innovative hardware for data transmission and reception; and the optimal planning and deployment of network infrastructure. In the present paper, we focus our attention on optimizing the average number of Base Stations (BSs) to be deployed (or to be kept operational) per unit area and their transmit power. Henceforth, this is referred to as ``system-level EE'' optimization, i.e., the EE across the entire (or a large portion of the) cellular network is the utility function of interest.

System-level analysis and optimization are useful when the network operators are interested in optimizing the average performance across the entire cellular network. Hence, they are relevant for optimally operating current networks, and for deploying and planning future networks. In the first case, given an average number of BSs per unit area already deployed, they may provide information on the average number of BSs that can be switched off based on the average load of the network, and on their optimal transmit power to avoid coverage holes. In the second case, they may guide the initial deployment of cellular infrastructure that employs new types of BSs (e.g., powered by renewable energy sources), new transmission technologies (e.g., large-scale antennas), or that operate in new frequency bands (e.g., the millimeter-wave spectrum).

In the last few years, the system-level modeling and analysis of cellular networks have been facilitated by capitalizing on the mathematical tool of stochastic geometry and, more precisely, on the theory of spatial point processes \cite{AndrewsNov2011}-\cite{MDR_IM}. It has been empirically validated that, from the system-level standpoint, the locations of the BSs can be abstracted as points of a homogeneous Poisson Point Process (PPP) whose intensity coincides with the average number of BSs per unit area \cite{MDR_ACM}. A comprehensive survey of recent results in this field of research is available in \cite{Hesham_ProcSurvey}.

A relevant performance metric for the design of cellular networks is the Potential Spectral Efficiency (PSE), which is the network information rate per unit area (measured in bit/sec/m$^2$) that corresponds to the minimum signal quality for reliable transmission. Under the PPP modeling assumption, the PSE can be obtained in two steps: i) first by computing the PSE of a randomly chosen Mobile Terminal (MT) and by assuming a given spatial realization for the locations of the BSs and ii) then by averaging the obtained conditional PSE with respect to all possible realizations for the locations of the BSs and MTs. In the interference-limited regime, this approach allows one to obtain a closed-form expression of the PSE under the (henceforth called) \textit{standard modeling assumptions}, i.e., single-antenna transmission, singular path-loss model, Rayleigh fading, fully-loaded BSs, cell association based on the highest average received power \cite{AndrewsNov2011}. Motivated by these results, the PPP modeling approach for the locations of the BSs has been widely used to analyze the trade-off between the network spectral efficiency and the network power consumption, e.g., \cite{MariosANDTony2013}, as well as to minimize the network power consumption given some constraints on the network spectral efficiency or to maximize the network spectral efficiency given some constraints on the network power consumption \cite{Unpublished2014}. The PPP modeling approach has been applied to optimize the EE of cellular networks as well. Notable examples for this field of research are \cite{EE_CellularRepulsive2013}-\cite{Alam_Voronoi}. A general study of the energy and spectral efficiencies of multi-tier cellular networks can be found in \cite{Mukherjee_Book}. In the authors' opinion, however, currently available approaches for modeling and optimizing the system-level EE of cellular networks are insufficient and/or unsuitable for mathematical analysis. This is further elaborated in the next section.
\subsection{Fundamental Limitations of Current Approaches for System-Level EE Optimization} \label{Introduction_EE}
We begin with an example that shows the limitations of the available analytical frameworks. In the interference-limited regime, under the standard modeling assumptions, the PSE is:
\setcounter{equation}{0}
\begin{equation}
\label{Eq_1}
\begin{split}
{\rm{PSE}} & = \lambda_{\rm{BS}}{{\rm{B}}_{\rm{W}}}{\log _2}\left( {1 + {\gamma _{\rm{D}}}} \right){{\rm{P}}_{{\rm{cov}}}}\left( {{\gamma _{\rm{D}}}} \right) \\ & \mathop  = \limits^{\left( a \right)} \frac{{{\lambda _{{\rm{BS}}}}{{\rm{B}}_{\rm{W}}}{{\log }_2}\left( {1 + {\gamma _{\rm{D}}}} \right)}}{{{}_2{F_1}\left( {1, - {2 \mathord{\left/
 {\vphantom {2 \beta }} \right.
 \kern-\nulldelimiterspace} \beta },1 - {2 \mathord{\left/
 {\vphantom {2 \beta }} \right.
 \kern-\nulldelimiterspace} \beta }, - {\gamma _{\rm{D}}}} \right)}}
 \end{split}
\end{equation}
\noindent where $\lambda_{\rm{BS}}$ is the density of BSs, ${{\rm{B}}_{\rm{W}}}$ is the transmission bandwidth, ${{\gamma _{\rm{D}}}}$ is the threshold for reliable decoding, $\beta >2$ is the path-loss exponent, ${}_2{F_1}\left( { \cdot , \cdot , \cdot , \cdot } \right)$ is the Gauss hypergeometric function, ${{\rm{P}}_{{\rm{cov}}}}\left( \cdot \right)$ is the coverage probability defined in \cite[Eq. (1)]{AndrewsNov2011}, and (a) follows from \cite[Eq. (8)]{AndrewsNov2011}.

The main strength of \eqref{Eq_1} is its simple closed-form formulation. This is, however, its main limitation as well, especially as far as formulating meaningful system-level EE optimization problems is concerned. Under the standard modeling assumptions, in fact, the network power consumption (Watt/m$^2$) is\footnote{In the present paper, this holds true for Load Model 1 that is introduced in Section \ref{LoadModeling}.} ${{\rm{P}}_{{\rm{grid}}}} = {\lambda _{{\rm{BS}}}}\left( {{{\rm{P}}_{{\rm{tx}}}} + {{\rm{P}}_{{\rm{circ}}}}} \right)$, where ${{{\rm{P}}_{{\rm{tx}}}}}$ is the transmit power of the BSs and ${{{\rm{P}}_{{\rm{circ}}}}}$ is the static power consumption of the BSs, which accounts for the power consumed in all hardware blocks, e.g., analog-to-digital and digital-to-analog converters, analog filters, cooling components, and digital signal processing \cite{AlessioSurvey2016}. The system-level EE (bit/Joule) is defined as the ratio between \eqref{Eq_1} and the network power consumption, i.e., ${\rm{EE}} = {{{\rm{PSE}}} \mathord{\left/ {\vphantom {{{\rm{PSE}}} {{{\rm{P}}_{{\rm{grid}}}}}}} \right. \kern-\nulldelimiterspace} {{{\rm{P}}_{{\rm{grid}}}}}}$. Since the PSE in \eqref{Eq_1} is independent of the transmit power of the BSs, ${{\rm{P}}_{{\rm{tx}}}}$, and the network power consumption, ${{\rm{P}}_{{\rm{grid}}}}$, linearly increases with ${{\rm{P}}_{{\rm{tx}}}}$, we conclude that any EE optimization problems formulated based on \eqref{Eq_1} would result in the trivial optimal solution consisting of turning all the BSs off (the optimal transmit power is zero). In the context of multi-tier cellular networks, a similar conclusion has been obtained in some early papers on system-level EE optimization, e.g., \cite{MariosANDTony2013}, where it is shown that the EE is maximized if all macro BSs operate in sleeping mode. A system-level EE optimization problem formulated based on \eqref{Eq_1} would result, in addition, in a physically meaningless utility function, which provides a non-zero benefit-cost ratio, i.e., a strictly positive EE while transmitting zero power (${\rm{EE}}\left( {{{\rm{P}}_{{\rm{tx}}}} = 0} \right) = {{{\rm{PSE}}} \mathord{\left/ {\vphantom {{{\rm{PSE}}} {\left( {{\lambda _{{\rm{BS}}}}{{\rm{P}}_{{\rm{circ}}}}} \right)}}} \right. \kern-\nulldelimiterspace} {\left( {{\lambda _{{\rm{BS}}}}{{\rm{P}}_{{\rm{circ}}}}} \right)}} >0$). In addition, the EE computed from \eqref{Eq_1} is independent of the density of BSs. We briefly mention here, but will detail it in Section \ref{PSE}, that the load model, i.e., the fully-loaded assumption, determines the conclusion that the EE does not depend on ${{\lambda _{{\rm{BS}}}}}$. This assumption, however, does not affect the conclusion that the optimal ${{{\rm{P}}_{{\rm{tx}}}}}$ is zero. This statement is made more formal in the sequel (see \textit{Proposition \ref{Proposition__PSE}} and \textit{Corollary \ref{Corollary__PSE}}). It is worth nothing that the conclusion that the PSE is independent of ${{{\rm{P}}_{{\rm{tx}}}}}$ is valid regardless of the specific path-loss model being used\footnote{The reader may verify this statement by direct inspection of \eqref{Eq_4}, where ${{{\rm{P}}_{{\rm{tx}}}}}$ cancels out for any path-loss models.}. It depends, on the other hand, on the assumptions of interference-limited operating regime and of having BSs that emit the same ${{{\rm{P}}_{{\rm{tx}}}}}$.

Based on these observations, we conclude that a new analytical formulation of the PSE that explicitly depends on the transmit power and density of the BSs, and that is tractable enough for system-level EE optimization is needed. From an optimization point of view, in particular, it is desirable that the PSE is formulated in a closed-form expression and that the resulting EE function is unimodal and strictly pseudo-concave in the transmit power (given the density) and in the density (given the transmit power) of the BSs. This would imply, e.g., that the first-order derivative of the EE with respect to the transmit power of the BSs (assuming the density given) would have a unique zero, which would be the unique optimal transmit power that maximizes the EE \cite{AlessioMonograph2015}. Similar conclusions would apply to the optimal density of the BSs for a given transmit power. Further details are provided in Section \ref{EEopt}. In this regard, a straightforward approach to overcome the limitations of \eqref{Eq_1} would be to abandon the interference-limited assumption and to take the receiver noise into account. In this case, the PSE would be formulated in terms of a single-integral that, in general, cannot be expressed in closed-form \cite{AndrewsNov2011}, \cite[Eq. (9)]{NoiseBiased2015}. This integral formulation, in particular, results in a system-level EE optimization problem that is not easy to tackle. This approach, in addition, has the inconvenience of formulating the optimization problem for an operating regime where cellular networks are unlikely to operate in practice.
\subsection{State-of-the-Art on System-Level EE Optimization} \label{Introduction_SOTA}
We briefly summarize the most relevant research contributions on energy-aware design and optimization of cellular networks. Due to space limitations, we discuss only the contributions that are closely related to ours. A state-of-the-art survey on EE optimization is available in \cite{AlessioMonograph2015}.

In \cite{MariosANDTony2013}, the authors study the impact of switching some macro BSs off in order to minimize the power consumption under some constraints on the coverage probability. Since the authors rely on the mathematical framework in \eqref{Eq_1}, they conclude that all macro BSs need to be switched off to maximize the EE. In \cite{Unpublished2014}, the author exploits geometric programming to minimize the power consumption of cellular networks given some constraints on the network coverage and capacity. The EE is not studied. A similar optimization problem is studied in \cite{OptBSdensity2013} and \cite{NoiseBiased2015} for two-tier cellular networks but the EE is not studied either. As far as multi-tier cellular networks are concerned, an important remark is necessary. In the interference-limited regime, optimal transmit powers and densities for the different tiers of BSs may exist if the tiers have different thresholds for reliably decoding the data. The PSE, otherwise, is the same as that of single-tier networks, i.e., it is independent of the transmit power and density of the BSs. In \cite{EE_MIMO2014}, the authors study the EE of small cell networks with multi-antenna BSs. For some parameter setups, it is shown that an optimal density of the BSs exists. The EE, however, still decreases monotonically with the transmit power of the BSs, which implies that the EE optimization problem is not well formulated from the transmit power standpoint. More general scenarios are considered in \cite{EE_CellularRepulsive2013}, \cite{PowerRangeAdaptation2013}, \cite{EE_CRNs2013}, \cite{UserBehaviorPPP2014}, \cite{ACM_OptimalCellActivation2014}, \cite{EE_RandomNets2015}-\cite{EE_DiscTx2017}, but similar limitations hold. In some cases, e.g., \cite{EE_WCL2016}, the existence and uniqueness of an optimal transmit power and density of the BSs are not mathematically proved or, e.g., in \cite{EE_Access2017}, the problem formulation has a prohibitive numerical complexity as it necessitates the computation of multiple integrals and infinite series. It is apparent, therefore, that a tractable approach for system-level EE optimization is missing in the open technical literature. In the present paper, we introduce a new definition of PSE that overcomes these limitations.
\subsection{Research Contribution and Novelty} \label{Introduction_Contribution}
In the depicted context, the specific novel contributions made by this paper are as follows:
\begin{itemize}
  \item We introduce a new closed-form analytical formulation of the PSE for interference-limited cellular networks (during data transmission), which depends on the transmit power and density of the BSs. The new expression of the PSE is obtained by taking into account the power sensitivity of the receiver not only for data transmission but for cell association as well.
  \item Based on the new expression of the PSE, a new system-level EE optimization problem is formulated and comprehensively studied. It is mathematically proved that the EE is a unimodal and strictly pseudo-concave function in the transmit power given the BSs' density and in the BSs' density given the transmit power. The dependency of the optimal power as a function of the density and of the optimal density as a function of the power is discussed.
  \item A first-order optimal pair of transmit power and density of the BSs is obtained by using a simple alternating optimization algorithm whose details are discussed in the sequel. Numerical evidence of the global optimality of this approach is provided as well.
  \item Two load models for the BSs are analyzed and compared against each other. It is shown that they provide the same PSE but have different network power consumptions. Hence, the optimal transmit power and density of the BSs that maximize their EEs are, in general, different. Their optimal EEs and PSEs are studied and compared against each other.
\end{itemize}

The paper is organized as follows. In Section \ref{SystemModel}, the system model is presented. In Section \ref{PSE}, the new definition of PSE is introduced. In Section \ref{EEopt}, the EE optimization problem is formulated and studied. In Section \ref{Results}, numerical results are shown. Finally, Section \ref{Conclusion} concludes the paper.

\textit{Notation}: The main symbols and functions used in the present paper are reported in Table \ref{Table_Notation}.
\begin{table*}[!t] 
\centering
\caption{Summary of main symbols and functions used throughout the paper.}
\newcommand{\tabincell}[2]{\begin{tabular}{@{}#1@{}}#2\end{tabular}}
 \begin{tabular}{|l||l|} \hline
\hspace{2.5cm} Symbol/Function & \hspace{3.75cm} Definition \\ \hline \hline
$\mathbb{E}\{\cdot\}$, $\Pr \left\{  \cdot  \right\}$ & Expectation operator, probability measure \\ \hline
$\lambda_{\rm{BS}}$, $\lambda_{\rm{MT}}$ & Density of base stations, mobile terminals \\ \hline
$\Psi_{\rm{BS}}$, $\Psi_{\rm{MT}}$, $\Psi_{\rm{BS}}^{\left( \rm{I} \right)}$ & PPP of base stations, mobile terminals, interfering base stations \\ \hline
${\rm{B}}{{\rm{S}}_{\rm{0}}}$, ${\rm{B}}{{\rm{S}}_{{i}}}$, ${\rm{B}}{{\rm{S}}_{{n}}}$ & Serving, interfering, generic base station \\ \hline
${\rm{P}}_{{\rm{tx}}}$, ${\rm{P}}_{{\rm{circ}}}$, ${\rm{P}}_{{\rm{idle}}}$ & Transmit, circuits, idle power consumption of base stations \\ \hline
$r_n$, $g_n$ & Distance, fading power gain of a generic link \\ \hline
$l \left( \cdot \right)$, $L_n$, $L_0$ & Path-loss, shorthand of path-loss, path-loss of intended link \\ \hline
$\kappa$, $\beta >0$ & Path-loss constant, slope (exponent) \\ \hline
${{{\rm{B}}_{\rm{W}}}}$, ${{\rm{N}}_{\rm{0}}}$ & Transmission bandwidth, noise power spectral density \\ \hline
$\sigma_{\rm{N}}^2 = {{{\rm{B}}_{\rm{W}}}{{\rm{N}}_{\rm{0}}}}$, $I_{{\rm{agg}}} \left(  \cdot  \right)$ & Noise variance, aggregate other-cell interference \\ \hline
${\gamma _{\rm{D}}}$, ${\gamma _{\rm{A}}}$ & Reliability threshold for decoding, cell association \\ \hline
$\mathcal{L}\left( x \right) = 1 - {\left( {1 + x/\alpha } \right)^{ - \alpha }}$, $\alpha=3.5$ & Probability that a base station is in transmission mode \\ \hline
$f_{X}(\cdot)$, $F_X(\cdot)$  & Probability density/mass, cumulative distribution/mass function of $X$ \\ \hline
$\mathbbm{1}\left( \cdot \right)$, ${}_2F_1 \left( { \cdot , \cdot , \cdot , \cdot } \right)$, $\Gamma(\cdot)$ & Indicator function, Gauss hypergeometric function, gamma function \\ \hline
$\max \left\{ {x,y} \right\}$, $\min \left\{ {x,y} \right\}$ & Maximum, minimum between $x$ and $y$ \\ \hline
$\Upsilon  = {}_2{F_1}\left( { - 2/\beta ,1,1 - 2/\beta , - {\gamma _{\rm{D}}}} \right) -1 \ge 0$ & Shorthand \\ \hline
$\mathcal{Q}\left( {x,y,z} \right) = 1 - \exp \left( { - \pi x{{\left( {y/\eta } \right)}^{2/\beta }}\left( {1 + \Upsilon \mathcal{L}\left( z \right)} \right)} \right)$ & Shorthand with $\eta  = \kappa \sigma _{\rm{N}}^2{\gamma _{\rm{A}}}$ \\ \hline
SIR, $\overline {{\rm{SNR}}}$ & Signal-to-interference-ratio, average signal-to-noise-ratio \\ \hline
${{\rm{P}}_{{\rm{cov}}}}$, PSE, ${{\rm{P}}_{{\rm{grid}}}}$ & Coverage, potential spectral efficiency, network power consumption \\ \hline
${{\dt z}_x}\left( {x,y} \right)$, ${{\ddt z}_x}\left( {x,y} \right)$ & First-order, second-order derivative with respect to $x$ \\ \hline
\end{tabular}
\label{Table_Notation}
\end{table*}
\section{System Model} \label{SystemModel}
In this section, the network model is introduced. With the exception of the load model, we focus our attention on a system where the \textit{standard modeling assumptions} hold. One of the main aims of the present paper is, in fact, to highlight the differences between currently available analytical frameworks and the new definition of PSE that is introduced. The proposed approach can be readily generalized to more advanced system models, such as that recently adopted in \cite{MDR_IM}.
\subsection{Cellular Network Modeling} \label{CellularNetworkModeling}
A downlink cellular network is considered. The BSs are modeled as points of a homogeneous PPP, denoted by $\Psi_{\rm{BS}}$, of density $\lambda_{\rm{BS}}$. The MTs are modeled as another homogeneous PPP, denoted by $\Psi_{\rm{MT}}$, of density $\lambda_{\rm{MT}}$. $\Psi_{\rm{BS}}$ and $\Psi_{\rm{MT}}$ are independent of each other. The BSs and MTs are equipped with a single omnidirectional antenna. Each BS transmits with a constant power denoted by ${\rm{P}}_{\rm{tx}}$. The analytical frameworks are developed for the typical MT, denoted by ${\rm{MT}}_{0}$, that is located at the origin (Slivnyak theorem \cite[Th. 1.4.5]{BaccelliBook2009}). The BS serving ${\rm{MT}}_{0}$ is denoted by ${\rm{BS}}_{0}$. The cell association criterion is introduced in Section \ref{CellAssociationCriterion}. The subscripts ${0}$, ${i}$ and ${n}$ identify the intended link, a generic interfering link, and a generic BS-to-MT link. The set of interfering BSs is denoted by $\Psi_{\rm{BS}} ^{\left( \rm{I} \right)}$. As for data transmission, the network operates in the interference-limited regime, i.e., the noise is negligible compared with the inter-cell interference.
\subsection{Channel Modeling} \label{ChannelModeling}
For each BS-to-MT link, path-loss and fast-fading are considered. Shadowing is not explicitly taken into account because its net effect lies in modifying the density of the BSs \cite{MDR_IM}. All BS-to-MT links are assumed to be mutually independent and identically distributed (i.i.d.).
\paragraph{Path-Loss}
Consider a generic BS-to-MT link of length $r_{n}$. The path-loss is $l \left( r_{n} \right) = \kappa r_n^{\beta}$, where $\kappa$ and $\beta$ are the path-loss constant and the path-loss slope (exponent). For simplicity, only the unbounded path-loss model is studied in the present paper. The analysis of more general path-loss models is an interesting but challenging generalization that is left to future research \cite{COMML_UnboundedPathLoss}.
\paragraph{Fast-Fading}
Consider a generic BS-to-MT link. The power gain due to small-scale fading is assumed to follow an exponential distribution with mean $\Omega$. Without loss of generality, $\Omega = 1$ is assumed. The power gain of a generic BS-to-MT link is denoted by $g_n$.
\subsection{Cell Association Criterion} \label{CellAssociationCriterion}
A cell association criterion based on the highest average received power is assumed. Let ${\rm{BS}}_n \in {\Psi _{{\rm{BS}}}}$ denote a generic BS of the network. The serving BS, ${\rm{BS}}_{0}$, is obtained as follows:
\setcounter{equation}{1}
\begin{equation}
\label{Eq_2}
\begin{split}
{\rm{B}}{{\rm{S}}_{\rm{0}}} & = {\arg \max }_{{{\rm{BS}}_n} \in {\Psi _{{\rm{BS}}}}} \left\{ {{1 \mathord{\left/
 {\vphantom {1 {l\left( {{r_n}} \right)}}} \right.
 \kern-\nulldelimiterspace} {l\left( {{r_n}} \right)}}} \right\} \\ & = {\arg \max }_{{{\rm{BS}}_n} \in {\Psi _{{\rm{BS}}}}} \left\{ {{1 \mathord{\left/
 {\vphantom {1 {{L_n}}}} \right.
 \kern-\nulldelimiterspace} {{L_n}}}} \right\}
\end{split}
\end{equation}
\noindent where the shorthand ${L_n} = l\left( {{r_n}} \right)$ is used. As for the intended link, ${L_0} = {\min }_{{r_n} \in {\Psi _{{\rm{BS}}}}} \left\{ {{L_n}} \right\}$ holds.
\subsection{Load Modeling} \label{LoadModeling}
Based on \eqref{Eq_2}, several or no MTs can be associated to a generic BS. In the latter case, the BS transmits zero power, i.e., ${{\mathop{\rm P}\nolimits} _{\rm{tx}}} = 0$, and, thus, it does not generate inter-cell interference. In the former case, on the other hand, two load models are studied and compared against each other. The main objective is to analyze the impact of the load model on the power consumption and EE of cellular networks. Further details are provided in the sequel. Let ${{\rm{N}}_{{\rm{MT}}}}$ denote the number of MTs associated to a generic BS and ${{\rm{B}}_{\rm{W}}}$ denote the transmission bandwidth available to each BS. If ${{\rm{N}}_{{\rm{MT}}}} = 1$, for both load models, the single MT associated to the BS is scheduled for transmission and the entire bandwidth, ${{\rm{B}}_{\rm{W}}}$, and transmit power, ${{\rm{P}}_{{\rm{tx}}}}$, are assigned to it.
\paragraph{Load Model 1: Exclusive Allocation of Bandwidth and Power to a Randomly Selected MT}
If ${{\rm{N}}_{{\rm{MT}}}} > 1$, the BS randomly selects, at each transmission instance, a single MT among the ${{\rm{N}}_{{\rm{MT}}}}$ associated to it. Also, the BS allocates the entire transmission bandwidth, ${{\rm{B}}_{\rm{W}}}$, and the total transmit power, ${{\rm{P}}_{{\rm{tx}}}}$, to it. The random scheduling of the MTs at each transmission instance ensures that, in the long term, all the MTs associated to a BS are scheduled for transmission.
\paragraph{Load Model 2: Equal Allocation of Bandwidth and Power Among All the MTs}
If ${{\rm{N}}_{{\rm{MT}}}} > 1$, the BS selects, at each transmission instance, all the ${{\rm{N}}_{{\rm{MT}}}}$ MTs associated to it. The BS equally splits the available transmission bandwidth, ${{\rm{B}}_{\rm{W}}}$, and evenly spreads the available transmit power, ${{\rm{P}}_{{\rm{tx}}}}$, among the ${{\rm{N}}_{{\rm{MT}}}}$ MTs. Thus, the bandwidth and power are viewed as continuous resources by the BS's scheduler: each MT is assigned a bandwidth equal to ${{{{\rm{B}}_{\rm{W}}}} \mathord{\left/ {\vphantom {{{{\rm{B}}_{\rm{W}}}} {{{\rm{N}}_{{\rm{MT}}}}}}} \right. \kern-\nulldelimiterspace} {{{\rm{N}}_{{\rm{MT}}}}}}$ and the power spectral density at the detector's (i.e., the typical MT, ${{\rm{MT}}}_0$) input is equal to ${{{{\rm{P}}_{{\rm{tx}}}}} \mathord{\left/ {\vphantom {{{{\rm{P}}_{{\rm{tx}}}}} {{{\rm{B}}_{\rm{W}}}}}} \right. \kern-\nulldelimiterspace} {{{\rm{B}}_{\rm{W}}}}}$.

In the sequel, we show that the main difference between the two load models lies in the power consumption of the BSs. In simple terms, the more MTs are scheduled for transmission the higher the static power consumption of the BSs is. The analysis of general load models, e.g., based on a discrete number of resource blocks \cite{MDR_IM}, is left to future research due to space limits.
\subsection{Power Consumption Modeling} \label{PowerConsumptionModeling}
In the considered system model, the BSs can operate in two different modes: i) they are in idle mode if no MTs are associated to them and ii) they are in transmission mode if at least one MT is associated to them. The widespread linear power consumption model for the BSs is adopted \cite{AlessioSurvey2016}, \cite{EE_EARTH}, which accounts for the power consumption due to the transmit power, ${{{\rm{P}}_{{\rm{tx}}}}}$, the static (circuit) power, ${{{\rm{P}}_{{\rm{circ}}}}}$, and the idle power, ${{{\rm{P}}_{{\rm{idle}}}}}$. If the BS is in idle mode, its power consumption is equal to ${{{\rm{P}}_{{\rm{idle}}}}}$. If the BS is in transmission mode, its power consumption is a function of ${{{\rm{P}}_{{\rm{tx}}}}}$, ${{{\rm{P}}_{{\rm{circ}}}}}$, and depends on the load model. Further details are provided in the sequel. In the present paper, based on physical considerations, the inequalities $0 \le {{\rm{P}}_{{\rm{idle}}}} \le {{\rm{P}}_{{\rm{circ}}}}$ are assumed.
\section{A New Analytical Formulation of the PSE} \label{PSE}
In this section, we introduce and motivate a new definition of coverage probability, ${{\rm{P}}_{{\rm{cov}}}}$, and PSE, which overcomes the limitations of currently available analytical frameworks and is suitable for system-level optimization (see Section \ref{Introduction_EE}). All symbols are defined in Table \ref{Table_Notation}.
\begin{definition} \label{Definition__Pcov}
Let ${{\gamma _{\rm{D}}}}$ and ${{\gamma _{\rm{A}}}}$ be the reliability thresholds for the successful decoding of information data and for the successful detection of the serving BS, ${\rm{B}}{{\rm{S}}_{\rm{0}}}$, respectively. The coverage probability, ${{\rm{P}}_{{\rm{cov}}}}$, of the typical MT, ${\rm{M}}{{\rm{T}}_{\rm{0}}}$, is defined as follows:
\setcounter{equation}{2}
\begin{equation}
\label{Eq_3}
\begin{split}
& {{\rm{P}}_{{\rm{cov}}}}\left( {{\gamma _{\rm{D}}},{\gamma _{\rm{A}}}} \right) \\ & = \begin{cases}
\Pr \left\{ {{\rm{SIR}} \ge {\gamma _{\rm{D}}},\overline {{\rm{SNR}}}  \ge {\gamma _{\rm{A}}}} \right\}\quad & {\rm{if}}\quad {\rm{M}}{{\rm{T}}_{\rm{0}}}{\rm{\; is \; selected}}\\
0\quad & {\rm{if}}\quad {\rm{M}}{{\rm{T}}_{\rm{0}}}{\rm{\; is \; not \; selected}}
\end{cases}
\end{split}
\end{equation}
\noindent where the Signal-to-Interference-Ratio (${{\rm{SIR}}}$) and the average Signal-to-Noise-Ratio (${\overline {{\rm{SNR}}} }$) can be formulated, for the network model under analysis, as follows:
\setcounter{equation}{3}
\begin{equation}
\label{Eq_4}
\begin{split}
& {\rm{SIR}} = \frac{{{{{{\rm{P}}_{{\rm{tx}}}}{g_0}} \mathord{\left/
 {\vphantom {{{{\rm{P}}_{{\rm{tx}}}}{g_0}} {{L_0}}}} \right.
 \kern-\nulldelimiterspace} {{L_0}}}}}{{\sum\nolimits_{{\rm{B}}{{\rm{S}}_i} \in {\Psi_{\rm{BS}}^{\left( \rm{I} \right)}}} {{{{{\rm{P}}_{{\rm{tx}}}}{g_i}} \mathord{\left/
 {\vphantom {{{{\rm{P}}_{{\rm{tx}}}}{g_i}} {{L_i}}}} \right.
 \kern-\nulldelimiterspace} {{L_i}}}{\mathbbm{1}}\left( {{L_i} > {L_0}} \right)} }} \\
 & \overline {{\rm{SNR}}}  = \frac{{{{{{\rm{P}}_{{\rm{tx}}}}} \mathord{\left/
 {\vphantom {{{{\rm{P}}_{{\rm{tx}}}}} {{L_0}}}} \right.
 \kern-\nulldelimiterspace} {{L_0}}}}}{{\sigma _{\rm{N}}^2}}.
 \end{split}
\end{equation}
\end{definition}
\begin{remark} \label{Remark__Pcov}
The definition of ${{\rm{P}}_{{\rm{cov}}}}$ in \eqref{Eq_3} reduces to the conventional one if ${\gamma _{\rm{A}}} = 0$ \cite{AndrewsNov2011}.  \hfill $\Box$
\end{remark}
\begin{remark} \label{Remark__AverageSNR}
The average SNR, $\overline {{\rm{SNR}}}$, in \eqref{Eq_4} is averaged with respect to the fast fading. The SIR depends, on the other hand, on fast fading. This choice is discussed in the sequel. \hfill $\Box$
\end{remark}
\begin{remark} \label{Remark__3GPP}
The new definition of coverage probability, ${{\rm{P}}_{{\rm{cov}}}}$, in \eqref{Eq_3} is in agreement with the cell selection criterion specified by the 3rd Generation Partnership Project (3GPP) \cite[Sec. 5.2.3.2]{3GPP_TS36.304}. \hfill $\Box$
\end{remark}
\begin{figure}[!t]
\centering
\includegraphics[width=\columnwidth]{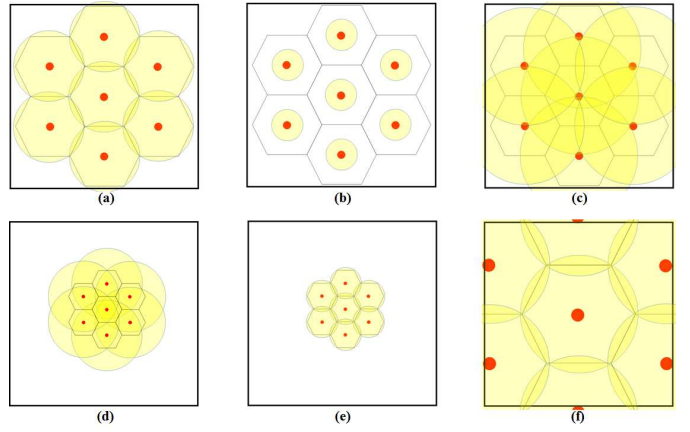}
\caption{Illustration of the interplay between ${{\rm{P}}_{{\rm{tx}}}}$ and ${{\lambda _{{\rm{BS}}}}}$. For simplicity, only a cluster of seven BSs is represented by keeping the size of the region of interest (square box) the same. The inter-site distance of the BSs (represented as red dots), i.e., the size of the hexagonal cells, is determined by ${{\lambda _{{\rm{BS}}}}}$. The shape of the cells depends on the cell association in \eqref{Eq_2}. The circular shaded disk (in light yellow) represents the actual coverage region of the BSs that is determined by ${{\rm{P}}_{{\rm{tx}}}}$: i) a MT inside the disk receives a sufficiently good signal to detect the BS and to get associated with it, ii) a MT outside the disk cannot detect the BS and is not in coverage. The sub-figures (a)-(c) are obtained by assuming the same ${{\lambda _{{\rm{BS}}}}}$ but a different ${{\rm{P}}_{{\rm{tx}}}}$. The sub-figures (d) and (e) are obtained by considering a ${{\lambda _{{\rm{BS}}}}}$ greater than that of sub-figures (a)-(c) but keeping the same ${{\rm{P}}_{{\rm{tx}}}}$ as sub-figures (a) and (b), respectively. The sub-figure (f) is obtained by considering a ${{\lambda _{{\rm{BS}}}}}$ smaller than that of sub-figure (c) but keeping the same ${{\rm{P}}_{{\rm{tx}}}}$ as it. We observe that, for a given ${{\lambda _{{\rm{BS}}}}}$, the transmit power ${{\rm{P}}_{{\rm{tx}}}}$ is appropriately chosen in sub-figures (a), (e) and (f). ${{\rm{P}}_{{\rm{tx}}}}$ is, on the other hand, under-provisioned in sub-figure (b) and over-provisioned in sub-figures (c) and (d). In the first case, the MTs are not capable of detecting the BS throughout the entire cell, i.e., a high outage probability is expected. In the second case, the BSs emit more power than what is actually needed, which results in a high power consumption.} \label{Fig_1}
\end{figure}
\paragraph{Motivation for the New Definition of ${{\rm{P}}_{{\rm{cov}}}}$}
The motivation for the new definition of coverage probability originates from the inherent limitations of the conventional definition (obtained by setting ${\gamma _{\rm{A}}} = 0$ in \eqref{Eq_3}), which prevents one from taking into account the strong interplay between the transmit power and the density of the BSs for optimal cellular networks planning. In fact, the authors of \cite{AndrewsNov2011} have shown that, in the interference-limited regime, ${{\rm{P}}_{{\rm{cov}}}}$ is independent of the transmit power of the BSs. If, in addition, a fully-loaded model is assumed, i.e., ${{{\lambda _{{\rm{MT}}}}} \mathord{\left/ {\vphantom {{{\lambda _{{\rm{MT}}}}} {{\lambda _{{\rm{BS}}}}}}} \right. \kern-\nulldelimiterspace} {{\lambda _{{\rm{BS}}}}}} \gg 1$, then ${{\rm{P}}_{{\rm{cov}}}}$ is independent of the density of BSs as well. This is known as the invariance property of ${{\rm{P}}_{{\rm{cov}}}}$ as a function of ${{\rm{P}}_{{\rm{tx}}}}$ and ${{\lambda _{{\rm{BS}}}}}$ \cite{MDR_IM}. The tight interplay between ${{\rm{P}}_{{\rm{tx}}}}$ and ${{\lambda _{{\rm{BS}}}}}$ is, on the other hand, illustrated in Fig. \ref{Fig_1}, where, for ease of representation, an hexagonal cellular layout is considered. Similar conclusions apply to the PPP-based cellular layout studied in the present paper. In Fig. \ref{Fig_1}, it is shown that, for a given ${{\lambda _{{\rm{BS}}}}}$, ${{\rm{P}}_{{\rm{tx}}}}$ needs to be appropriately chosen in order to guarantee that, for any possible location of ${\rm{MT}_0}$ in the cell, two conditions are fulfilled: i) the MT receives a sufficiently good signal quality, i.e., the average SNR is above a given threshold, ${\gamma _{\rm{A}}}$, that ensures a successful cell association, i.e., to detect the presence (pilot signal) of the serving BS and ii) the BSs do not over-provision ${{\rm{P}}_{{\rm{tx}}}}$, which results in an unnecessary increase of the power consumption. It is expected, therefore, that an optimal value of ${{\rm{P}}_{{\rm{tx}}}}$ given ${{\lambda _{{\rm{BS}}}}}$ and an optimal value of ${{\lambda _{{\rm{BS}}}}}$ given ${{\rm{P}}_{{\rm{tx}}}}$ that optimize EE exist \cite{CellBreathing2010}.
\paragraph{Advantages of the New Definition of ${{\rm{P}}_{{\rm{cov}}}}$} The new definition of ${{\rm{P}}_{{\rm{cov}}}}$ allows one to overcome the limitations of the conventional definition and brings about two main advantages. The first advantage originates from direct inspection of \eqref{Eq_4}. In the conventional definition of ${{\rm{P}}_{{\rm{cov}}}}$, only the SIR is considered and the transmit power of the BSs, ${{\rm{P}}_{{\rm{tx}}}}$, cancels out between numerator and denominator. This is the reason why ${{\rm{P}}_{{\rm{cov}}}}$ is independent of ${{\rm{P}}_{{\rm{tx}}}}$. In the proposed new definition, on the other hand, ${{\rm{P}}_{{\rm{tx}}}}$ explicitly appears in the second constraint and does not cancel out. The density of the BSs, ${{\lambda _{{\rm{BS}}}}}$, appears implicitly in the distribution of the path-loss of the intended link, $L_0$. The mathematical details are provided in the sequel. The second inequality, as a result, allows one to explicitly account for the interplay between ${{\rm{P}}_{{\rm{tx}}}}$ and ${{\lambda _{{\rm{BS}}}}}$ (shown in Fig. \ref{Fig_1}). If ${{\lambda _{{\rm{BS}}}}}$ increases (decreases), in particular, $L_0$ decreases (increases) in statistical terms. This implies that ${{\rm{P}}_{{\rm{tx}}}}$ can be decreased (increased) while still ensuring that the average SNR is above ${{\gamma _{\rm{A}}}}$. The second advantage is that the new definition of ${{\rm{P}}_{{\rm{cov}}}}$ is still mathematically tractable and the PSE is formulated in a closed-form expression. This is detailed in \textit{Proposition \ref{Proposition__PSE}}.
\begin{remark} \label{Remark__L0}
The new definition of ${{\rm{P}}_{{\rm{cov}}}}$ in \eqref{Eq_3} is based on the actual value of $L_0$ because a necessary condition for the typical MT to be in coverage is that it can detect the pilot signal of at least one BS during the cell association. If the BS that provides the highest average received power cannot be detected, then any other BSs cannot be detected either. The second constraint on the definition of ${{\rm{P}}_{{\rm{cov}}}}$, in addition, is based on the average SNR, i.e., the SNR averaged with respect to the fast fading, because the cell association is performed based on long-term statistics, i.e., based on the path-loss in the present paper, in order to prevent too frequent handovers. \hfill $\Box$
\end{remark}
\begin{remark} \label{Remark__ApproximationSINR}
Compared with the conventional definition of coverage based on the Signal-to-Interference+Noise-Ratio (SINR) \cite{AndrewsNov2011}, the new definition in \eqref{Eq_3} is conceptually different. Equation \eqref{Eq_3} accounts for the signal quality during both the cell association and data transmission phases. The definition of coverage based on the SINR, on the other hand, accounts for the signal quality only during the data transmission phase. In spite of this fundamental difference, ${{\rm{P}}_{{\rm{cov}}}}$ in \eqref{Eq_3} may be interpreted as an approximation for the coverage probability based on the SINR, and, more precisely, as an alternative method to incorporate the thermal noise into the problem formulation. Compared with the coverage based on the SINR, however, the new definition in \eqref{Eq_3} accounts for the impact of thermal noise when it is the dominant factor, i.e., during the cell association phase when the inter-cell interference can be ignored as orthogonal pilot signals are used. \hfill $\Box$
\end{remark}
\begin{remark} \label{Remark__3GPP_BroaderScope}
Figure \ref{Fig_1} highlights that the new definition of coverage in \eqref{Eq_3} is not only compliant with \cite{3GPP_TS36.304} but it has a more profound motivation and wider applicability. In PPP-based cellular networks, in contrast to regular grid-based network layouts, the size and shape of the cells are random. This implies that it is not possible to identify a relation, based on pure geometric arguments, between the cell size and the transmit power of the BSs that makes the constraint on $\overline {{\rm{SNR}}}$ in \eqref{Eq_3} ineffective in practice. In equivalent terms, in this case, the threshold ${\gamma _{\rm{A}}}$ may turn out to be sufficiently small to render the constraint on $\overline {{\rm{SNR}}}$ ineffective. This is, e.g., the approach employed in \cite[Eq. (1)]{CellBreathing2010}, where the relation between the transmit power and density of BSs is imposed a priori based on the path-loss. In practice, however, cellular networks are irregularly deployed, which makes the optimal relation between the transmit power and density of BSs difficult to identify because of the coexistence of cells of small and large sizes. The constraint on $\overline {{\rm{SNR}}}$ in \eqref{Eq_3} allows one to take into account the interplay between the transmit power and density of BSs in irregular (realistic) cellular network deployments. \hfill $\Box$
\end{remark}
\subsection{Analytical Formulation of the PSE} \label{PSE_Definition}
In this section, we provide the mathematical definitions of the PSE for the two load models introduced in Section \ref{LoadModeling}. They are summarized in the following two lemmas, which constitute the departing point to obtain the closed-form analytical frameworks derived in Section \ref{PSE_Pgrid_ClosedFormFrameworks}.
\begin{remark} \label{Remark__CroftonCell}
The PSE is defined from the perspective of the typical MT, ${{\rm{MT}}_{\rm{0}}}$ rather than from the perspective of the typical cell (or BS). This implies that the proposed approach allows one to characterize the PSE of the so-called Crofton cell, which is the cell that contains ${{\rm{MT}}_{\rm{0}}}$. This approach is commonly used in the literature and is motivated by the lack of results on the explicit distribution of the main geometrical characteristics of the typical cell of a Voronoi tessellation. Further details on the Crofton and typical cells are available in \cite{Sayan_Crofton} and \cite{Martin_Crofton}. \hfill $\Box$
\end{remark}

Let ${\bar {\rm{ N}}_{{\rm{MT}}}}$ be the number of MTs that lie in the cell of the typical MT, ${{\rm{MT}}_{\rm{0}}}$, with the exception of ${{\rm{MT}}_{\rm{0}}}$. ${\bar {\rm{ N}}_{{\rm{MT}}}}$ is a discrete random variable whose probability mass function in the considered system model can be formulated, in an approximated closed-form expression, as \cite[Eq. (3)]{WiOPT_2013}:
\setcounter{equation}{4}
\begin{equation}
\label{Eq_5}
\begin{split}
{f_{{{{\rm{\bar N}}}_{{\rm{MT}}}}}}\left( u \right) & = \Pr \left\{ {{{{\rm{\bar N}}}_{{\rm{MT}}}} = u} \right\} \\ & \approx \frac{{{{3.5}^{4.5}}\Gamma \left( {u + 4.5} \right){{\left( {{{{\lambda _{{\rm{MT}}}}} \mathord{\left/
 {\vphantom {{{\lambda _{{\rm{MT}}}}} {{\lambda _{{\rm{BS}}}}}}} \right.
 \kern-\nulldelimiterspace} {{\lambda _{{\rm{BS}}}}}}} \right)}^u}}}{{\Gamma \left( {4.5} \right)\Gamma \left( {u + 1} \right){{\left( {{{3.5 + {\lambda _{{\rm{MT}}}}} \mathord{\left/
 {\vphantom {{3.5 + {\lambda _{{\rm{MT}}}}} {{\lambda _{{\rm{BS}}}}}}} \right.
 \kern-\nulldelimiterspace} {{\lambda _{{\rm{BS}}}}}}} \right)}^{u + 4.5}}}}.
\end{split}
\end{equation}
\begin{remark} \label{Remark__ExactAreaOfVoronoiCell}
The probability mass function in \eqref{Eq_5} is an approximation because it is based on the widely used empirical expression of the probability density function of the area of the Voronoi cells in \cite[Eq. (1)]{CurveFitting_Voronoi}. A precise formula for the latter probability density function is available in \cite{Calka_Voronoi}. It is, however, not used in the present paper due to its mathematical intractability, as recently remarked in \cite{Alam_Voronoi}. Throughout the rest of the paper, for simplicity, we employ the sign of equality (``='') in all the analytical formulas that rely \textit{solely} on the approximation in \eqref{Eq_5}. This is to make explicit that our analytical frameworks are not based on any other hidden approximations. \hfill $\Box$
\end{remark}
\begin{figure*}[!t]
\setcounter{equation}{5}
\begin{equation}
\label{Eq_6}
\begin{split}
{\rm{PSE}}\left( {{\gamma _{\rm{D}}},{\gamma _{\rm{A}}}} \right) & = {{\mathbb{E}}_{{{{\rm{\bar N}}}_{{\rm{MT}}}}}}\left\{ {{\rm{PSE}}\left( {\left. {{\gamma _{\rm{D}}},{\gamma _{\rm{A}}}} \right|{{{\rm{\bar N}}}_{{\rm{MT}}}}} \right)} \right\}\\
& \mathop  = \limits^{\left( a \right)} {\lambda _{{\rm{MT}}}}{{\rm{B}}_{\rm{W}}}{\log _2}\left( {1 + {\gamma _{\rm{D}}}} \right)\Pr \left\{ {{\rm{SIR}} \ge {\gamma _{\rm{D}}},\overline {{\rm{SNR}}}  \ge {\gamma _{\rm{A}}}} \right\}\Pr \left\{ {{{{\rm{\bar N}}}_{{\rm{MT}}}} = 0} \right\}\\
& + \sum\limits_{u = 1}^{ + \infty } {{\lambda _{{\rm{MT}}}}{{\rm{B}}_{\rm{W}}}{{\log }_2}\left( {1 + {\gamma _{\rm{D}}}} \right)\frac{1}{{u + 1}}\Pr \left\{ {{\rm{SIR}} \ge {\gamma _{\rm{D}}},\overline {{\rm{SNR}}}  \ge {\gamma _{\rm{A}}}} \right\}\Pr \left\{ {{{{\rm{\bar N}}}_{{\rm{MT}}}} = u} \right\}} \\
& = {\lambda _{{\rm{MT}}}}{{\rm{B}}_{\rm{W}}}{\log _2}\left( {1 + {\gamma _{\rm{D}}}} \right)\Pr \left\{ {{\rm{SIR}} \ge {\gamma _{\rm{D}}},\overline {{\rm{SNR}}}  \ge {\gamma _{\rm{A}}}} \right\}\sum\limits_{u = 0}^{ + \infty } {\frac{{\Pr \left\{ {{{{\rm{\bar N}}}_{{\rm{MT}}}} = u} \right\}}}{{u + 1}}}.
\end{split}
\end{equation}
\normalsize \hrulefill \vspace*{-0pt}
\end{figure*}
\begin{figure*}[!t]
\setcounter{equation}{6}
\begin{equation}
\label{Eq_7}
\begin{split}
{\rm{PSE}}\left( {{\gamma _{\rm{D}}},{\gamma _{\rm{A}}}} \right) & = {{\mathbb{E}}_{{{{\rm{\bar N}}}_{{\rm{MT}}}}}}\left\{ {{\rm{PSE}}\left( {\left. {{\gamma _{\rm{D}}},{\gamma _{\rm{A}}}} \right|{{{\rm{\bar N}}}_{{\rm{MT}}}}} \right)} \right\}\\
& \mathop  = \limits^{\left( b \right)} \sum\limits_{u = 0}^{ + \infty } {{\lambda _{{\rm{MT}}}}\frac{{{{\rm{B}}_{\rm{W}}}}}{{u + 1}}{{\log }_2}\left( {1 + {\gamma _{\rm{D}}}} \right)\Pr \left\{ {{\rm{SIR}} \ge {\gamma _{\rm{D}}},\overline {{\rm{SNR}}}  \ge {\gamma _{\rm{A}}}} \right\}\Pr \left\{ {{{{\rm{\bar N}}}_{{\rm{MT}}}} = u} \right\}} \\
& = {\lambda _{{\rm{MT}}}}{{\rm{B}}_{\rm{W}}}{\log _2}\left( {1 + {\gamma _{\rm{D}}}} \right)\Pr \left\{ {{\rm{SIR}} \ge {\gamma _{\rm{D}}},\overline {{\rm{SNR}}}  \ge {\gamma _{\rm{A}}}} \right\}\sum\limits_{u = 0}^{ + \infty } {\frac{{\Pr \left\{ {{{{\rm{\bar N}}}_{{\rm{MT}}}} = u} \right\}}}{{u + 1}}}.
\end{split}
\end{equation}
\normalsize \hrulefill \vspace*{-0pt}
\end{figure*}
Based on \eqref{Eq_5}, a formal mathematical formulation for the PSE is given as follows.
\begin{lemma} \label{Lemma__PSE_LoadModel1}
Let \textit{Load Model 1} be assumed. The PSE (bit/sec/m$^2$) can be formulated as shown in \eqref{Eq_6} at the top of this page.

\emph{Proof}: It follows from the definition of PSE \cite{MDR_IM}, where (a) originates from the fact that ${{\rm{MT}}_{\rm{0}}}$ is scheduled for transmission with unit probability if it is the only MT in the cell, while it is scheduled for transmission with probability ${1 \mathord{\left/ {\vphantom {1 {\left( {u + 1} \right)}}} \right. \kern-\nulldelimiterspace} {\left( {u + 1} \right)}}$ if there are other $u$ MTs in the cell. \hfill $\Box$
\end{lemma}
\begin{lemma} \label{Lemma__PSE_LoadModel2}
Let \textit{Load Model 2} be assumed. The PSE (bit/sec/m$^2$) can be formulated as shown in \eqref{Eq_7} at the top of this page.

\emph{Proof}: It follows from the definition of PSE \cite{MDR_IM}, where (b) originates from the fact that ${{\rm{MT}}_{\rm{0}}}$ is scheduled for transmission with unit probability but the bandwidth is equally allocated among the MTs in the cell, i.e., each of the $u+1$ MTs is given a bandwidth equal to ${{{{\rm{B}}_{\rm{W}}}} \mathord{\left/ {\vphantom {{{{\rm{B}}_{\rm{W}}}} {\left( {u + 1} \right)}}} \right. \kern-\nulldelimiterspace} {\left( {u + 1} \right)}}$. \hfill $\Box$
\end{lemma}
\begin{remark} \label{Remark__SamePSE}
By comparing \eqref{Eq_6} and \eqref{Eq_7}, we note that the same PSE is obtained for both load models. This originates from the fact that ${{\rm{P}}_{{\rm{cov}}}}$ in \eqref{Eq_3} is independent of the number of MTs in the cell. This property follows by direct inspection of \eqref{Eq_4} and has been used in the proof of \textit{Lemma \ref{Lemma__PSE_LoadModel1}} and \textit{Lemma \ref{Lemma__PSE_LoadModel2}}. As far as the first load model is concerned, this property originates from the fact that a single MT is scheduled at every transmission instance. It is, however, less intuitive for the second load model. In this latter case, as mentioned in Section \ref{LoadModeling}, ${{\rm{P}}_{{\rm{tx}}}}$ and ${{{\rm{B}}_{\rm{W}}}}$ are viewed as continuous resources by the BS's scheduler. The transmit power per unit bandwidth of both intended and interfering links is equal to ${{{{\rm{P}}_{{\rm{tx}}}}} \mathord{\left/ {\vphantom {{{{\rm{P}}_{{\rm{tx}}}}} {{{\rm{B}}_{\rm{W}}}}}} \right. \kern-\nulldelimiterspace} {{{\rm{B}}_{\rm{W}}}}}$. Regardless of the number of MTs available in the interfering cells, ${{\rm{MT}}_{\rm{0}}}$ ``integrates'' this transmit power per unit bandwidth over the bandwidth allocated to it, which depends on the total number of MTs in its own cell. Let the number of these MTs be $u+1$. Thus, the receiver bandwidth of ${{\rm{MT}}_{\rm{0}}}$ is ${{{{\rm{B}}_{\rm{W}}}} \mathord{\left/ {\vphantom {{{{\rm{B}}_{\rm{W}}}} {\left( {u + 1} \right)}}} \right. \kern-\nulldelimiterspace} {\left( {u + 1} \right)}}$. This implies that the received power (neglecting path-loss and fast-fading) of both intended and interfering links is ${{\rm{P}}_{{\rm{rx}}}} = \left( {{{{{\rm{P}}_{{\rm{tx}}}}} \mathord{\left/{\vphantom {{{{\rm{P}}_{{\rm{tx}}}}} {{{\rm{B}}_{\rm{W}}}}}} \right. \kern-\nulldelimiterspace} {{{\rm{B}}_{\rm{W}}}}}} \right)\left( {{{{{\rm{B}}_{\rm{W}}}} \mathord{\left/ {\vphantom {{{{\rm{B}}_{\rm{W}}}} {\left( {u + 1} \right)}}} \right. \kern-\nulldelimiterspace} {\left( {u + 1} \right)}}} \right) = {{{{\rm{P}}_{{\rm{tx}}}}} \mathord{\left/ {\vphantom {{{{\rm{P}}_{{\rm{tx}}}}} {\left( {u + 1} \right)}}} \right. \kern-\nulldelimiterspace} {\left( {u + 1} \right)}}$. As a result, the number of MTs, $u+1$, cancels out in the SIR of \eqref{Eq_4}. Likewise, the received average SNR (neglecting the path-loss) is equal to ${{{{\rm{P}}_{{\rm{rx}}}}} \mathord{\left/ {\vphantom {{{{\rm{P}}_{{\rm{rx}}}}} {\left( {{{{{\rm{N}}_{\rm{0}}}{{\rm{B}}_{\rm{W}}}} \mathord{\left/ {\vphantom {{{{\rm{N}}_{\rm{0}}}{{\rm{B}}_{\rm{W}}}} {\left( {u + 1} \right)}}} \right. \kern-\nulldelimiterspace} {\left( {u + 1} \right)}}} \right)}}} \right. \kern-\nulldelimiterspace} {\left( {{{{{\rm{N}}_{\rm{0}}}{{\rm{B}}_{\rm{W}}}} \mathord{\left/ {\vphantom {{{{\rm{N}}_{\rm{0}}}{{\rm{B}}_{\rm{W}}}} {\left( {u + 1} \right)}}} \right. \kern-\nulldelimiterspace} {\left( {u + 1} \right)}}} \right)}} = {{\left( {{{{{\rm{P}}_{{\rm{tx}}}}} \mathord{\left/ {\vphantom {{{{\rm{P}}_{{\rm{tx}}}}} {\left( {u + 1} \right)}}} \right. \kern-\nulldelimiterspace} {\left( {u + 1} \right)}}} \right)} \mathord{\left/ {\vphantom {{\left( {{{{{\rm{P}}_{{\rm{tx}}}}} \mathord{\left/ {\vphantom {{{{\rm{P}}_{{\rm{tx}}}}} {\left( {u + 1} \right)}}} \right. \kern-\nulldelimiterspace} {\left( {u + 1} \right)}}} \right)} {\left( {{{{{\rm{N}}_{\rm{0}}}{{\rm{B}}_{\rm{W}}}} \mathord{\left/ {\vphantom {{{{\rm{N}}_{\rm{0}}}{{\rm{B}}_{\rm{W}}}} {\left( {u + 1} \right)}}} \right. \kern-\nulldelimiterspace} {\left( {u + 1} \right)}}} \right)}}} \right. \kern-\nulldelimiterspace} {\left( {{{{{\rm{N}}_{\rm{0}}}{{\rm{B}}_{\rm{W}}}} \mathord{\left/ {\vphantom {{{{\rm{N}}_{\rm{0}}}{{\rm{B}}_{\rm{W}}}} {\left( {u + 1} \right)}}} \right. \kern-\nulldelimiterspace} {\left( {u + 1} \right)}}} \right)}} = {{{{\rm{P}}_{{\rm{tx}}}}} \mathord{\left/ {\vphantom {{{{\rm{P}}_{{\rm{tx}}}}} {\sigma _{\rm{N}}^2}}} \right. \kern-\nulldelimiterspace} {\sigma _{\rm{N}}^2}}$, which is independent of the number of MTs, $u+1$, and agrees with the definition of average SNR in \eqref{Eq_4}. In the next section, we show that the load models are not equivalent in terms of network power consumption. \hfill $\Box$
\end{remark}
\begin{figure*}[!t]
\setcounter{equation}{13}
\begin{equation}
\label{Eq_14}
{\rm{EE}}\left( {{{\rm{P}}_{{\rm{tx}}}},{\lambda _{{\rm{BS}}}}} \right) = \frac{{{\rm{PSE}}}}{{{{\rm{P}}_{{\rm{grid}}}}}} = \frac{{{{\rm{B}}_{\rm{W}}}{{\log }_2}\left( {1 + {\gamma _{\rm{D}}}} \right){\mathcal{L}}\left( {{{{\lambda _{{\rm{MT}}}}} \mathord{\left/
 {\vphantom {{{\lambda _{{\rm{MT}}}}} {{\lambda _{{\rm{BS}}}}}}} \right.
 \kern-\nulldelimiterspace} {{\lambda _{{\rm{BS}}}}}}} \right){\mathcal{Q}}\left( {{\lambda _{{\rm{BS}}}},{{\rm{P}}_{{\rm{tx}}}},{{{\lambda _{{\rm{MT}}}}} \mathord{\left/
 {\vphantom {{{\lambda _{{\rm{MT}}}}} {{\lambda _{{\rm{BS}}}}}}} \right.
 \kern-\nulldelimiterspace} {{\lambda _{{\rm{BS}}}}}}} \right)}}{{\left[ {1 + \Upsilon \mathcal{L}\left( {{{{\lambda _{{\rm{MT}}}}} \mathord{\left/
 {\vphantom {{{\lambda _{{\rm{MT}}}}} {{\lambda _{{\rm{BS}}}}}}} \right.
 \kern-\nulldelimiterspace} {{\lambda _{{\rm{BS}}}}}}} \right)} \right]\left[ {{\mathcal{L}}\left( {{{{\lambda _{{\rm{MT}}}}} \mathord{\left/
 {\vphantom {{{\lambda _{{\rm{MT}}}}} {{\lambda _{{\rm{BS}}}}}}} \right.
 \kern-\nulldelimiterspace} {{\lambda _{{\rm{BS}}}}}}} \right)\left( {{{\rm{P}}_{{\rm{tx}}}} + {{\rm{P}}_{{\rm{circ}}}} - {{\rm{P}}_{{\rm{idle}}}}} \right) + {{\rm{P}}_{{\rm{idle}}}} + {\mathcal{M}}\left( {{{{\lambda _{{\rm{MT}}}}} \mathord{\left/
 {\vphantom {{{\lambda _{{\rm{MT}}}}} {{\lambda _{{\rm{BS}}}}}}} \right.
 \kern-\nulldelimiterspace} {{\lambda _{{\rm{BS}}}}}}} \right){{\rm{P}}_{{\rm{circ}}}}} \right]}}.
\end{equation}
\normalsize \hrulefill \vspace*{-0pt}
\end{figure*}
\subsection{Closed-Form Expressions of PSE and ${{\rm{P}}_{{\rm{grid}}}}$} \label{PSE_Pgrid_ClosedFormFrameworks}
In this section, we introduce new closed-form analytical frameworks for computing the PSE. We provide, in addition, closed-form expressions of the network power consumption for the two load models under analysis. These results are summarized in the following three propositions.

Let ${{\rm{ N}}_{{\rm{MT}}}}$ be the number of MTs that lie in an arbitrary cell. The probability that the BS is in idle mode, ${\mathbb{P}}_{{\rm{BS}}}^{\left( {{\rm{idle}}} \right)}$, and in transmission mode, ${\mathbb{P}}_{{\rm{BS}}}^{\left( {{\rm{tx}}} \right)}$, can be formulated as follows \cite[Prop. 1]{WiOPT_2013}:
\setcounter{equation}{7}
\begin{equation}
\label{Eq_8}
\begin{split}
& {\mathbb{P}}_{{\rm{BS}}}^{\left( {{\rm{idle}}} \right)} = \Pr \left\{ {{{\rm{N}}_{{\rm{MT}}}} = 0} \right\} = 1 - {\mathcal{L}}\left( {{{{\lambda _{{\rm{MT}}}}} \mathord{\left/
 {\vphantom {{{\lambda _{{\rm{MT}}}}} {{\lambda _{{\rm{BS}}}}}}} \right.
 \kern-\nulldelimiterspace} {{\lambda _{{\rm{BS}}}}}}} \right)\\
& {\mathbb{P}}_{{\rm{BS}}}^{\left( {{\rm{tx}}} \right)} = \Pr \left\{ {{{\rm{N}}_{{\rm{MT}}}} \ge 1} \right\} = 1 - {\mathbb{P}}_{{\rm{BS}}}^{\left( {{\rm{idle}}} \right)} = {\mathcal{L}}\left( {{{{\lambda _{{\rm{MT}}}}} \mathord{\left/
 {\vphantom {{{\lambda _{{\rm{MT}}}}} {{\lambda _{{\rm{BS}}}}}}} \right.
 \kern-\nulldelimiterspace} {{\lambda _{{\rm{BS}}}}}}} \right)
\end{split}
\end{equation}
\noindent where $\mathcal{L}\left( \cdot \right)$ is defined in Table \ref{Table_Functions}. Using \eqref{Eq_8}, PSE and ${{\rm{P}}_{{\rm{grid}}}}$ are given in the following propositions.
\begin{proposition} \label{Proposition__PSE}
Consider either \textit{Load Model 1} or \textit{Load Model 2}. Assume notation and functions given in Tables \ref{Table_Notation} and \ref{Table_Functions}. The PSE (bit/sec/m$^2$) can be formulated, in closed-form, as follows:
\setcounter{equation}{8}
\begin{equation}
\label{Eq_9}
\begin{split}
{\rm{PSE}}\left( {{\gamma _{\rm{D}}},{\gamma _{\rm{A}}}} \right) &= {{\rm{B}}_{\rm{W}}}{\log _2}\left( {1 + {\gamma _{\rm{D}}}} \right)\frac{{{\lambda _{{\rm{BS}}}}{\mathcal{L}}\left( {{{{\lambda _{{\rm{MT}}}}} \mathord{\left/
 {\vphantom {{{\lambda _{{\rm{MT}}}}} {{\lambda _{{\rm{BS}}}}}}} \right.
 \kern-\nulldelimiterspace} {{\lambda _{{\rm{BS}}}}}}} \right)}}{{1 + \Upsilon \mathcal{L}\left( {{{{\lambda _{{\rm{MT}}}}} \mathord{\left/
 {\vphantom {{{\lambda _{{\rm{MT}}}}} {{\lambda _{{\rm{BS}}}}}}} \right.
 \kern-\nulldelimiterspace} {{\lambda _{{\rm{BS}}}}}}} \right)}} \\ & \times {\mathcal{Q}}\left( {{\lambda _{{\rm{BS}}}},{{\rm{P}}_{{\rm{tx}}}},{{{\lambda _{{\rm{MT}}}}} \mathord{\left/
 {\vphantom {{{\lambda _{{\rm{MT}}}}} {{\lambda _{{\rm{BS}}}}}}} \right.
 \kern-\nulldelimiterspace} {{\lambda _{{\rm{BS}}}}}}} \right).
\end{split}
\end{equation}

\emph{Proof}: See Appendix \ref{Appendix_PropPSE}. \hfill $\Box$
\end{proposition}
\begin{corollary} \label{Corollary__PSE}
If ${{\gamma _{\rm{A}}}}=0$, i.e., the conventional definition of ${{\rm{P}}_{{\rm{cov}}}}$ is used, the PSE in \eqref{Eq_9} simplifies as follows:
\setcounter{equation}{9}
\begin{equation}
\label{Eq_10}
{\rm{PSE}}\left( {{\gamma _{\rm{D}}},{\gamma _{\rm{A}}} = 0} \right) = {{\rm{B}}_{\rm{W}}}{\log _2}\left( {1 + {\gamma _{\rm{D}}}} \right)\frac{{{\lambda _{{\rm{BS}}}}{\mathcal{L}}\left( {{{{\lambda _{{\rm{MT}}}}} \mathord{\left/
 {\vphantom {{{\lambda _{{\rm{MT}}}}} {{\lambda _{{\rm{BS}}}}}}} \right.
 \kern-\nulldelimiterspace} {{\lambda _{{\rm{BS}}}}}}} \right)}}{{1 + \Upsilon \mathcal{L}\left( {{{{\lambda _{{\rm{MT}}}}} \mathord{\left/
 {\vphantom {{{\lambda _{{\rm{MT}}}}} {{\lambda _{{\rm{BS}}}}}}} \right.
 \kern-\nulldelimiterspace} {{\lambda _{{\rm{BS}}}}}}} \right)}}.
\end{equation}

If, in addition, ${{{\lambda _{{\rm{MT}}}}} \mathord{\left/ {\vphantom {{{\lambda _{{\rm{MT}}}}} {{\lambda _{{\rm{BS}}}}}}} \right. \kern-\nulldelimiterspace} {{\lambda _{{\rm{BS}}}}}} \gg 1$, the PSE in \eqref{Eq_9} reduces to \eqref{Eq_1}.

\emph{Proof}: It follows because ${\mathcal{Q}}\left( { \cdot , \cdot , \cdot } \right) = 1$ if ${{\gamma _{\rm{A}}}}=0$ and ${\mathcal{L}}\left( {{{{\lambda _{{\rm{MT}}}}} \mathord{\left/ {\vphantom {{{\lambda _{{\rm{MT}}}}} {{\lambda _{{\rm{BS}}}}}}} \right. \kern-\nulldelimiterspace} {{\lambda _{{\rm{BS}}}}}} \gg 1} \right) \to 1$. \hfill $\Box$
\end{corollary}
\begin{remark} \label{Remark__CorollaryPSE}
\textit{Corollary \ref{Corollary__PSE}} substantiates the comments made above in this section about the need of a new definition of PSE, as well as the advantages of the proposed analytical formulation. In particular, \eqref{Eq_10} confirms that the PSE is independent of ${{\rm{P}}_{{\rm{tx}}}}$ if ${{\gamma _{\rm{A}}}}=0$ and that the PSE is independent of ${{\rm{P}}_{{\rm{tx}}}}$ and ${{\lambda _{{\rm{BS}}}}}$ if fully-loaded conditions hold, i.e., ${{{\lambda _{{\rm{MT}}}}} \mathord{\left/ {\vphantom {{{\lambda _{{\rm{MT}}}}} {{\lambda _{{\rm{BS}}}}}}} \right. \kern-\nulldelimiterspace} {{\lambda _{{\rm{BS}}}}}} \gg 1$. \hfill $\Box$
\end{remark}
\begin{proposition} \label{Proposition__Pgrid_LoadModel1}
Let \textit{Load Model 1} be assumed. ${{\rm{P}}_{{\rm{grid}}}}$ (Watt/m$^2$) can be formulated as follows:
\setcounter{equation}{10}
\begin{equation}
\label{Eq_11}
\begin{split}
{{\rm{P}}^{(1)}_{{\rm{grid}}}} & = {\lambda _{{\rm{BS}}}}\left( {{{\rm{P}}_{{\rm{tx}}}} + {{\rm{P}}_{{\rm{circ}}}}} \right){\mathcal{L}}\left( {{{{\lambda _{{\rm{MT}}}}} \mathord{\left/
 {\vphantom {{{\lambda _{{\rm{MT}}}}} {{\lambda _{{\rm{BS}}}}}}} \right.
 \kern-\nulldelimiterspace} {{\lambda _{{\rm{BS}}}}}}} \right) \\ & + {\lambda _{{\rm{BS}}}}{{\rm{P}}_{{\rm{idle}}}}\left( {1 - {\mathcal{L}}\left( {{{{\lambda _{{\rm{MT}}}}} \mathord{\left/
 {\vphantom {{{\lambda _{{\rm{MT}}}}} {{\lambda _{{\rm{BS}}}}}}} \right.
 \kern-\nulldelimiterspace} {{\lambda _{{\rm{BS}}}}}}} \right)} \right).
\end{split}
\end{equation}

\emph{Proof}: The network power consumption is obtained by multiplying the average number of BSs per unit area, i.e., ${{\lambda _{{\rm{BS}}}}}$, and the average power consumption of a generic BS, which is ${{{\rm{P}}_{{\rm{tx}}}} + {{\rm{P}}_{{\rm{circ}}}}}$ if the BS operates in transmission mode, i.e., with probability ${\mathcal{L}}\left( {{{{\lambda _{{\rm{MT}}}}} \mathord{\left/ {\vphantom {{{\lambda _{{\rm{MT}}}}} {{\lambda _{{\rm{BS}}}}}}} \right. \kern-\nulldelimiterspace} {{\lambda _{{\rm{BS}}}}}}} \right)$, and ${{{\rm{P}}_{{\rm{idle}}}}}$ if the BS operates in idle mode, i.e., with probability $1-{\mathcal{L}}\left( {{{{\lambda _{{\rm{MT}}}}} \mathord{\left/ {\vphantom {{{\lambda _{{\rm{MT}}}}} {{\lambda _{{\rm{BS}}}}}}} \right. \kern-\nulldelimiterspace} {{\lambda _{{\rm{BS}}}}}}} \right)$. \hfill $\Box$
\end{proposition}
\begin{proposition} \label{Proposition__Pgrid_LoadModel2}
Let \textit{Load Model 2} be assumed. ${{\rm{P}}_{{\rm{grid}}}}$ (Watt/m$^2$) can be formulated as follows:
\setcounter{equation}{11}
\begin{equation}
\label{Eq_12}
\begin{split}
{{\rm{P}}^{(2)}_{{\rm{grid}}}} & = {\lambda _{{\rm{BS}}}}{{\rm{P}}_{{\rm{tx}}}}{\mathcal{L}}\left( {{{{\lambda _{{\rm{MT}}}}} \mathord{\left/
 {\vphantom {{{\lambda _{{\rm{MT}}}}} {{\lambda _{{\rm{BS}}}}}}} \right.
 \kern-\nulldelimiterspace} {{\lambda _{{\rm{BS}}}}}}} \right) \\ & + {\lambda _{{\rm{MT}}}}{{\rm{P}}_{{\rm{circ}}}} + {\lambda _{{\rm{BS}}}}{{\rm{P}}_{{\rm{idle}}}}\left( {1 - {\mathcal{L}}\left( {{{{\lambda _{{\rm{MT}}}}} \mathord{\left/
 {\vphantom {{{\lambda _{{\rm{MT}}}}} {{\lambda _{{\rm{BS}}}}}}} \right.
 \kern-\nulldelimiterspace} {{\lambda _{{\rm{BS}}}}}}} \right)} \right).
\end{split}
\end{equation}

\emph{Proof}: It is similar to the proof of \textit{Proposition \ref{Proposition__Pgrid_LoadModel1}}. The difference is that the power dissipation of a generic BS that operates in transmission mode is, in this case, equal to ${{\rm{P}}_{{\rm{tx}}}} + {{\rm{P}}_{{\rm{circ}}}}\sum\nolimits_{u = 1}^{ + \infty } {u\Pr \left\{ {{{\rm{N}}_{{\rm{MT}}}} = u} \right\}}  = {{\rm{P}}_{{\rm{tx}}}} + {{\rm{P}}_{{\rm{circ}}}}\left( {{{{\lambda _{{\rm{MT}}}}} \mathord{\left/
{\vphantom {{{\lambda _{{\rm{MT}}}}} {{\lambda _{{\rm{BS}}}}}}} \right.
\kern-\nulldelimiterspace} {{\lambda _{{\rm{BS}}}}}}} \right)$, where ${{{\rm{N}}_{{\rm{MT}}}}}$ is the number of MTs in the cell and the last equality follows from \cite[Lemma 1]{WiOPT_2013}. \hfill $\Box$
\end{proposition}
\begin{remark} \label{Remark__PgridOtherPapers}
The power consumption models obtained in \eqref{Eq_11} and \eqref{Eq_12}, which account for the transmit, circuits, and idle power consumption of the BSs, have been used, under some simplifying assumptions, in previous research works focused on the analysis of the EE of cellular networks. Among the many research works, an early paper that has adopted this approach under the assumption of fully-loaded BSs and of having a single active MT per cell is \cite{MariosANDTony2013}. \hfill $\Box$
\end{remark}
\begin{remark} \label{Remark__Pgrid}
Since ${\mathcal{L}}\left( {{{{\lambda _{{\rm{MT}}}}} \mathord{\left/ {\vphantom {{{\lambda _{{\rm{MT}}}}} {{\lambda _{{\rm{BS}}}}}}} \right. \kern-\nulldelimiterspace} {{\lambda _{{\rm{BS}}}}}}} \right) \le {{{\lambda _{{\rm{MT}}}}} \mathord{\left/ {\vphantom {{{\lambda _{{\rm{MT}}}}} {{\lambda _{{\rm{BS}}}}}}} \right. \kern-\nulldelimiterspace} {{\lambda _{{\rm{BS}}}}}}$ for every ${{{\lambda _{{\rm{MT}}}}} \mathord{\left/{\vphantom {{{\lambda _{{\rm{MT}}}}} {{\lambda _{{\rm{BS}}}}}}} \right.\kern-\nulldelimiterspace} {{\lambda _{{\rm{BS}}}}}} \ge 0$, we conclude that ${{\rm{P}}^{(2)}_{{\rm{grid}}}} \ge {{\rm{P}}^{(1)}_{{\rm{grid}}}}$ by assuming the same ${{\rm{P}}_{{\rm{tx}}}}$ and ${{\lambda _{{\rm{BS}}}}}$ for both load models. This originates from the fact that, in the present paper, we assume that the circuits power consumption increases with the number of MTs that are served by the BSs. It is unclear, however, the best load model to be used from the EE standpoint, especially if ${{\rm{P}}_{{\rm{tx}}}}$ and ${{\lambda _{{\rm{BS}}}}}$ are optimized to maximize their respective EEs. In other words, the optimal ${{\rm{P}}_{{\rm{tx}}}}$ and ${{\lambda _{{\rm{BS}}}}}$ that maximize the EE of each load model may be different, which may lead to different optimal EEs. The trade-off between the optimal PSE and the optimal EE is analyzed numerically in Section \ref{Results} for both load models. \hfill $\Box$
\end{remark}
\section{System-Level EE Optimization: Formulation and Solution} \label{EEopt}
In this section, we formulate a system-level EE optimization problem and comprehensively analyze its properties. For convenience of analysis, we introduce the following auxiliary function (LM = Load Model):
\setcounter{equation}{12}
\begin{equation}
\label{Eq_14pre}
\begin{split}
& {\mathcal{M}}\left( {{{{\lambda _{{\rm{MT}}}}} \mathord{\left/
 {\vphantom {{{\lambda _{{\rm{MT}}}}} {{\lambda _{{\rm{BS}}}}}}} \right.
 \kern-\nulldelimiterspace} {{\lambda _{{\rm{BS}}}}}}} \right) \\ & = \begin{cases}
0\quad & {\rm{if}}\quad {\rm{LM-1 \; is \; assumed}}\\
{{{\lambda _{{\rm{MT}}}}} \mathord{\left/
 {\vphantom {{{\lambda _{{\rm{MT}}}}} {{\lambda _{{\rm{BS}}}}}}} \right.
 \kern-\nulldelimiterspace} {{\lambda _{{\rm{BS}}}}}} - {\mathcal{L}}\left( {{{{\lambda _{{\rm{MT}}}}} \mathord{\left/
 {\vphantom {{{\lambda _{{\rm{MT}}}}} {{\lambda _{{\rm{BS}}}}}}} \right.
 \kern-\nulldelimiterspace} {{\lambda _{{\rm{BS}}}}}}} \right)\quad & {\rm{if}}\quad {\rm{LM-2 \; is \; assumed}}.
\end{cases}
\end{split}
\end{equation}

A unified formulation of the EE (bit/Joule) for the cellular network under analysis is provided in \eqref{Eq_14} shown at the top of this page, where the parameters of interest from the optimization standpoint, i.e., ${{{\rm{P}}_{{\rm{tx}}}}}$ and ${{\lambda _{{\rm{BS}}}}}$, are explicitly highlighted. In the rest of the present paper, all the other parameters are assumed to be given.
\begin{table*}[!t] 
\centering
\caption{Summary of main auxiliary functions used throughout the paper.}
\newcommand{\tabincell}[2]{\begin{tabular}{@{}#1@{}}#2\end{tabular}}
\begin{tabular}{|l|} \hline
\hspace{5.00cm} Function Definition \\ \hline \hline
$\begin{array}{l}{\mathcal{L}}\left( {{{{\lambda _{{\rm{MT}}}}} \mathord{\left/
 {\vphantom {{{\lambda _{{\rm{MT}}}}} {{\lambda _{{\rm{BS}}}}}}} \right.
 \kern-\nulldelimiterspace} {{\lambda _{{\rm{BS}}}}}}} \right) = 1 - {\left( {1 + {{\left( {{1 \mathord{\left/
 {\vphantom {1 \alpha }} \right.
 \kern-\nulldelimiterspace} \alpha }} \right){\lambda _{{\rm{MT}}}}} \mathord{\left/
 {\vphantom {{\left( {{1 \mathord{\left/
 {\vphantom {1 \alpha }} \right.
 \kern-\nulldelimiterspace} \alpha }} \right){\lambda _{{\rm{MT}}}}} {{\lambda _{{\rm{BS}}}}}}} \right.
 \kern-\nulldelimiterspace} {{\lambda _{{\rm{BS}}}}}}} \right)^{ - \alpha }} \end{array}$ \\
$\begin{array}{l}{\mathcal{M}}\left( {{{{\lambda _{{\rm{MT}}}}} \mathord{\left/
 {\vphantom {{{\lambda _{{\rm{MT}}}}} {{\lambda _{{\rm{BS}}}}}}} \right.
 \kern-\nulldelimiterspace} {{\lambda _{{\rm{BS}}}}}}} \right) = {{{\lambda _{{\rm{MT}}}}} \mathord{\left/
 {\vphantom {{{\lambda _{{\rm{MT}}}}} {{\lambda _{{\rm{BS}}}}}}} \right.
 \kern-\nulldelimiterspace} {{\lambda _{{\rm{BS}}}}}} - {\mathcal{L}}\left( {{{{\lambda _{{\rm{MT}}}}} \mathord{\left/
 {\vphantom {{{\lambda _{{\rm{MT}}}}} {{\lambda _{{\rm{BS}}}}}}} \right.
 \kern-\nulldelimiterspace} {{\lambda _{{\rm{BS}}}}}}} \right) \end{array}$ \\
$\begin{array}{l}{\mathcal{Q}}\left( {{\lambda _{{\rm{BS}}}},{{\rm{P}}_{{\rm{tx}}}},{{{\lambda _{{\rm{MT}}}}} \mathord{\left/
 {\vphantom {{{\lambda _{{\rm{MT}}}}} {{\lambda _{{\rm{BS}}}}}}} \right.
 \kern-\nulldelimiterspace} {{\lambda _{{\rm{BS}}}}}}} \right) = 1 - \exp \left( { - \pi {\lambda _{{\rm{BS}}}}{{\left( {{{{{\rm{P}}_{{\rm{tx}}}}} \mathord{\left/
 {\vphantom {{{{\rm{P}}_{{\rm{tx}}}}} \eta }} \right.
 \kern-\nulldelimiterspace} \eta }} \right)}^{{2 \mathord{\left/
 {\vphantom {2 \beta }} \right.
 \kern-\nulldelimiterspace} \beta }}}\left( {1 + \Upsilon {\mathcal{L}}\left( {{{{\lambda _{{\rm{MT}}}}} \mathord{\left/
 {\vphantom {{{\lambda _{{\rm{MT}}}}} {{\lambda _{{\rm{BS}}}}}}} \right.
 \kern-\nulldelimiterspace} {{\lambda _{{\rm{BS}}}}}}} \right)} \right)} \right) \end{array}$ \\
$\begin{array}{l} {{{\mathcal{\dt Q}}}_{{{\rm{P}}_{{\rm{tx}}}}}}\left( {{\lambda _{{\rm{BS}}}},{{\rm{P}}_{{\rm{tx}}}},{{{\lambda _{{\rm{MT}}}}} \mathord{\left/
 {\vphantom {{{\lambda _{{\rm{MT}}}}} {{\lambda _{{\rm{BS}}}}}}} \right.
 \kern-\nulldelimiterspace} {{\lambda _{{\rm{BS}}}}}}} \right) = \pi {\lambda _{{\rm{BS}}}}{\left( {{{\rm{1}} \mathord{\left/
 {\vphantom {{\rm{1}} \eta }} \right.
 \kern-\nulldelimiterspace} \eta }} \right)^{{2 \mathord{\left/
 {\vphantom {2 \beta }} \right.
 \kern-\nulldelimiterspace} \beta }}}\left( {{2 \mathord{\left/
 {\vphantom {2 \beta }} \right.
 \kern-\nulldelimiterspace} \beta }} \right)\left( {1 + \Upsilon {\mathcal{L}}\left( {{{{\lambda _{{\rm{MT}}}}} \mathord{\left/
 {\vphantom {{{\lambda _{{\rm{MT}}}}} {{\lambda _{{\rm{BS}}}}}}} \right.
 \kern-\nulldelimiterspace} {{\lambda _{{\rm{BS}}}}}}} \right)} \right){\rm{P}}_{{\rm{tx}}}^{{2 \mathord{\left/
 {\vphantom {2 \beta }} \right.
 \kern-\nulldelimiterspace} \beta } - 1} \\ \hspace{4.0cm} \times \exp \left( { - \pi {\lambda _{{\rm{BS}}}}{{\left( {{{{{\rm{P}}_{{\rm{tx}}}}} \mathord{\left/
 {\vphantom {{{{\rm{P}}_{{\rm{tx}}}}} \eta }} \right.
 \kern-\nulldelimiterspace} \eta }} \right)}^{{2 \mathord{\left/
 {\vphantom {2 \beta }} \right.
 \kern-\nulldelimiterspace} \beta }}}\left( {1 + \Upsilon {\mathcal{L}}\left( {{{{\lambda _{{\rm{MT}}}}} \mathord{\left/
 {\vphantom {{{\lambda _{{\rm{MT}}}}} {{\lambda _{{\rm{BS}}}}}}} \right.
 \kern-\nulldelimiterspace} {{\lambda _{{\rm{BS}}}}}}} \right)} \right)} \right) \end{array}$ \\
$\begin{array}{l}
{{{\rm{\ddt Q}}}_{{{\rm{P}}_{{\rm{tx}}}}}}\left( {{\lambda _{{\rm{BS}}}},{{\rm{P}}_{{\rm{tx}}}},{{{\lambda _{{\rm{MT}}}}} \mathord{\left/
 {\vphantom {{{\lambda _{{\rm{MT}}}}} {{\lambda _{{\rm{BS}}}}}}} \right.
 \kern-\nulldelimiterspace} {{\lambda _{{\rm{BS}}}}}}} \right) = \pi {\lambda _{{\rm{BS}}}}{\left( {{{\rm{1}} \mathord{\left/
 {\vphantom {{\rm{1}} \eta }} \right.
 \kern-\nulldelimiterspace} \eta }} \right)^{{2 \mathord{\left/
 {\vphantom {2 \beta }} \right.
 \kern-\nulldelimiterspace} \beta }}}\left( {{2 \mathord{\left/
 {\vphantom {2 \beta }} \right.
 \kern-\nulldelimiterspace} \beta }} \right)\left( {1 + \Upsilon {\mathcal{L}}\left( {{{{\lambda _{{\rm{MT}}}}} \mathord{\left/
 {\vphantom {{{\lambda _{{\rm{MT}}}}} {{\lambda _{{\rm{BS}}}}}}} \right.
 \kern-\nulldelimiterspace} {{\lambda _{{\rm{BS}}}}}}} \right)} \right){\rm{P}}_{{\rm{tx}}}^{{2 \mathord{\left/
 {\vphantom {2 \beta }} \right.
 \kern-\nulldelimiterspace} \beta } - 1}\\
 \hspace{4.0cm} \times \left[ { - {{{\mathcal{\dot Q}}}_{{{\rm{P}}_{{\rm{tx}}}}}}\left( {{\lambda _{{\rm{BS}}}},{{\rm{P}}_{{\rm{tx}}}},{{{\lambda _{{\rm{MT}}}}} \mathord{\left/
 {\vphantom {{{\lambda _{{\rm{MT}}}}} {{\lambda _{{\rm{BS}}}}}}} \right.
 \kern-\nulldelimiterspace} {{\lambda _{{\rm{BS}}}}}}} \right)} \right]\\
 \hspace{4.0cm} + {\pi}\lambda _{{\rm{BS}}}{\left( {{{\rm{1}} \mathord{\left/
 {\vphantom {{\rm{1}} \eta }} \right.
 \kern-\nulldelimiterspace} \eta }} \right)^{{2 \mathord{\left/
 {\vphantom {2 \beta }} \right.
 \kern-\nulldelimiterspace} \beta }}}\left( {{2 \mathord{\left/
 {\vphantom {2 \beta }} \right.
 \kern-\nulldelimiterspace} \beta }} \right)\left( {{2 \mathord{\left/
 {\vphantom {2 \beta }} \right.
 \kern-\nulldelimiterspace} \beta } - 1} \right){\left( {1 + \Upsilon {\mathcal{L}}\left( {{{{\lambda _{{\rm{MT}}}}} \mathord{\left/
 {\vphantom {{{\lambda _{{\rm{MT}}}}} {{\lambda _{{\rm{BS}}}}}}} \right.
 \kern-\nulldelimiterspace} {{\lambda _{{\rm{BS}}}}}}} \right)} \right)}{\rm{P}}_{{\rm{tx}}}^{{2 \mathord{\left/
 {\vphantom {2 \beta }} \right.
 \kern-\nulldelimiterspace} \beta } - 2}\\
 \hspace{4.0cm} \times \exp \left( { - \pi {\lambda _{{\rm{BS}}}}{{\left( {{{{{\rm{P}}_{{\rm{tx}}}}} \mathord{\left/
 {\vphantom {{{{\rm{P}}_{{\rm{tx}}}}} \eta }} \right.
 \kern-\nulldelimiterspace} \eta }} \right)}^{{2 \mathord{\left/
 {\vphantom {2 \beta }} \right.
 \kern-\nulldelimiterspace} \beta }}}\left( {1 + \Upsilon {\mathcal{L}}\left( {{{{\lambda _{{\rm{MT}}}}} \mathord{\left/
 {\vphantom {{{\lambda _{{\rm{MT}}}}} {{\lambda _{{\rm{BS}}}}}}} \right.
 \kern-\nulldelimiterspace} {{\lambda _{{\rm{BS}}}}}}} \right)} \right)} \right)
\end{array}$ \\
$\begin{array}{l}{{{\mathcal{\dt L}}}_{{\lambda _{{\rm{BS}}}}}}\left( {{{{\lambda _{{\rm{MT}}}}} \mathord{\left/
 {\vphantom {{{\lambda _{{\rm{MT}}}}} {{\lambda _{{\rm{BS}}}}}}} \right.
 \kern-\nulldelimiterspace} {{\lambda _{{\rm{BS}}}}}}} \right) =  - \left( {{{{\lambda _{{\rm{MT}}}}} \mathord{\left/
 {\vphantom {{{\lambda _{{\rm{MT}}}}} {\lambda _{{\rm{BS}}}^2}}} \right.
 \kern-\nulldelimiterspace} {\lambda _{{\rm{BS}}}^2}}} \right){\left( {1 + {{\left( {{1 \mathord{\left/
 {\vphantom {1 \alpha }} \right.
 \kern-\nulldelimiterspace} \alpha }} \right){\lambda _{{\rm{MT}}}}} \mathord{\left/
 {\vphantom {{\left( {{1 \mathord{\left/
 {\vphantom {1 \alpha }} \right.
 \kern-\nulldelimiterspace} \alpha }} \right){\lambda _{{\rm{MT}}}}} {{\lambda _{{\rm{BS}}}}}}} \right.
 \kern-\nulldelimiterspace} {{\lambda _{{\rm{BS}}}}}}} \right)^{ - \left( {\alpha  + 1} \right)}} \end{array}$ \\
$\begin{array}{l}{{{\mathcal{\dt M}}}_{{\lambda _{{\rm{BS}}}}}}\left( {{{{\lambda _{{\rm{MT}}}}} \mathord{\left/
 {\vphantom {{{\lambda _{{\rm{MT}}}}} {{\lambda _{{\rm{BS}}}}}}} \right.
 \kern-\nulldelimiterspace} {{\lambda _{{\rm{BS}}}}}}} \right) =  - \left( {{{{\lambda _{{\rm{MT}}}}} \mathord{\left/
 {\vphantom {{{\lambda _{{\rm{MT}}}}} {\lambda _{{\rm{BS}}}^2}}} \right.
 \kern-\nulldelimiterspace} {\lambda _{{\rm{BS}}}^2}}} \right)\left[ {1 - {{\left( {1 + {{\left( {{1 \mathord{\left/
 {\vphantom {1 \alpha }} \right.
 \kern-\nulldelimiterspace} \alpha }} \right){\lambda _{{\rm{MT}}}}} \mathord{\left/
 {\vphantom {{\left( {{1 \mathord{\left/
 {\vphantom {1 \alpha }} \right.
 \kern-\nulldelimiterspace} \alpha }} \right){\lambda _{{\rm{MT}}}}} {{\lambda _{{\rm{BS}}}}}}} \right.
 \kern-\nulldelimiterspace} {{\lambda _{{\rm{BS}}}}}}} \right)}^{ - \left( {\alpha  + 1} \right)}}} \right] \end{array}$ \\
$\begin{array}{l}{{{\mathcal{\dt Q}}}_{{\lambda _{{\rm{BS}}}}}}\left( {{\lambda _{{\rm{BS}}}},{{\rm{P}}_{{\rm{tx}}}},{{{\lambda _{{\rm{MT}}}}} \mathord{\left/
 {\vphantom {{{\lambda _{{\rm{MT}}}}} {{\lambda _{{\rm{BS}}}}}}} \right.
 \kern-\nulldelimiterspace} {{\lambda _{{\rm{BS}}}}}}} \right) = \pi {\left( {{{{{\rm{P}}_{{\rm{tx}}}}} \mathord{\left/
 {\vphantom {{{{\rm{P}}_{{\rm{tx}}}}} \eta }} \right.
 \kern-\nulldelimiterspace} \eta }} \right)^{{2 \mathord{\left/
 {\vphantom {2 \beta }} \right.
 \kern-\nulldelimiterspace} \beta }}}\left[ {1 + \Upsilon {\mathcal{L}}\left( {{{{\lambda _{{\rm{MT}}}}} \mathord{\left/
 {\vphantom {{{\lambda _{{\rm{MT}}}}} {{\lambda _{{\rm{BS}}}}}}} \right.
 \kern-\nulldelimiterspace} {{\lambda _{{\rm{BS}}}}}}} \right) + \Upsilon {\lambda _{{\rm{BS}}}}{{{\mathcal{\dt L}}}_{{\lambda _{{\rm{BS}}}}}}\left( {{{{\lambda _{{\rm{MT}}}}} \mathord{\left/
 {\vphantom {{{\lambda _{{\rm{MT}}}}} {{\lambda _{{\rm{BS}}}}}}} \right.
 \kern-\nulldelimiterspace} {{\lambda _{{\rm{BS}}}}}}} \right)} \right] \\ \hspace{4.05cm} \times \exp \left( { - \pi {\lambda _{{\rm{BS}}}}{{\left( {{{{{\rm{P}}_{{\rm{tx}}}}} \mathord{\left/
 {\vphantom {{{{\rm{P}}_{{\rm{tx}}}}} \eta }} \right.
 \kern-\nulldelimiterspace} \eta }} \right)}^{{2 \mathord{\left/
 {\vphantom {2 \beta }} \right.
 \kern-\nulldelimiterspace} \beta }}}\left( {1 + \Upsilon {\mathcal{L}}\left( {{{{\lambda _{{\rm{MT}}}}} \mathord{\left/
 {\vphantom {{{\lambda _{{\rm{MT}}}}} {{\lambda _{{\rm{BS}}}}}}} \right.
 \kern-\nulldelimiterspace} {{\lambda _{{\rm{BS}}}}}}} \right)} \right)} \right) \end{array}$ \\
$\begin{array}{l}
{{{\mathcal{\ddt L}}}_{{\lambda _{{\rm{BS}}}}}}\left( {{{{\lambda _{{\rm{MT}}}}} \mathord{\left/
 {\vphantom {{{\lambda _{{\rm{MT}}}}} {{\lambda _{{\rm{BS}}}}}}} \right.
 \kern-\nulldelimiterspace} {{\lambda _{{\rm{BS}}}}}}} \right) = \left( {{{{\lambda _{{\rm{MT}}}}} \mathord{\left/
 {\vphantom {{{\lambda _{{\rm{MT}}}}} {\lambda _{{\rm{BS}}}^3}}} \right.
 \kern-\nulldelimiterspace} {\lambda _{{\rm{BS}}}^3}}} \right){\left( {1 + {{\left( {{1 \mathord{\left/
 {\vphantom {1 \alpha }} \right.
 \kern-\nulldelimiterspace} \alpha }} \right){\lambda _{{\rm{MT}}}}} \mathord{\left/
 {\vphantom {{\left( {{1 \mathord{\left/
 {\vphantom {1 \alpha }} \right.
 \kern-\nulldelimiterspace} \alpha }} \right){\lambda _{{\rm{MT}}}}} {{\lambda _{{\rm{BS}}}}}}} \right.
 \kern-\nulldelimiterspace} {{\lambda _{{\rm{BS}}}}}}} \right)^{ - \left( {\alpha  + 1} \right)}}\\
 \hspace{2.55cm} \times \left[ {2 - \left( {1 + \alpha } \right)\left( {{1 \mathord{\left/
 {\vphantom {1 \alpha }} \right.
 \kern-\nulldelimiterspace} \alpha }} \right){{{\lambda _{{\rm{MT}}}}} \mathord{\left/
 {\vphantom {{{\lambda _{{\rm{MT}}}}} {{\lambda _{{\rm{BS}}}}}}} \right.
 \kern-\nulldelimiterspace} {{\lambda _{{\rm{BS}}}}}}{{\left( {1 + (1/\alpha){{{\lambda _{{\rm{MT}}}}} \mathord{\left/
 {\vphantom {{{\lambda _{{\rm{MT}}}}} {{\lambda _{{\rm{BS}}}}}}} \right.
 \kern-\nulldelimiterspace} {{\lambda _{{\rm{BS}}}}}}} \right)}^{ - 1}}} \right]
\end{array}$ \\
$\begin{array}{l}
{{{\mathcal{\ddt M}}}_{{\lambda _{{\rm{BS}}}}}}\left( {{{{\lambda _{{\rm{MT}}}}} \mathord{\left/
 {\vphantom {{{\lambda _{{\rm{MT}}}}} {{\lambda _{{\rm{BS}}}}}}} \right.
 \kern-\nulldelimiterspace} {{\lambda _{{\rm{BS}}}}}}} \right) = 2\left( {{{{\lambda _{{\rm{MT}}}}} \mathord{\left/
 {\vphantom {{{\lambda _{{\rm{MT}}}}} {\lambda _{{\rm{BS}}}^3}}} \right.
 \kern-\nulldelimiterspace} {\lambda _{{\rm{BS}}}^3}}} \right)\left[ {1 - {{\left( {1 + {{\left( {{1 \mathord{\left/
 {\vphantom {1 \alpha }} \right.
 \kern-\nulldelimiterspace} \alpha }} \right){\lambda _{{\rm{MT}}}}} \mathord{\left/
 {\vphantom {{\left( {{1 \mathord{\left/
 {\vphantom {1 \alpha }} \right.
 \kern-\nulldelimiterspace} \alpha }} \right){\lambda _{{\rm{MT}}}}} {{\lambda _{{\rm{BS}}}}}}} \right.
 \kern-\nulldelimiterspace} {{\lambda _{{\rm{BS}}}}}}} \right)}^{ - \left( {\alpha  + 1} \right)}}} \right]\\
 \hspace{2.75cm} + \left( {1 + \alpha } \right)\left( {{1 \mathord{\left/
 {\vphantom {1 \alpha }} \right.
 \kern-\nulldelimiterspace} \alpha }} \right)\left( {{{\lambda _{{\rm{MT}}}^2} \mathord{\left/
 {\vphantom {{\lambda _{{\rm{MT}}}^2} {\lambda _{{\rm{BS}}}^4}}} \right.
 \kern-\nulldelimiterspace} {\lambda _{{\rm{BS}}}^4}}} \right){\left( {1 + {{\left( {{1 \mathord{\left/
 {\vphantom {1 \alpha }} \right.
 \kern-\nulldelimiterspace} \alpha }} \right){\lambda _{{\rm{MT}}}}} \mathord{\left/
 {\vphantom {{\left( {{1 \mathord{\left/
 {\vphantom {1 \alpha }} \right.
 \kern-\nulldelimiterspace} \alpha }} \right){\lambda _{{\rm{MT}}}}} {{\lambda _{{\rm{BS}}}}}}} \right.
 \kern-\nulldelimiterspace} {{\lambda _{{\rm{BS}}}}}}} \right)^{ - \left( {\alpha  + 2} \right)}}
\end{array}$ \\
$\begin{array}{l} {{\mathcal{S}}_{\mathcal{P}}}\left( {{{\rm{P}}_{{\rm{tx}}}}} \right) = {\mathcal{L}}\left( {\frac{{{\lambda _{{\rm{MT}}}}}}{{{\lambda _{{\rm{BS}}}}}}} \right)\left[ {\frac{{{\mathcal{Q}}\left( {{\lambda _{{\rm{BS}}}},{\rm{P}}_{{\rm{tx}}},{{{\lambda _{{\rm{MT}}}}} \mathord{\left/
 {\vphantom {{{\lambda _{{\rm{MT}}}}} {{\lambda _{{\rm{BS}}}}}}} \right.
 \kern-\nulldelimiterspace} {{\lambda _{{\rm{BS}}}}}}} \right)}}{{{{{\mathcal{\dt Q}}}_{{{\rm{P}}_{{\rm{tx}}}}}}\left( {{\lambda _{{\rm{BS}}}},{\rm{P}}_{{\rm{tx}}},{{{\lambda _{{\rm{MT}}}}} \mathord{\left/
 {\vphantom {{{\lambda _{{\rm{MT}}}}} {{\lambda _{{\rm{BS}}}}}}} \right.
 \kern-\nulldelimiterspace} {{\lambda _{{\rm{BS}}}}}}} \right)}} - \left( {{\rm{P}}_{{\rm{tx}}} + \Delta {\rm{P}}} \right)} \right] - {{\rm{P}}_{{\rm{circ}}}}{\mathcal{M}}\left( {{{{\lambda _{{\rm{MT}}}}} \mathord{\left/ {\vphantom {{{\lambda _{{\rm{MT}}}}} {{\lambda _{{\rm{BS}}}}}}} \right. \kern-\nulldelimiterspace} {{\lambda _{{\rm{BS}}}}}}} \right) \end{array}$ \\
$\begin{array}{l} {{\mathcal{S}}_{\mathcal{D}}}\left( {{\lambda _{{\rm{BS}}}}} \right) = \frac{{{{\rm{P}}_{{\rm{circ}}}}}}{{{{{\mathcal{\dt L}}}_{{\lambda _{{\rm{BS}}}}}}\left( {{{{\lambda _{{\rm{MT}}}}} \mathord{\left/
 {\vphantom {{{\lambda _{{\rm{MT}}}}} {\lambda _{{\rm{BS}}}}}} \right.
 \kern-\nulldelimiterspace} {\lambda _{{\rm{BS}}}}}} \right)}}\left( {{\mathcal{L}}\left( {\frac{{{\lambda _{{\rm{MT}}}}}}{{\lambda _{{\rm{BS}}}}}} \right){{{\mathcal{\dt M}}}_{{\lambda _{{\rm{BS}}}}}}\left( {\frac{{{\lambda _{{\rm{MT}}}}}}{{\lambda _{{\rm{BS}}}}}} \right) - {{{\mathcal{\dt L}}}_{{\lambda _{{\rm{BS}}}}}}\left( {\frac{{{\lambda _{{\rm{MT}}}}}}{{\lambda _{{\rm{BS}}}}}} \right){\mathcal{M}}\left( {\frac{{{\lambda _{{\rm{MT}}}}}}{{\lambda _{{\rm{BS}}}
 }}} \right)} \right)\\
\hspace{1.45cm} + \Upsilon {{\mathcal{L}}^2}\left( {\frac{{{\lambda _{{\rm{MT}}}}}}{{\lambda _{{\rm{BS}}}}}} \right)\left( {{{\rm{P}}_{{\rm{tx}}}} + \Delta {\rm{P}}} \right) + \Upsilon {{\rm{P}}_{{\rm{circ}}}}{{\mathcal{L}}^2}\left( {\frac{{{\lambda _{{\rm{MT}}}}}}{{\lambda _{{\rm{BS}}}}}} \right)\frac{{{{{\mathcal{\dt M}}}_{{\lambda _{{\rm{BS}}}}}}\left( {{{{\lambda _{{\rm{MT}}}}} \mathord{\left/
 {\vphantom {{{\lambda _{{\rm{MT}}}}} {\lambda _{{\rm{BS}}}}}} \right.
 \kern-\nulldelimiterspace} {\lambda _{{\rm{BS}}}}}} \right)}}{{{{{\mathcal{\dt L}}}_{{\lambda _{{\rm{BS}}}}}}\left( {{{{\lambda _{{\rm{MT}}}}} \mathord{\left/
 {\vphantom {{{\lambda _{{\rm{MT}}}}} {\lambda _{{\rm{BS}}}}}} \right.
 \kern-\nulldelimiterspace} {\lambda _{{\rm{BS}}}}}} \right)}}\\
\hspace{1.45cm} - \frac{{{\mathcal{L}}\left( {{{{\lambda _{{\rm{MT}}}}} \mathord{\left/
 {\vphantom {{{\lambda _{{\rm{MT}}}}} {\lambda _{{\rm{BS}}}}}} \right.
 \kern-\nulldelimiterspace} {\lambda _{{\rm{BS}}}}}} \right){{{\mathcal{\dt Q}}}_{{\lambda _{{\rm{BS}}}}}}\left( {\lambda _{{\rm{BS}}},{{\rm{P}}_{{\rm{tx}}}},{{{\lambda _{{\rm{MT}}}}} \mathord{\left/
 {\vphantom {{{\lambda _{{\rm{MT}}}}} {\lambda _{{\rm{BS}}}}}} \right.
 \kern-\nulldelimiterspace} {\lambda _{{\rm{BS}}}}}} \right)}}{{{{{\mathcal{\dt L}}}_{{\lambda _{{\rm{BS}}}}}}\left( {{{{\lambda _{{\rm{MT}}}}} \mathord{\left/
 {\vphantom {{{\lambda _{{\rm{MT}}}}} {\lambda _{{\rm{BS}}}}}} \right.
 \kern-\nulldelimiterspace} {\lambda _{{\rm{BS}}}}}} \right){\mathcal{Q}}\left( {\lambda _{{\rm{BS}}},{{\rm{P}}_{{\rm{tx}}}},{{{\lambda _{{\rm{MT}}}}} \mathord{\left/
 {\vphantom {{{\lambda _{{\rm{MT}}}}} {\lambda _{{\rm{BS}}}}}} \right.
 \kern-\nulldelimiterspace} {\lambda _{{\rm{BS}}}}}} \right)}}\left( {1 + \Upsilon {\mathcal{L}}\left( {\frac{{{\lambda _{{\rm{MT}}}}}}{{\lambda _{{\rm{BS}}}}}} \right)} \right)\\
\hspace{1.45cm} \times \left[ {{\mathcal{L}}\left( {{{{{\lambda _{{\rm{MT}}}}} \mathord{\left/ {\vphantom {{{\lambda _{{\rm{MT}}}}} {\lambda _{{\rm{BS}}}}}} \right. \kern-\nulldelimiterspace} {\lambda _{{\rm{BS}}}}}}} \right)\left( {{{\rm{P}}_{{\rm{tx}}}} + \Delta {\rm{P}}} \right) + {{\rm{P}}_{{\rm{idle}}}} + {{\rm{P}}_{{\rm{circ}}}}{\mathcal{M}}\left( {{{{{\lambda _{{\rm{MT}}}}} \mathord{\left/ {\vphantom {{{\lambda _{{\rm{MT}}}}} {\lambda _{{\rm{BS}}}}}} \right. \kern-\nulldelimiterspace} {\lambda _{{\rm{BS}}}}}}} \right)} \right] \end{array}$ \\ \hline
\end{tabular}
\label{Table_Functions}
\end{table*}
\subsection{Preliminaries} \label{Preliminaries}
For ease of presentation, we report some lemmas that summarize structural properties of the main functions that constitute \eqref{Eq_14}. Some lemmas are stated without proof because they are obtained by simply studying the sign of the first-order and second-order derivatives of the function with respect to the variable of interest and by keeping all the other variables fixed. Functions of interest for this section are given in Table \ref{Table_Functions}. Also, we define $\Delta {\rm{P}} = {{\rm{P}}_{{\rm{circ}}}} - {{\rm{P}}_{{\rm{idle}}}} \ge 0$.
\begin{lemma} \label{Lemma__LoadFunction}
The function ${{\mathcal{L}}\left( {{{{\lambda _{{\rm{MT}}}}} \mathord{\left/{\vphantom {{{\lambda _{{\rm{MT}}}}} {{\lambda _{{\rm{BS}}}}}}} \right. \kern-\nulldelimiterspace} {{\lambda _{{\rm{BS}}}}}}} \right)}$ fulfills the following properties with respect to ${{\lambda _{{\rm{BS}}}}}$ (assuming ${{\lambda _{{\rm{MT}}}}}$ fixed):
i) ${\mathcal{L}}\left( {{{{\lambda _{{\rm{MT}}}}} \mathord{\left/ {\vphantom {{{\lambda _{{\rm{MT}}}}} {{\lambda _{{\rm{BS}}}}}}} \right.\kern-\nulldelimiterspace} {{\lambda _{{\rm{BS}}}}}}} \right) \ge 0$ for ${{\lambda _{{\rm{BS}}}}} \ge 0$;
ii) ${\mathcal{L}}\left( {{{{\lambda _{{\rm{MT}}}}} \mathord{\left/ {\vphantom {{{\lambda _{{\rm{MT}}}}} {{\lambda _{{\rm{BS}}}}}}} \right.\kern-\nulldelimiterspace} {{\lambda _{{\rm{BS}}}}}}} \right) = 1$ if ${{\lambda _{{\rm{BS}}}}} \to 0$;
iii) ${\mathcal{L}}\left( {{{{\lambda _{{\rm{MT}}}}} \mathord{\left/ {\vphantom {{{\lambda _{{\rm{MT}}}}} {{\lambda _{{\rm{BS}}}}}}} \right.\kern-\nulldelimiterspace} {{\lambda _{{\rm{BS}}}}}}} \right) = 0$ if ${{\lambda _{{\rm{BS}}}}} \to \infty$;
iv) ${{{\mathcal{\dt L}}}_{{\lambda _{{\rm{BS}}}}}}\left( {{{{\lambda _{{\rm{MT}}}}} \mathord{\left/ {\vphantom {{{\lambda _{{\rm{MT}}}}} {{\lambda _{{\rm{BS}}}}}}} \right. \kern-\nulldelimiterspace} {{\lambda _{{\rm{BS}}}}}}} \right) \le 0$ for ${{\lambda _{{\rm{BS}}}}} \ge 0$;
v) ${{{\mathcal{\ddt L}}}_{{\lambda _{{\rm{BS}}}}}}\left( {{{{\lambda _{{\rm{MT}}}}} \mathord{\left/ {\vphantom {{{\lambda _{{\rm{MT}}}}} {{\lambda _{{\rm{BS}}}}}}} \right.\kern-\nulldelimiterspace} {{\lambda _{{\rm{BS}}}}}}} \right) \le 0$ for ${{{\lambda _{{\rm{MT}}}}} \mathord{\left/ {\vphantom {{{\lambda _{{\rm{MT}}}}} {{\lambda _{{\rm{BS}}}}}}} \right. \kern-\nulldelimiterspace} {{\lambda _{{\rm{BS}}}}}} \ge {{2\alpha } \mathord{\left/ {\vphantom {{2\alpha } {\left( {\alpha  - 1} \right) = 2.8}}} \right. \kern-\nulldelimiterspace} {\left( {\alpha  - 1} \right) = 2.8}}$; and
vi) ${{{\mathcal{\ddt L}}}_{{\lambda _{{\rm{BS}}}}}}\left( {{{{\lambda _{{\rm{MT}}}}} \mathord{\left/ {\vphantom {{{\lambda _{{\rm{MT}}}}} {{\lambda _{{\rm{BS}}}}}}} \right. \kern-\nulldelimiterspace} {{\lambda _{{\rm{BS}}}}}}} \right) \ge 0$ for ${{{\lambda _{{\rm{MT}}}}} \mathord{\left/ {\vphantom {{{\lambda _{{\rm{MT}}}}} {{\lambda _{{\rm{BS}}}}}}} \right. \kern-\nulldelimiterspace} {{\lambda _{{\rm{BS}}}}}} \le {{2\alpha } \mathord{\left/ {\vphantom {{2\alpha } {\left( {\alpha  - 1} \right) = 2.8}}} \right. \kern-\nulldelimiterspace} {\left( {\alpha  - 1} \right) = 2.8}}$.
\end{lemma}
\begin{lemma} \label{Lemma__LoadFunctionModified}
As far as Load Model 2 is concerned, the function ${{\mathcal{M}}\left( {{{{\lambda _{{\rm{MT}}}}} \mathord{\left/{\vphantom {{{\lambda _{{\rm{MT}}}}} {{\lambda _{{\rm{BS}}}}}}} \right. \kern-\nulldelimiterspace} {{\lambda _{{\rm{BS}}}}}}} \right)}$ fulfills the following properties with respect to ${{\lambda _{{\rm{BS}}}}}$ (assuming ${{\lambda _{{\rm{MT}}}}}$ fixed):
i) ${\mathcal{M}}\left( {{{{\lambda _{{\rm{MT}}}}} \mathord{\left/ {\vphantom {{{\lambda _{{\rm{MT}}}}} {{\lambda _{{\rm{BS}}}}}}} \right.\kern-\nulldelimiterspace} {{\lambda _{{\rm{BS}}}}}}} \right) \ge 0$ for ${{\lambda _{{\rm{BS}}}}} \ge 0$;
ii) ${\mathcal{M}}\left( {{{{\lambda _{{\rm{MT}}}}} \mathord{\left/ {\vphantom {{{\lambda _{{\rm{MT}}}}} {{\lambda _{{\rm{BS}}}}}}} \right.\kern-\nulldelimiterspace} {{\lambda _{{\rm{BS}}}}}}} \right) \to \infty$ if ${{\lambda _{{\rm{BS}}}}} \to 0$;
iii) ${\mathcal{M}}\left( {{{{\lambda _{{\rm{MT}}}}} \mathord{\left/ {\vphantom {{{\lambda _{{\rm{MT}}}}} {{\lambda _{{\rm{BS}}}}}}} \right.\kern-\nulldelimiterspace} {{\lambda _{{\rm{BS}}}}}}} \right) = 0$ if ${{\lambda _{{\rm{BS}}}}} \to \infty$;
iv) ${{{\mathcal{\dt M}}}_{{\lambda _{{\rm{BS}}}}}}\left( {{{{\lambda _{{\rm{MT}}}}} \mathord{\left/ {\vphantom {{{\lambda _{{\rm{MT}}}}} {{\lambda _{{\rm{BS}}}}}}} \right. \kern-\nulldelimiterspace} {{\lambda _{{\rm{BS}}}}}}} \right) \le 0$ for ${{\lambda _{{\rm{BS}}}}} \ge 0$; and
v) ${{{\mathcal{\ddt M}}}_{{\lambda _{{\rm{BS}}}}}}\left( {{{{\lambda _{{\rm{MT}}}}} \mathord{\left/ {\vphantom {{{\lambda _{{\rm{MT}}}}} {{\lambda _{{\rm{BS}}}}}}} \right.\kern-\nulldelimiterspace} {{\lambda _{{\rm{BS}}}}}}} \right) \ge 0$ for ${{\lambda _{{\rm{BS}}}}} \ge 0$.
\end{lemma}
\begin{lemma} \label{Lemma__PilotFunctionPower}
The function ${\mathcal{Q}}\left( {{\lambda _{{\rm{BS}}}},{{\rm{P}}_{{\rm{tx}}}},{{{\lambda _{{\rm{MT}}}}} \mathord{\left/ {\vphantom {{{\lambda _{{\rm{MT}}}}} {{\lambda _{{\rm{BS}}}}}}} \right. \kern-\nulldelimiterspace} {{\lambda _{{\rm{BS}}}}}}} \right)$ fulfills the following properties with respect to ${{{\rm{P}}_{{\rm{tx}}}}}$:
i) ${\mathcal{Q}}\left( {{\lambda _{{\rm{BS}}}},{{\rm{P}}_{{\rm{tx}}}},{{{\lambda _{{\rm{MT}}}}} \mathord{\left/ {\vphantom {{{\lambda _{{\rm{MT}}}}} {{\lambda _{{\rm{BS}}}}}}} \right. \kern-\nulldelimiterspace} {{\lambda _{{\rm{BS}}}}}}} \right) \ge 0$ for ${{\rm{P}}_{{\rm{tx}}}} \ge 0$;
ii) ${\mathcal{Q}}\left( {{\lambda _{{\rm{BS}}}},{{\rm{P}}_{{\rm{tx}}}},{{{\lambda _{{\rm{MT}}}}} \mathord{\left/ {\vphantom {{{\lambda _{{\rm{MT}}}}} {{\lambda _{{\rm{BS}}}}}}} \right. \kern-\nulldelimiterspace} {{\lambda _{{\rm{BS}}}}}}} \right) = 0$ if ${{\rm{P}}_{{\rm{tx}}}} \to 0$;
iii) ${\mathcal{Q}}\left( {{\lambda _{{\rm{BS}}}},{{\rm{P}}_{{\rm{tx}}}},{{{\lambda _{{\rm{MT}}}}} \mathord{\left/ {\vphantom {{{\lambda _{{\rm{MT}}}}} {{\lambda _{{\rm{BS}}}}}}} \right. \kern-\nulldelimiterspace} {{\lambda _{{\rm{BS}}}}}}} \right) = 1$ if ${{\rm{P}}_{{\rm{tx}}}} \to \infty$;
iv) ${{{\mathcal{\dt Q}}}_{{{\rm{P}}_{{\rm{tx}}}}}}\left( {{\lambda _{{\rm{BS}}}},{{\rm{P}}_{{\rm{tx}}}},{{{\lambda _{{\rm{MT}}}}} \mathord{\left/ {\vphantom {{{\lambda _{{\rm{MT}}}}} {{\lambda _{{\rm{BS}}}}}}} \right. \kern-\nulldelimiterspace} {{\lambda _{{\rm{BS}}}}}}} \right) \ge 0$ for ${{\rm{P}}_{{\rm{tx}}}} \ge 0$; and
v) ${{{\mathcal{\ddt Q}}}_{{{\rm{P}}_{{\rm{tx}}}}}}\left( {{\lambda _{{\rm{BS}}}},{{\rm{P}}_{{\rm{tx}}}},{{{\lambda _{{\rm{MT}}}}} \mathord{\left/ {\vphantom {{{\lambda _{{\rm{MT}}}}} {{\lambda _{{\rm{BS}}}}}}} \right. \kern-\nulldelimiterspace} {{\lambda _{{\rm{BS}}}}}}} \right) \le 0$ for ${{\rm{P}}_{{\rm{tx}}}} \ge 0$.

\emph{Proof}: The result in v) follows from ${{{\mathcal{\ddt Q}}}_{{{\rm{P}}_{{\rm{tx}}}}}}\left( \cdot, \cdot, \cdot  \right)$ in Table \ref{Table_Functions}, because iv) and $\beta > 2$ hold. \hfill $\Box$
\end{lemma}
\begin{lemma} \label{Lemma__PilotFunctionDensity}
The function ${\mathcal{Q}}\left( {{\lambda _{{\rm{BS}}}},{{\rm{P}}_{{\rm{tx}}}},{{{\lambda _{{\rm{MT}}}}} \mathord{\left/ {\vphantom {{{\lambda _{{\rm{MT}}}}} {{\lambda _{{\rm{BS}}}}}}} \right. \kern-\nulldelimiterspace} {{\lambda _{{\rm{BS}}}}}}} \right)$ fulfills the following properties with respect to ${{\lambda _{{\rm{BS}}}}}$ (assuming ${{\lambda _{{\rm{MT}}}}}$ fixed):
i) ${\mathcal{Q}}\left( {{\lambda _{{\rm{BS}}}},{{\rm{P}}_{{\rm{tx}}}},{{{\lambda _{{\rm{MT}}}}} \mathord{\left/ {\vphantom {{{\lambda _{{\rm{MT}}}}} {{\lambda _{{\rm{BS}}}}}}} \right. \kern-\nulldelimiterspace} {{\lambda _{{\rm{BS}}}}}}} \right) \ge 0$ for ${{\lambda _{{\rm{BS}}}}} \ge 0$;
ii) ${\mathcal{Q}}\left( {{\lambda _{{\rm{BS}}}},{{\rm{P}}_{{\rm{tx}}}},{{{\lambda _{{\rm{MT}}}}} \mathord{\left/ {\vphantom {{{\lambda _{{\rm{MT}}}}} {{\lambda _{{\rm{BS}}}}}}} \right. \kern-\nulldelimiterspace} {{\lambda _{{\rm{BS}}}}}}} \right) = 0$ if ${{\lambda _{{\rm{BS}}}}} \to 0$;
iii) ${\mathcal{Q}}\left( {{\lambda _{{\rm{BS}}}},{{\rm{P}}_{{\rm{tx}}}},{{{\lambda _{{\rm{MT}}}}} \mathord{\left/ {\vphantom {{{\lambda _{{\rm{MT}}}}} {{\lambda _{{\rm{BS}}}}}}} \right. \kern-\nulldelimiterspace} {{\lambda _{{\rm{BS}}}}}}} \right) = 1$ if ${{\lambda _{{\rm{BS}}}}} \to \infty$;
iv) ${{{\mathcal{\dt Q}}}_{{{\lambda _{{\rm{BS}}}}}}}\left( {{\lambda _{{\rm{BS}}}},{{\rm{P}}_{{\rm{tx}}}},{{{\lambda _{{\rm{MT}}}}} \mathord{\left/ {\vphantom {{{\lambda _{{\rm{MT}}}}} {{\lambda _{{\rm{BS}}}}}}} \right. \kern-\nulldelimiterspace} {{\lambda _{{\rm{BS}}}}}}} \right) \ge 0$ for ${{\lambda _{{\rm{BS}}}}} \ge 0$; and
v) ${{{\mathcal{\ddt Q}}}_{{{\lambda _{{\rm{BS}}}}}}}\left( {{\lambda _{{\rm{BS}}}},{{\rm{P}}_{{\rm{tx}}}},{{{\lambda _{{\rm{MT}}}}} \mathord{\left/ {\vphantom {{{\lambda _{{\rm{MT}}}}} {{\lambda _{{\rm{BS}}}}}}} \right. \kern-\nulldelimiterspace} {{\lambda _{{\rm{BS}}}}}}} \right) \le 0$ for ${{\lambda _{{\rm{BS}}}}} \ge 0$.

\emph{Proof}: The result in iv) follows from ${{{\mathcal{\dt Q}}}_{{{\lambda _{{\rm{BS}}}}}}}\left( \cdot, \cdot, \cdot \right)$ in Table \ref{Table_Functions} because, for ${\lambda _{{\rm{BS}}}} \ge 0$, ${\mathcal{L}}\left( {{{{\lambda _{{\rm{MT}}}}} \mathord{\left/ {\vphantom {{{\lambda _{{\rm{MT}}}}} {{\lambda _{{\rm{BS}}}}}}} \right. \kern-\nulldelimiterspace} {{\lambda _{{\rm{BS}}}}}}} \right) + {\lambda _{{\rm{BS}}}}{{{\mathcal{\dt L}}}_{{\lambda _{{\rm{BS}}}}}}\left( {{{{\lambda _{{\rm{MT}}}}} \mathord{\left/ {\vphantom {{{\lambda _{{\rm{MT}}}}} {{\lambda _{{\rm{BS}}}}}}} \right. \kern-\nulldelimiterspace} {{\lambda _{{\rm{BS}}}}}}} \right) \ge 0$. This latter inequality holds true because $1 + x\left( {1 + {1 \mathord{\left/ {\vphantom {1 \alpha }} \right. \kern-\nulldelimiterspace} \alpha }} \right) \le {\left( {1 + {x \mathord{\left/ {\vphantom {x \alpha }} \right. \kern-\nulldelimiterspace} \alpha }} \right)^{\left( {\alpha  + 1} \right)}}$ for $x \ge 0$. The result in v) follows without explicitly computing ${{{\mathcal{\ddt Q}}}_{{{\lambda _{{\rm{BS}}}}}}}\left( \cdot, \cdot, \cdot \right)$ because ${{{\mathcal{\dt Q}}}_{{{\lambda _{{\rm{BS}}}}}}}\left( \cdot, \cdot, \cdot \right)$ in Table \ref{Table_Functions} is the composition of two increasing and concave functions in ${\lambda _{{\rm{BS}}}}$, i.e., the function in the square brackets in the first row and the exponential function in the second row. \hfill $\Box$
\end{lemma}
\begin{lemma} \label{Lemma__EE}
The EE in \eqref{Eq_14} fulfills the following properties with respect to ${{\rm{P}}_{{\rm{tx}}}}$ and ${{\lambda _{{\rm{BS}}}}}$:
i) ${\rm{EE}}\left( {{{\rm{P}}_{{\rm{tx}}}},{\lambda _{{\rm{BS}}}}} \right) = 0$ if ${{\rm{P}}_{{\rm{tx}}}} \to 0 $ or ${{\lambda _{{\rm{BS}}}}} \to 0$; and
ii) ${\rm{EE}}\left( {{{\rm{P}}_{{\rm{tx}}}},{\lambda _{{\rm{BS}}}}} \right) = 0$ if ${{\rm{P}}_{{\rm{tx}}}} \to \infty$ or ${{\lambda _{{\rm{BS}}}}} = \to \infty$.

\emph{Proof}: This immediately follows from \textit{Lemmas \ref{Lemma__LoadFunction}-\ref{Lemma__PilotFunctionDensity}}. \hfill $\Box$
\end{lemma}
\subsection{Optimal Transmit Power Given the Density of the BSs} \label{OptimalPower_FixedDensity}
In this section, we analyze whether there exists an optimal and unique transmit power, ${\rm{P}}_{{\rm{tx}}}^{\left( {{\rm{opt}}} \right)}$, that maximizes the EE formulated in \eqref{Eq_14}, while all the other parameters, including ${{\lambda _{{\rm{BS}}}}}$, are fixed and given. In mathematical terms, the optimization problem can be formulated as follows:
\setcounter{equation}{14}
\begin{equation}
\label{Eq_15pre}
\begin{split}
& {\rm{ma}}{{\rm{x}}_{{{\rm{P}}_{{\rm{tx}}}}}}\quad {\rm{EE}}\left( {{{\rm{P}}_{{\rm{tx}}}},{\lambda _{{\rm{BS}}}}} \right) \\
& {\rm{subject \; to}}\quad{{\rm{P}}_{{\rm{tx}}}} \in \left[ {{\rm{P}}_{{\rm{tx}}}^{\left( {\min } \right)},{\rm{P}}_{{\rm{tx}}}^{\left( {\max } \right)}} \right].
\end{split}
\end{equation}
\noindent where ${{\rm{P}}_{{\rm{tx}}}^{\left( {\min } \right)}} \ge 0$ and ${{\rm{P}}_{{\rm{tx}}}^{\left( {\max } \right)}} \ge 0$ are the minimum and maximum power budget of the BSs, respectively. One may assume, without loss of generality, ${{\rm{P}}_{{\rm{tx}}}^{\left( {\min } \right)}} \to 0$ and ${{\rm{P}}_{{\rm{tx}}}^{\left( {\max } \right)}} \to \infty$.

The following theorem completely characterizes the solution of \eqref{Eq_15pre}.
\begin{theorem} \label{Theorem__OptimalPower}
Let ${{\mathcal{S}}_{\mathcal{P}}}\left( \cdot \right)$ be the function defined in Table \ref{Table_Functions}. The EE in \eqref{Eq_14} is a unimodal and strictly pseudo-concave function in ${{{\rm{P}}_{{\rm{tx}}}}}$. The optimization problem in \eqref{Eq_15pre} has a unique solution given by ${\rm{P}}_{{\rm{tx}}}^{\left( {{\rm{opt}}} \right)} = \max \left\{ {{\rm{P}}_{{\rm{tx}}}^{\left( {\min } \right)},\min \left\{ {{{{{\rm P}}}^*_{{\rm{tx}}}},{\rm{P}}_{{\rm{tx}}}^{\left( {\max } \right)}} \right\}} \right\}$, where ${{{{\rm{P}}}^*_{{\rm{tx}}}}}$ is the only stationary point of the EE in \eqref{Eq_14} that is obtained as the unique solution of the following equation:
\setcounter{equation}{15}
\begin{equation}
\label{Eq_15}
\begin{split}
{\dt {\rm{EE}}_{{{\rm{P}}_{{\rm{tx}}}}}}\left( {{\rm{P}}^*_{{\rm{tx}}},{\lambda _{{\rm{BS}}}}} \right) & = {{\rm{P}}_{{\rm{idle}}}} - {{\mathcal{S}}_{\mathcal{P}}}\left( {{\rm{P}}^*_{{\rm{tx}}}} \right) = 0 \\ & \Leftrightarrow  {{\mathcal{S}}_{\mathcal{P}}}\left( {{{\rm{P}}^*_{{\rm{tx}}}}} \right)= {{\rm{P}}_{{\rm{idle}}}}.
\end{split}
\end{equation}

\emph{Proof}: See Appendix \ref{Appendix_TheoPtx}. \hfill $\Box$
\end{theorem}
\subsection{Optimal Density Given the Transmit Power of the BSs} \label{OptimalDensity_GivenPower}
In this section, we analyze whether there exists an optimal and unique density of BSs, $\lambda _{{\rm{BS}}}^{\left( {{\rm{opt}}} \right)}$, that maximizes the EE formulated in \eqref{Eq_14}, while all the other parameters, including ${{{\rm{P}}_{{\rm{tx}}}}}$, are fixed and given. In mathematical terms, the optimization problem can be formulated as follows:
\setcounter{equation}{16}
\begin{equation}
\label{Eq_16pre}
\begin{split}
& {\rm{ma}}{{\rm{x}}_{{\lambda _{{\rm{BS}}}}}}\quad{\rm{EE}}\left( {{{\rm{P}}_{{\rm{tx}}}},{\lambda _{{\rm{BS}}}}} \right) \\ & {\rm{subject}}\;{\rm{to}}\quad{\lambda _{{\rm{BS}}}} \in \left[ {\lambda _{{\rm{BS}}}^{\left( {\min } \right)},\lambda _{{\rm{BS}}}^{\left( {\max } \right)}} \right]
\end{split}
\end{equation}
\noindent where ${\lambda _{{\rm{BS}}}^{\left( {\min } \right)}} \ge 0$ and ${\lambda _{{\rm{BS}}}^{\left( {\max } \right)}} \ge 0$ are the minimum and maximum allowed density of the BSs, respectively. One may assume, without loss of generality, ${\lambda _{{\rm{BS}}}^{\left( {\min } \right)}} \to 0$ and ${\lambda _{{\rm{BS}}}^{\left( {\min } \right)}} \to \infty$.

The following theorem completely characterizes the solution of \eqref{Eq_16pre}.
\begin{theorem} \label{Theorem__OptimalLambda}
Let ${{\mathcal{S}}_{\mathcal{D}}}\left( \cdot \right)$ be the function defined in Table \ref{Table_Functions}. The EE in \eqref{Eq_14} is a unimodal and strictly pseudo-concave function in ${{\lambda _{{\rm{BS}}}}}$. The optimization problem in \eqref{Eq_16pre} has a unique solution given by ${\lambda}_{{\rm{BS}}}^{\left( {{\rm{opt}}} \right)} = \max \left\{ {{\lambda}_{{\rm{BS}}}^{\left( {\min } \right)},\min \left\{ {{{{\rm{ \lambda}}}_{{\rm{BS}}}^*},{\lambda}_{{\rm{BS}}}^{\left( {\max } \right)}} \right\}} \right\}$, where ${{{{\rm{\lambda}}}^*_{{\rm{BS}}}}}$ is the only stationary point of the EE in \eqref{Eq_14} that is obtained as the unique solution of the following equation:
\setcounter{equation}{17}
\begin{equation}
\label{Eq_16}
\begin{split}
{\dt {\rm{EE}}_{{\lambda _{{\rm{BS}}}}}}\left( {{{\rm{P}}_{{\rm{tx}}}}, \lambda^* _{{\rm{BS}}}} \right) & = {{\mathcal{S}}_{\mathcal{D}}}\left( {\lambda^* _{{\rm{BS}}}} \right) - {{\rm{P}}_{{\rm{idle}}}} = 0 \\ & \Leftrightarrow {{\mathcal{S}}_{\mathcal{D}}}\left( {{\lambda^* _{{\rm{BS}}}}} \right) = {{\rm{P}}_{{\rm{idle}}}}.
\end{split}
\end{equation}

\emph{Proof}: See Appendix \ref{Appendix_TheoLambda}. \hfill $\Box$
\end{theorem}
\subsection{On the Dependency of Optimal Transmit Power and Density of the BSs} \label{Dependency}
The optimal transmit power and BSs' density that maximize the EE are obtained from the unique solutions of \eqref{Eq_15} and \eqref{Eq_16}, respectively. These equations, however, cannot be further simplified and, therefore, explicit analytical expressions for ${{\rm{P}}_{{\rm{tx}}}^{\left( {{\rm{opt}}} \right)}}$ and ${\lambda _{{\rm{BS}}}^{\left( {{\rm{opt}}} \right)}}$ cannot, in general, be obtained. This is an inevitable situation when dealing with EE optimization problems, and, indeed, a closed-form expression of the optimal transmit power for simpler EE optimization problems does not exist either \cite{AlessioSurvey2016}. In some special cases, the transmit power can be implicitly expressed in terms of the Lambert-W function, which, however, is the solution of a transcendental equation \cite{AlessioMonograph2015}. Notable examples of these case studies include even basic point-to-point communication systems without interference \cite{LambertW}. Based on these considerations, it seems hopeless to attempt finding explicit analytical expressions from \eqref{Eq_15} and \eqref{Eq_16}, respectively. However, thanks to the properties of the EE function, i.e., unimodality and strict pseudo-concavity, proved in \textit{Theorem \ref{Theorem__OptimalPower}} and \textit{Theorem \ref{Theorem__OptimalLambda}}, ${{\rm{P}}_{{\rm{tx}}}^{\left( {{\rm{opt}}} \right)}}$ and ${\lambda _{{\rm{BS}}}^{\left( {{\rm{opt}}} \right)}}$ can be efficiently computed with the aid of numerical methods that are routinely employed to obtain the roots of non-linear scalar equations, e.g., the Newton's method \cite{NewtonMethod}. For example, the unique solutions of \eqref{Eq_15} and \eqref{Eq_16} may be obtained by using the functions \texttt{FSolve} in Matlab and \texttt{NSolve} in Mathematica. \textit{Theorem \ref{Theorem__OptimalPower}} and \textit{Theorem \ref{Theorem__OptimalLambda}} are, however, of paramount importance, since they state that an optimum maximizer exists and is unique.

Even though explicit analytical formulas for ${{\rm{P}}_{{\rm{tx}}}^{\left( {{\rm{opt}}} \right)}}$ and ${\lambda _{{\rm{BS}}}^{\left( {{\rm{opt}}} \right)}}$ cannot be obtained, it is important to understand how these optimal values change if any other system parameter changes. For instance, two worthwhile questions to answer are: ``\textit{How does ${{\rm{P}}_{{\rm{tx}}}^{\left( {{\rm{opt}}} \right)}}$ change as a function of ${{\lambda _{{\rm{BS}}}}}$}?'' and ``\textit{How does ${{\lambda}_{{\rm{BS}}}^{\left( {{\rm{opt}}} \right)}}$ change as a function of ${{\rm{P}}_{{\rm{tx}}}}$}?''. These questions are relevant to optimize the deployment of cellular networks from the EE standpoint, since they unveil the inherent interplay between transmit power and density of BSs discussed in Section \ref{PSE} and illustrated in Fig. \ref{Fig_1}. A general answer to these two questions is provided in the following two propositions.
\begin{proposition} \label{Proposition__DependencyPower}
Let $\overline {\rm{P}}^* _{{\rm{tx}}}$ be the unique solution of \eqref{Eq_15} if ${\lambda _{{\rm{BS}}}} = {\overline \lambda  _{{\rm{BS}}}}$. Let the optimal ${{\rm{P}}_{{\rm{tx}}}}$ according to \textit{Theorem \ref{Theorem__OptimalPower}} be ${\rm{\overline P}}_{{\rm{tx}}}^{\left( {{\rm{opt}}} \right)} = \max \left\{ {{\rm{P}}_{{\rm{tx}}}^{\left( {\min } \right)},\min \left\{ {{{{{\overline {\rm{P}}}}}^*_{{\rm{tx}}}},{\rm{P}}_{{\rm{tx}}}^{\left( {\max } \right)}} \right\}} \right\}$. Let ${\overline{\overline \lambda } _{{\rm{BS}}}} \lessgtr {\overline \lambda  _{{\rm{BS}}}}$ be another BSs' density. Let ${\dt{\rm{EE}}_{{{\rm{P}}_{{\rm{tx}}}}}}\left( { \cdot , \cdot } \right)$ be the first-order derivative in \eqref{Eq_15}. The following holds:
\setcounter{equation}{18}
\begin{equation}
\label{Eq_17}
\overline{\overline {\rm{P}}} _{{\rm{tx}}}^{\left( {{\rm{opt}}} \right)} \lesseqgtr \overline {\rm{P}} _{{\rm{tx}}}^{\left( {{\rm{opt}}} \right)} \Leftrightarrow {\dt{\rm{EE}}_{{{\rm{P}}_{{\rm{tx}}}}}}\left( {\overline {\rm{P}} _{{\rm{tx}}}^{\left( {{\rm{opt}}} \right)},{{\overline{\overline \lambda } }_{{\rm{BS}}}}} \right) \lesseqgtr 0.
\end{equation}

\emph{Proof}: \textit{Theorem \ref{Theorem__OptimalPower}} states that the EE function has a single stationary point that is its unique global maximizer. In mathematical terms, this implies ${\dt{\rm{EE}}_{{{\rm{P}}_{{\rm{tx}}}}}}\left( {{\rm{P}}_{{\rm{tx}}},{\lambda _{{\rm{BS}}}}} \right) > 0$ if ${{\rm{P}}_{{\rm{tx}}}} < {\rm{P}}^*_{{\rm{tx}}}$ and ${\dt{\rm{EE}}_{{{\rm{P}}_{{\rm{tx}}}}}}\left( {{\rm{P}}_{{\rm{tx}}},{\lambda _{{\rm{BS}}}}} \right) < 0$ if ${{\rm{P}}_{{\rm{tx}}}} > {\rm{P}}^*_{{\rm{tx}}}$ for every ${\lambda _{{\rm{BS}}}} \ge 0$. Therefore, the optimal transmit power needs to be increased (decreased) if the first-order derivative of the EE is positive (negative). Based on this, \eqref{Eq_17} follows because $\min \left\{ { \cdot , \cdot } \right\}$ and $\max \left\{ { \cdot , \cdot } \right\}$ are increasing functions. \hfill $\Box$
\end{proposition}
\begin{proposition} \label{Proposition__DependencyLambda}
Let $\overline \lambda^* _{{\rm{BS}}}$ be the unique solution of \eqref{Eq_16} if ${{\rm{P}}_{{\rm{tx}}}} = {\overline {\rm{P}} _{{\rm{tx}}}}$. Let the optimal ${{\lambda_{{\rm{BS}}}}}$ according to \textit{Theorem \ref{Theorem__OptimalLambda}} be ${{\overline \lambda}}_{{\rm{BS}}}^{\left( {{\rm{opt}}} \right)} = \max \left\{ {{\lambda}_{{\rm{BS}}}^{\left( {\min } \right)},\min \left\{ {{{{{\overline {\lambda}}}}^*_{{\rm{BS}}}},{\lambda}_{{\rm{BS}}}^{\left( {\max } \right)}} \right\}} \right\}$. Let ${\overline{\overline {\rm{P}}} _{{\rm{tx}}}} \lessgtr {\overline {\rm{P}} _{{\rm{tx}}}}$ be another transmit power. Let ${\dt{\rm{EE}}_{{\lambda _{{\rm{BS}}}}}}\left( { \cdot , \cdot } \right)$ be the first-order derivative in \eqref{Eq_16}. The following holds:
\setcounter{equation}{19}
\begin{equation}
\label{Eq_18}
\overline{\overline \lambda } _{{\rm{BS}}}^{\left( {{\rm{opt}}} \right)} \lesseqgtr \overline \lambda  _{{\rm{BS}}}^{\left( {{\rm{opt}}} \right)} \Leftrightarrow {\dt{\rm{EE}}_{{\lambda _{{\rm{BS}}}}}}\left( {{{\overline{\overline {\rm{P}}} }_{{\rm{tx}}}},\overline \lambda  _{{\rm{BS}}}^{\left( {{\rm{opt}}} \right)}} \right) \lesseqgtr 0.
\end{equation}

\emph{Proof}: It follows from \textit{Theorem \ref{Theorem__OptimalLambda}}, similar to the proof of \textit{Proposition \ref{Proposition__DependencyPower}}. \hfill $\Box$
\end{proposition}
\begin{remark} \label{Remark__Dependency}
It is worth mentioning that the approach utilized to prove \textit{Proposition \ref{Proposition__DependencyPower}} and \textit{Proposition \ref{Proposition__DependencyLambda}} is applicable to study the dependency of ${{\rm{P}}_{{\rm{tx}}}^{\left( {{\rm{opt}}} \right)}}$ and ${\lambda _{{\rm{BS}}}^{\left( {{\rm{opt}}} \right)}}$, respectively, with respect to any other system parameters. The findings in \textit{Proposition \ref{Proposition__DependencyPower}} and \textit{Proposition \ref{Proposition__DependencyLambda}} are especially relevant for cellular network planning. Let us consider, e.g., \eqref{Eq_17}. By simply studying the sign of the first-order derivative ${\dt{\rm{EE}}_{{{\rm{P}}_{{\rm{tx}}}}}}\left( { \cdot , \cdot } \right)$, one can identify, with respect to an optimally deployed cellular network, the set of BSs' densities that would require to increase or decrease the transmit power while still operating at the optimum. In Section \ref{Results}, numerical examples are shown to highlight that ${{\rm{P}}_{{\rm{tx}}}^{\left( {{\rm{opt}}} \right)}}$ may either decrease or increase as ${{\lambda _{{\rm{BS}}}}}$ increases or decreases. \hfill $\Box$
\end{remark}
\begin{table*}[!t] 
\centering
\caption{Alternating optimization of the EE.}
\begin{tabular}{l} \hline
\textbf{Algorithm} \\ \hline
\texttt{Let} ${{\rm{P}}_{{\rm{tx}}}} \in \left[ {{\rm{P}}_{{\rm{tx}}}^{\left( {{\rm{min}}} \right)},{\rm{P}}_{{\rm{tx}}}^{\left( {{\rm{max}}} \right)}} \right]$; ${\lambda  _{{\rm{BS}}}} \in \left[ {\lambda _{{\rm{BS}}}^{\left( {{\rm{min}}} \right)},\lambda _{{\rm{BS}}}^{\left( {{\rm{max}}} \right)}} \right]$; \\
\texttt{Set} ${\lambda _{{\rm{BS}}}} = {\overline \lambda^{\left( {{\rm{opt}}} \right)}  _{{\rm{BS}}}} \in \left[ {\lambda _{{\rm{BS}}}^{\left( {{\rm{min}}} \right)},\lambda _{{\rm{BS}}}^{\left( {{\rm{max}}} \right)}} \right]$ (initial guess); $V = 0$; $\epsilon > 0$; \\
\textbf{Repeat} \\
\quad ${V_0} = V$; \\
\quad ${\overline {\rm{P}}^* _{{\rm{tx}}}} \leftarrow {\dt{\rm{EE}}_{{{\rm{P}}_{{\rm{tx}}}}}}\left( {{{ {\rm{P}} }_{{\rm{tx}}}},{{\overline \lambda}^{\left( {{\rm{opt}}} \right)}_{{\rm{BS}}}}} \right) = 0$; \quad ${\rm{\overline P}}_{{\rm{tx}}}^{\left( {{\rm{opt}}} \right)} = \max \left\{ {{\rm{P}}_{{\rm{tx}}}^{\left( {{\rm{min}}} \right)},\min \left\{ {{\rm{\overline P}}_{{\rm{tx}}}^*,{\rm{P}}_{{\rm{tx}}}^{\left( {{\rm{max}}} \right)}} \right\}} \right\}$; \, \, \, \quad \, \hspace{0.02mm} \eqref{Eq_15} \\
\quad ${\overline \lambda^*  _{{\rm{BS}}}} \leftarrow {\dt{\rm{EE}}_{{\lambda _{{\rm{BS}}}}}}\left( {{{\overline {\rm{P}} }^{\left( {{\rm{opt}}} \right)}_{{\rm{tx}}}},{{ \lambda  }_{{\rm{BS}}}}} \right) = 0$; \quad $\overline \lambda _{{\rm{BS}}}^{\left( {{\rm{opt}}} \right)} = \max \left\{ {\lambda _{{\rm{BS}}}^{\left( {{\rm{min}}} \right)},\min \left\{ {\overline \lambda _{{\rm{BS}}}^*,\lambda _{{\rm{BS}}}^{\left( {{\rm{max}}} \right)}} \right\}} \right\}$; \quad \quad \quad \eqref{Eq_16} \\
\quad $V = {\rm{EE}}\left( {{{\overline {\rm{P}} }^{\left( {{\rm{opt}}} \right)}_{{\rm{tx}}}},{{\overline \lambda  }^{\left( {{\rm{opt}}} \right)}_{{\rm{BS}}}}} \right)$; \quad \quad \quad \quad \quad \quad \quad \quad \quad \quad \quad \quad \quad \quad \quad \quad \quad \quad \quad \quad \quad \quad \quad \quad \quad \, \, \hspace{0.0001mm} \eqref{Eq_14} \\
\textbf{Until} $\left| {V - {V_0}} \right|/V \le \epsilon$; \\
\texttt{Return} ${\rm{P}}_{{\rm{tx}}}^{\left( {{\rm{opt}}} \right)} = {\overline {\rm{P}}^{\left( {{\rm{opt}}} \right)} _{{\rm{tx}}}}$; $\lambda _{{\rm{BS}}}^{\left( {{\rm{opt}}} \right)} = {\overline \lambda^{\left( {{\rm{opt}}} \right)}  _{{\rm{BS}}}}$. \\ \hline
\end{tabular}
\label{Table_AlternatingOpt}
\end{table*}
\subsection{Joint Optimization of Transmit Power and Density of the BSs} \label{JointOptimization}
In Sections \ref{OptimalPower_FixedDensity} and \ref{OptimalDensity_GivenPower}, either ${{\lambda _{{\rm{BS}}}}}$ or ${{{\rm{P}}_{{\rm{tx}}}}}$ are assumed to be given, respectively. In practical applications, however, it is important to identify the optimal pair $\left( {{\rm{P}}_{{\rm{tx}}}^{\left( {{\rm{opt}}} \right)},\lambda _{{\rm{BS}}}^{\left( {{\rm{opt}}} \right)}} \right)$ that \textit{jointly} maximizes the EE in \eqref{Eq_14}. This joint optimization problem can be formulated as follows:
\setcounter{equation}{20}
\begin{equation}
\label{Eq_19}
\begin{split}
&{\rm{ma}}{{\rm{x}}_{{{\rm{P}}_{{\rm{tx}}}},{\lambda _{{\rm{BS}}}}}}\quad{\rm{EE}}\left( {{{\rm{P}}_{{\rm{tx}}}},{\lambda _{{\rm{BS}}}}} \right)\\
& {\rm{subject}}\;{\rm{to}}\quad{{\rm{P}}_{{\rm{tx}}}} \in \left[ {{\rm{P}}_{{\rm{tx}}}^{\left( {\min } \right)},{\rm{P}}_{{\rm{tx}}}^{\left( {\max } \right)}} \right], \;{\lambda _{{\rm{BS}}}} \in \left[ {\lambda _{{\rm{BS}}}^{\left( {\min } \right)},\lambda _{{\rm{BS}}}^{\left( {\max } \right)}} \right]
\end{split}
\end{equation}
\noindent where a notation similar to that used in \eqref{Eq_15pre} and \eqref{Eq_16pre} is adopted.

In \textit{Theorem \ref{Theorem__OptimalPower}} and \textit{Theorem \ref{Theorem__OptimalLambda}}, we have solved the optimization problem formulated in \eqref{Eq_19} with respect to ${{\rm{P}}_{{\rm{tx}}}}$ for a given ${\lambda _{{\rm{BS}}}}$ and with respect to ${\lambda _{{\rm{BS}}}}$ for a given ${{\rm{P}}_{{\rm{tx}}}}$, respectively. By leveraging these results, a convenient approach for tackling \eqref{Eq_19} with respect to ${{\rm{P}}_{{\rm{tx}}}}$ and ${\lambda _{{\rm{BS}}}}$ is to utilize the alternating optimization method, which iteratively optimizes ${{{\rm{P}}_{{\rm{tx}}}}}$ for a given ${{\lambda _{{\rm{BS}}}}}$ and ${{\lambda _{{\rm{BS}}}}}$ for a given ${{{\rm{P}}_{{\rm{tx}}}}}$ until convergence of the EE in \eqref{Eq_14} within a desired level of accuracy \cite[Proposition 2.7.1]{BertsekasBook}. The algorithm that solves \eqref{Eq_19} based on the alternating optimization method is reported in Table \ref{Table_AlternatingOpt}. Its convergence and optimality properties are summarized as follows.
\begin{proposition} \label{Proposition__Alessio}
Let ${{\rm{\overline P}}_{{\rm{tx}}}^{\left( {{\rm{opt}}} \right)}\left( m \right)}$, ${\overline \lambda _{{\rm{BS}}}^{\left( {{\rm{opt}}} \right)}\left( m \right)}$, and ${\rm EE}(m)$ be ${{\rm{P}}_{{\rm{tx}}}}$, $\lambda_{{\rm{BS}}}$ and EE obtained from the algorithm in Table \ref{Table_AlternatingOpt} at the $m$th iteration, respectively. The sequence ${\rm EE}(m)$ is monotonically increasing and converges. In addition, every limit point of the sequence $\left( {{\rm{\overline P}}_{{\rm{tx}}}^{\left( {{\rm{opt}}} \right)}\left( m \right),\overline \lambda _{{\rm{BS}}}^{\left( {{\rm{opt}}} \right)}\left( m \right)} \right)$ fulfills the Karush-Kuhn-Tucker (KKT) first-order optimality conditions of the problem in \eqref{Eq_19}.

\emph{Proof}: At the end of each iteration of the algorithm in Table \ref{Table_AlternatingOpt}, the value of EE does not decrease. The sequence ${\rm EE}(m)$, hence, converges, because the EE in \eqref{Eq_14} is a continuous function over the compact feasible set of the problem in \eqref{Eq_19} and, thus, it admits a finite maximum by virtue of the Weierstrass extreme value theorem \cite{BertsekasBook}. From \cite[Proposition 2.7.1]{BertsekasBook}, the alternating optimization method fulfills the KKT optimality conditions, provided that i) the objective and constraint functions are differentiable, ii) each constraint function depends on a single variable, and iii) each subproblem has a unique solution. The first and second requirements follow by direct inspection of \eqref{Eq_19}. The third requirement is ensured by \textit{Theorem \ref{Theorem__OptimalPower}} and \textit{Theorem \ref{Theorem__OptimalLambda}}. \hfill $\Box$
\end{proposition}
\begin{remark} \label{Remark__ConvergeRate}
The optimization problems in \textit{Theorem \ref{Theorem__OptimalPower}} and \textit{Theorem \ref{Theorem__OptimalLambda}} can be efficiently solved by using the Newton's method, which allows one to find the root of real-valued objective functions via multiple iterations of increasing accuracy and at a super-linear (i.e., quadratic if the initial guess is sufficiently close to the actual root) convergence rate \cite{NewtonMethod}. The properties of convergence of the alternating maximization algorithm in Table \ref{Table_AlternatingOpt} to a stationary point of the objective function in \eqref{Eq_19} are discussed in \cite[Proposition 2.7.1]{BertsekasBook}. Under mild assumptions that hold for the specific problem at hand, the algorithm in Table \ref{Table_AlternatingOpt} is locally q-linearly convergent to a local maximizer of the objective function provided that the initial guess is sufficiently close to the actual root \cite[Section 2]{JointOptimizationPaper}. Further details can be found in \cite{JointOptimizationPaper}. \hfill $\Box$
\end{remark}

In Section \ref{Results}, numerical evidence of the global optimality of the algorithm in Table \ref{Table_AlternatingOpt} is given as well. In addition, numerical results on the average (with respect to the initial guess) number of iterations as a function of the tolerance of convergence, $\epsilon > 0$, are illustrated.
\begin{table}[!t] 
\centering
\caption{Setup of parameters (unless otherwise stated). It is worth nothing that the setup $\gamma_{\rm{D}} = \gamma_{\rm{A}}$ constitutes just a case study and that the main findings of the present paper hold true for every $\gamma_{\rm{A}} > 0$.}
\begin{tabular}{l|l} \hline
Parameter & Value \\ \hline
$\beta$ & 3.5 \\
$\kappa  = {\left( {4\pi {f_c}/3 \cdot {{10}^8}} \right)^2}$ & $f_c$ = 2.1 GHz \\
$\rm{N}_0$ & -174 dBm/Hz \\
$\rm{B_W}$ & 20 MHz \\
$\rm{P_{circ}}$ & 51.14 dBm \cite{MariosANDTony2013} \\
$\rm{P_{idle}}$ & 48.75 dBm \cite{MariosANDTony2013} \\
$\rm{P_{tx}}$ & 43 dBm \cite{MariosANDTony2013} \\
${\lambda _{{\rm{BS}}}} = 1/\left( {\pi {\rm{R}}_{{\rm{cell}}}^{\rm{2}}} \right)$ BSs/m$^2$ & ${{\rm{R}}_{{\rm{cell}}}}$ = 250 m \\
${\lambda _{{\rm{MT}}}} = 1/\left( {\pi {\rm{R}}_{{\rm{MT}}}^{\rm{2}}} \right)$ = 121 MTs/km$^2$ & ${{\rm{R}}_{{\rm{MT}}}}$ = 51.29 \\
$\gamma_{\rm{D}} = \gamma_{\rm{A}}$ & 5 dB \\ \hline
\end{tabular}
\label{Table_SimulationSetup}
\end{table}
\begin{figure}[!t]
\centering
\includegraphics[width=\columnwidth]{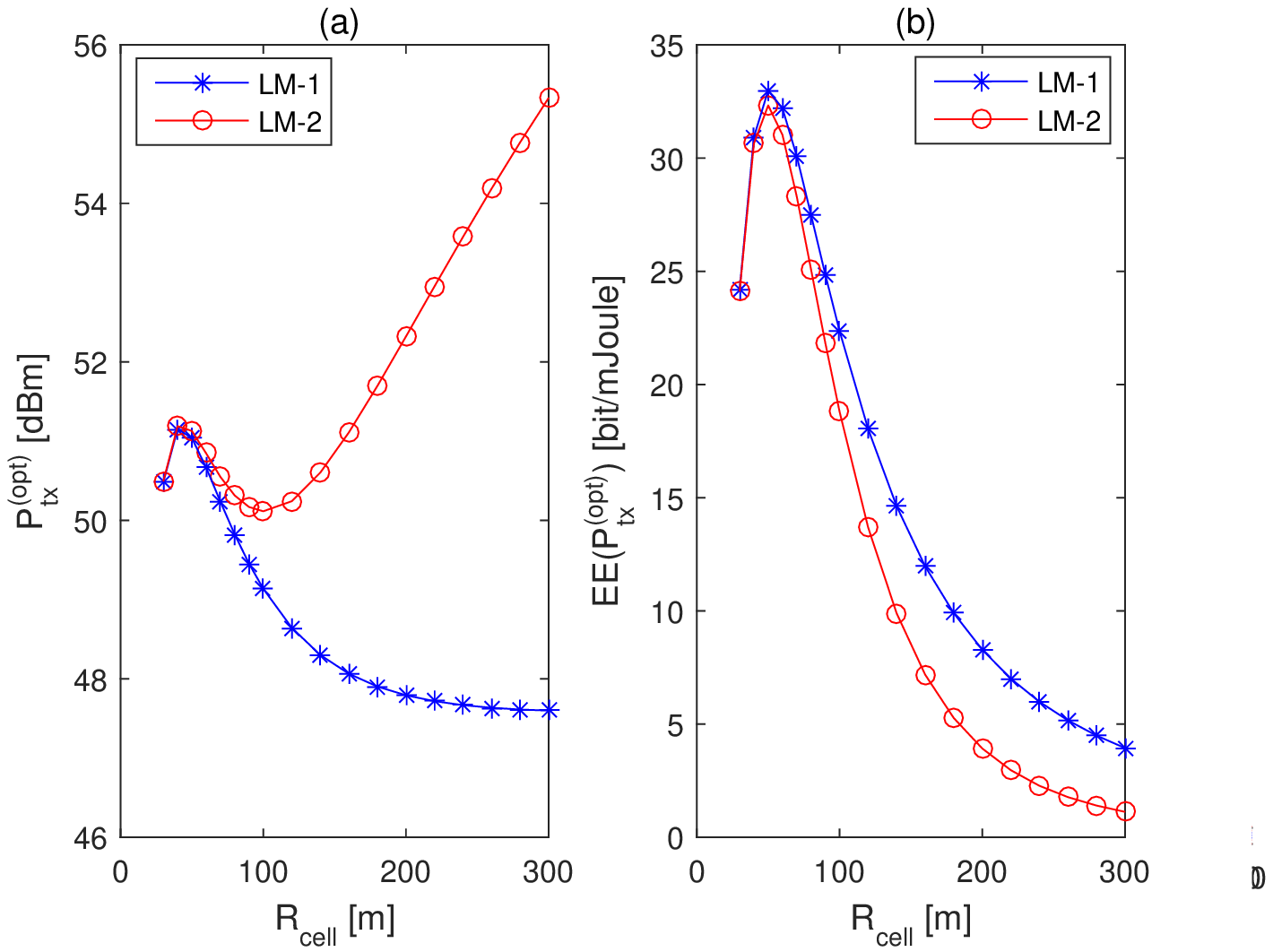}
\caption{Optimal transmit power (a) and energy efficiency (b) versus $\rm{R_{\rm{cell}}}$. Solid lines: Optimum from \textit{Theorem \ref{Theorem__OptimalPower}}. Markers: Optimum from a brute-force search of \eqref{Eq_15pre}. Special case with $\beta = 6.5$ and ${\lambda _{{\rm{MT}}}} = 21$ MTs/km$^2$.} \label{Fig_Theo1bis}
\end{figure}
\begin{figure}[!t]
\centering
\includegraphics[width=\columnwidth]{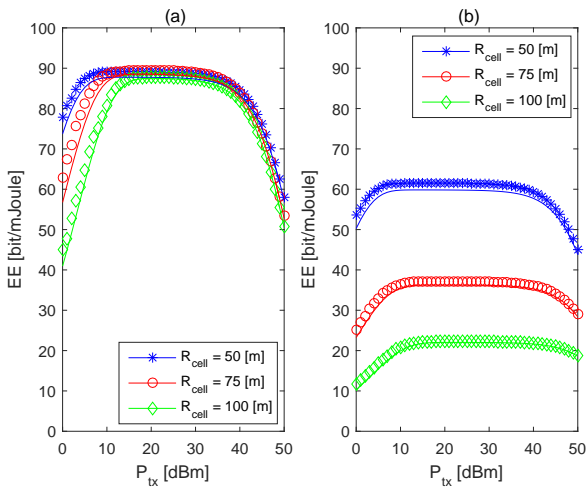}
\caption{Energy efficiency versus the transmit power for Load Model 1 (a) and Load Model 2 (b). Solid lines: Framework from \eqref{Eq_14}. Markers: Monte Carlo simulations.} \label{Fig_MC1}
\end{figure}
\begin{figure}[!t]
\centering
\includegraphics[width=\columnwidth]{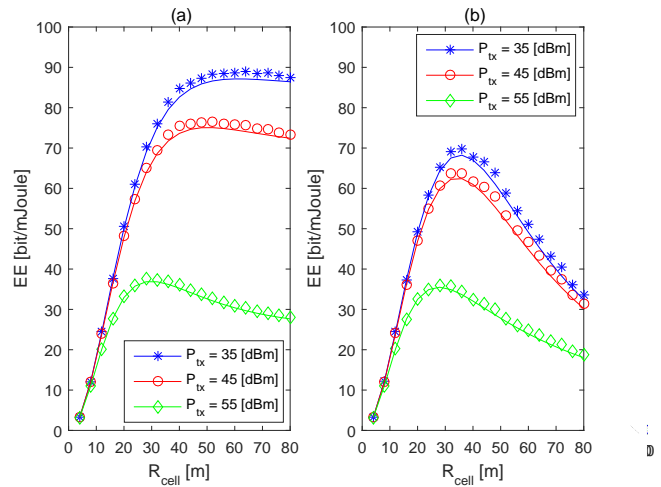}
\caption{Energy efficiency versus $\rm{R_{\rm{cell}}}$ for Load Model 1 (a) and Load Model 2 (b). Solid lines: Mathematical Framework from \eqref{Eq_14}. Markers: Monte Carlo simulations.} \label{Fig_MC2}
\end{figure}
\section{Numerical Results} \label{Results}
In this section, we show numerical results to validate the proposed analytical framework for computing the PSE and EE, as well as to substantiate the findings originating from the analysis of the system-level EE optimization problems as a function of the transmit power and density of the BSs. Unless otherwise stated, the simulation setup is summarized in Table IV. For ease of understanding, the BSs' density is represented via the inter-site distance ($\rm R_{cell}$) defined in Table IV. A similar comment applies to the density of the MTs that is expressed in terms of their average distance ($\rm R_{MT}$). As far as the choice of the setup of parameters is concerned, it is worth mentioning that the power consumption model is in agreement with \cite{MariosANDTony2013} and \cite{EE_EARTH}. The density of the MTs coincides with the average density of inhabitants in France.
\paragraph{Validation Against Monte Carlo Simulations} In Figs. \ref{Fig_MC1} and \ref{Fig_MC2}, we validate the correctness of \eqref{Eq_14} against Monte Carlo simulations. Monte Carlo results are obtained by simulating several realizations, according to the PPP model, of the cellular network and by empirically computing the PSE according to its definition in \eqref{Eq_6} and \eqref{Eq_7}, as well as the power consumption based on the operating principle described in the proofs of \textit{Proposition \ref{Proposition__Pgrid_LoadModel1}} and \textit{Proposition \ref{Proposition__Pgrid_LoadModel2}}. It is worth mentioning that, to estimate the PSE, only the definitions in the first line of \eqref{Eq_6} and \eqref{Eq_7} are used. The results depicted in Figs. \ref{Fig_MC1} and \ref{Fig_MC2} confirm the good accuracy of the proposed mathematical approach. They highlight, in addition, the unimodal and pseudo-concave shape of the EE as a function of the transmit power, given the BSs' density, and of the BSs' density, given the transmit power. If the same transmit power and BSs' density are assumed for both load models, we observe, as expected, that the first load model provides a better EE than the second load model.
\begin{figure}[!t]
\centering
\includegraphics[width=\columnwidth]{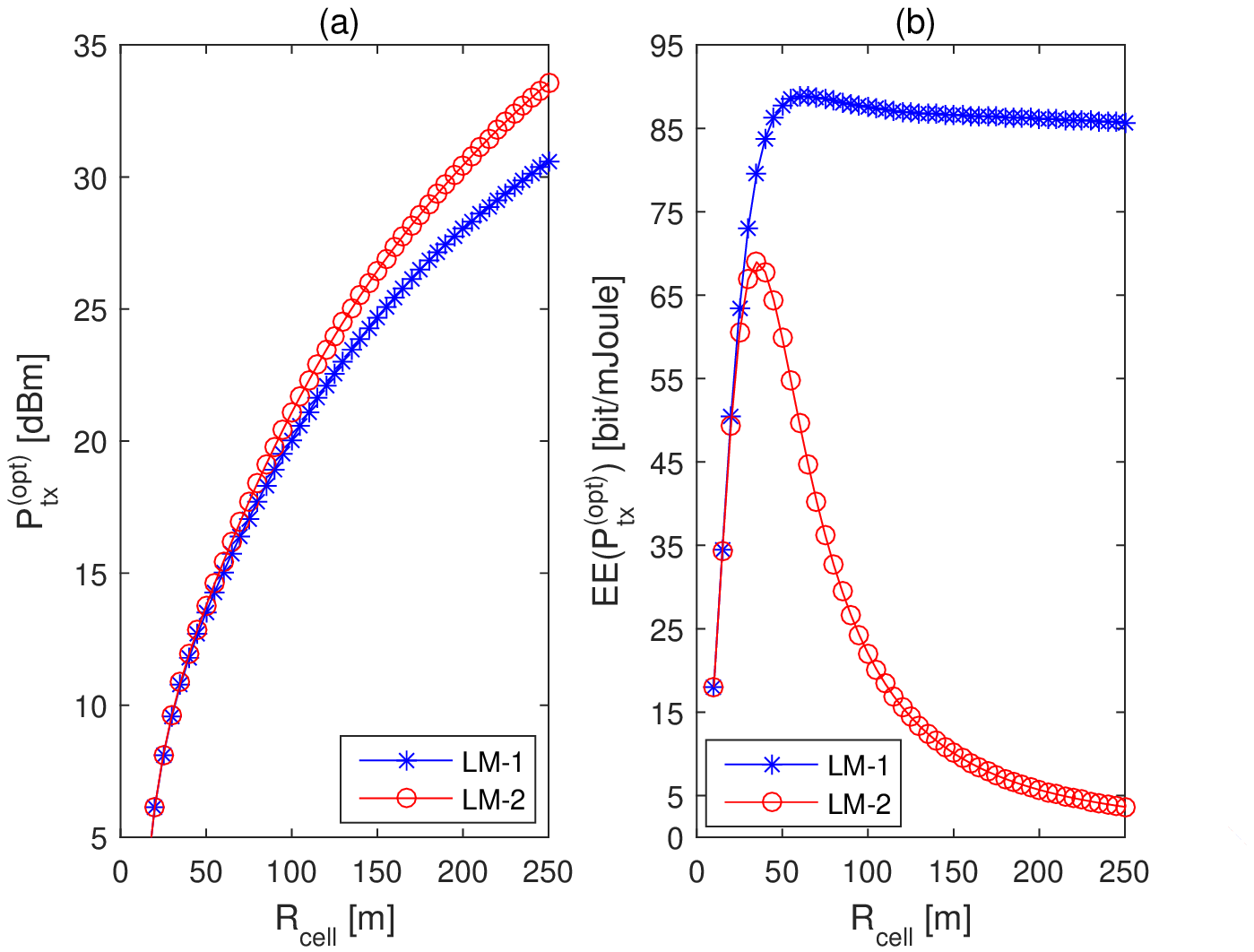}
\caption{Optimal transmit power (a) and energy efficiency (b) versus $\rm{R_{\rm{cell}}}$. Solid lines: Optimum from \textit{Theorem \ref{Theorem__OptimalPower}}. Markers: Optimum from a brute-force search of \eqref{Eq_15pre}. LM-1: Load Model 1 and LM-2: Load Model 2.} \label{Fig_Theo1}
\end{figure}
\begin{figure}[!t]
\centering
\includegraphics[width=\columnwidth]{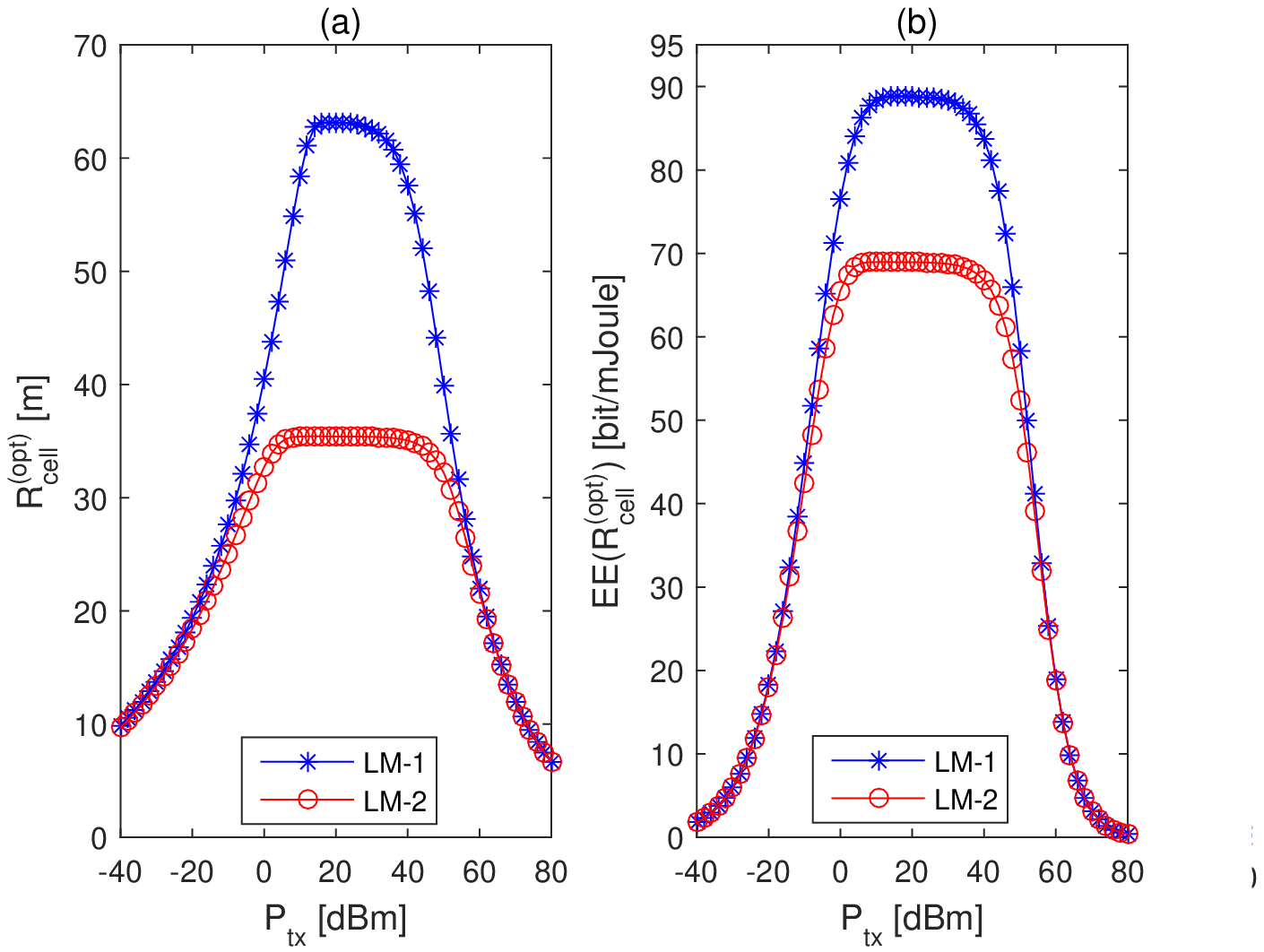}
\caption{Optimal density of BSs ($\rm{R_{\rm{cell}}}$) (a) and energy efficiency (b) versus the transmit power. Solid lines: Optimum from \textit{Theorem \ref{Theorem__OptimalLambda}}. Markers: Optimum from a brute-force search of \eqref{Eq_16pre}. LM-1: Load Model 1, LM-2: Load Model 2.} \label{Fig_Theo2}
\end{figure}
\begin{figure}[!t]
\centering
\includegraphics[width=\columnwidth]{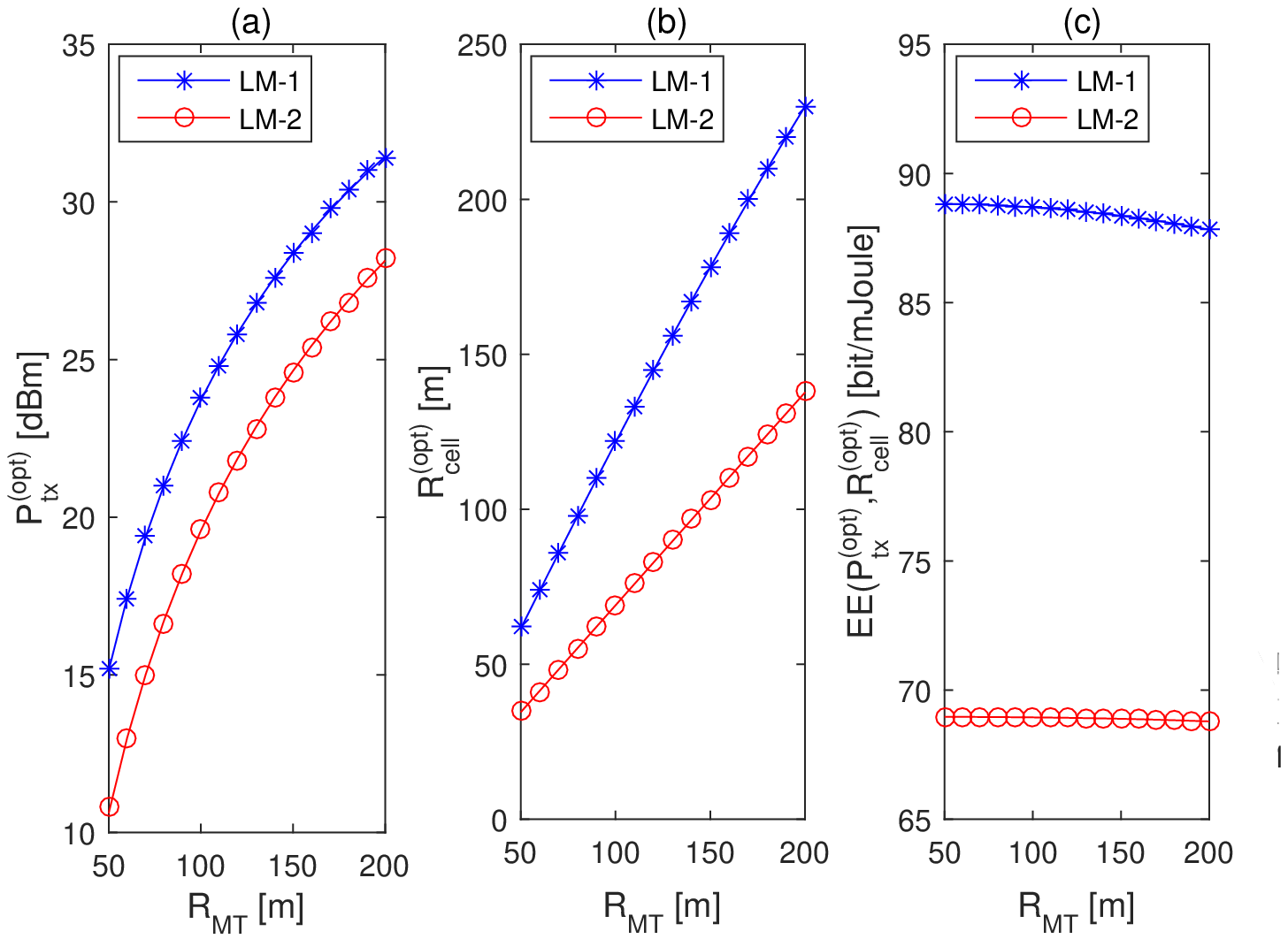}
\caption{Optimal transmit power (a), density of BSs ($\rm{R_{\rm{cell}}}$) (b), and energy efficiency (c) versus the density of MTs ($\rm{R_{\rm{MT}}}$). Solid lines: Optimum from the algorithm in Table \ref{Table_AlternatingOpt}. Markers: Optimum from a brute-force search of \eqref{Eq_19}. LM-1: Load Model 1 and LM-2: Load Model 2.} \label{Fig_DensityMT}
\end{figure}
\begin{figure}[!t]
\centering
\includegraphics[width=\columnwidth]{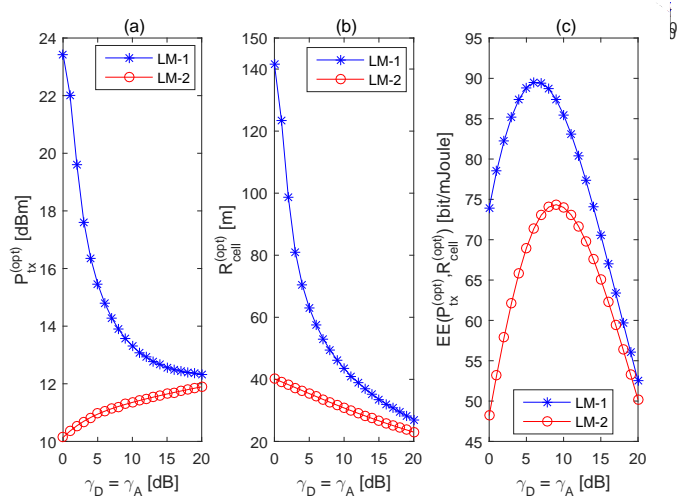}
\caption{Optimal transmit power (a), density of BSs ($\rm{R_{\rm{cell}}}$) (b), and energy efficiency (c) versus the reliability thresholds ($\gamma_{\rm D} = \gamma_{\rm A}$). Solid lines: Optimum from the algorithm in Table \ref{Table_AlternatingOpt}. Markers: Optimum from a brute-force search of \eqref{Eq_19}. LM-1: Load Model 1 and LM-2: Load Model 2.} \label{Fig_GammaIA}
\end{figure}
\begin{figure}[!t]
\centering
\includegraphics[width=\columnwidth]{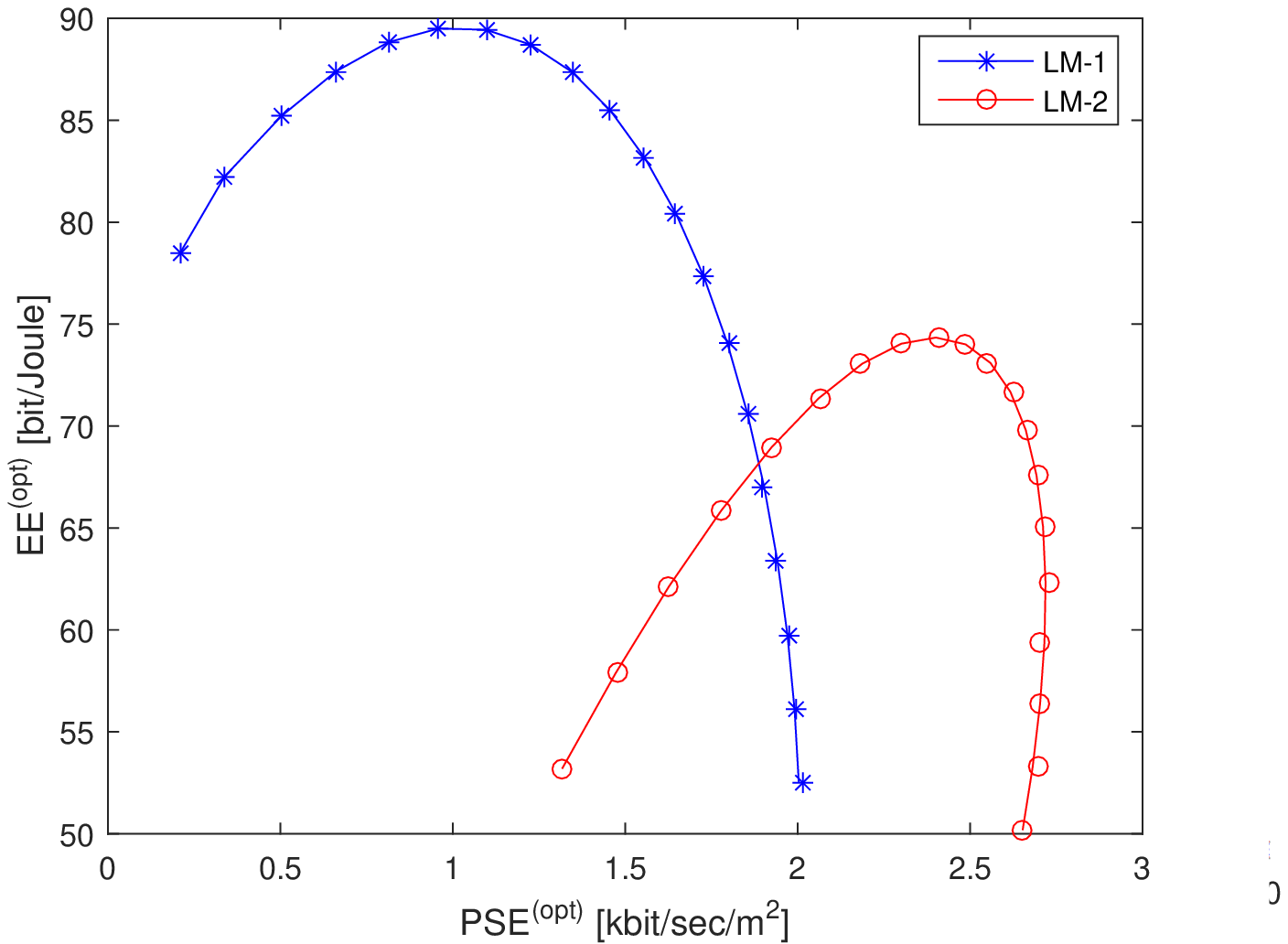}
\caption{Analysis of the EE vs. PSE trade-off. Solid lines: Optimum from the algorithm in Table \ref{Table_AlternatingOpt}. Markers: Optimum from a brute-force search of \eqref{Eq_19}. LM-1: Load Model 1 and LM-2: Load Model 2.} \label{Fig_EEvsPSE}
\end{figure}
\begin{figure}[!t]
\centering
\includegraphics[width=\columnwidth]{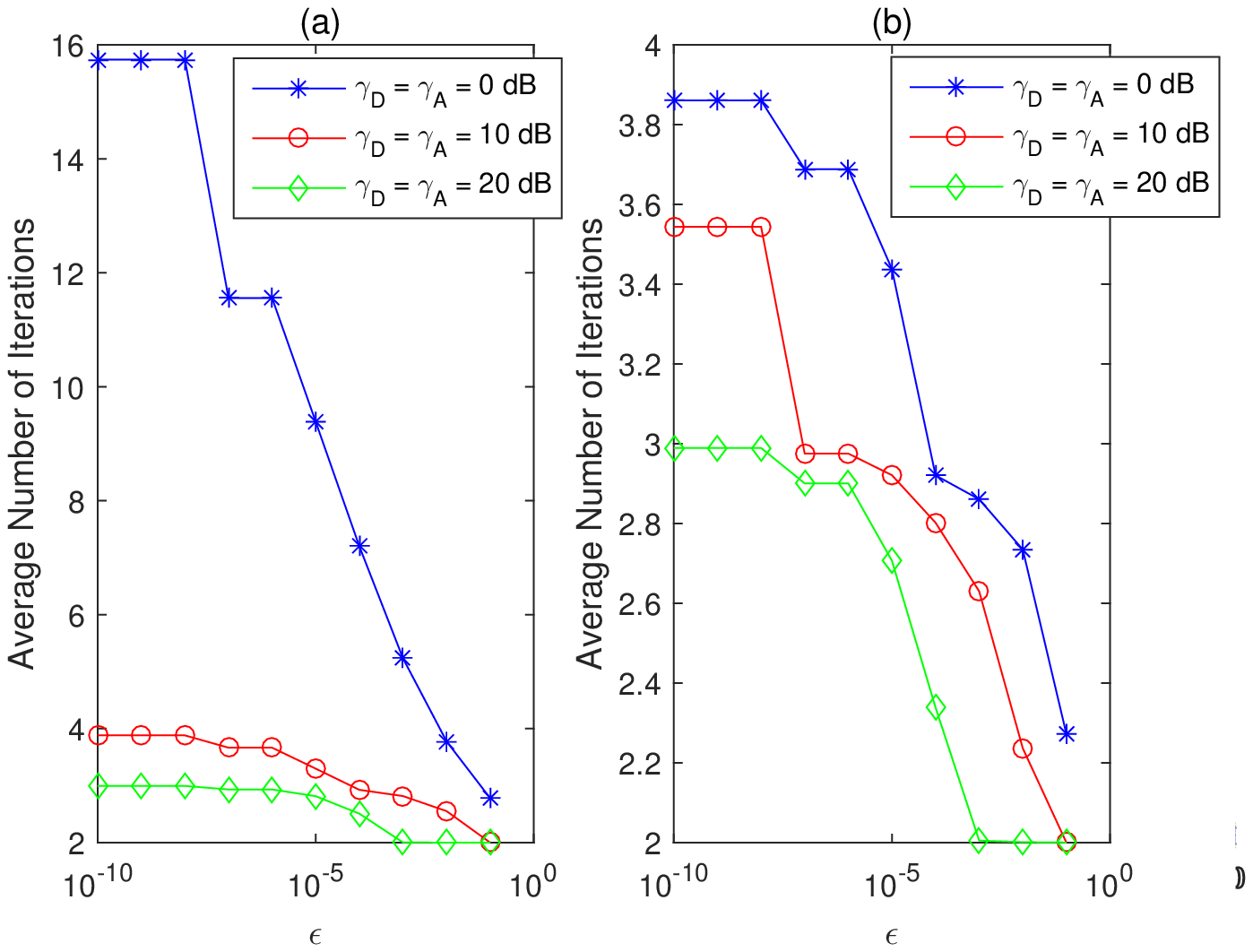}
\caption{Number of iterations of the algorithm in Table \ref{Table_AlternatingOpt} as a function of $\epsilon > 0$. The number of iterations is averaged (15000 trials) over the initial guess ${\lambda _{{\rm{BS}}}} = {\overline \lambda^{\left( {{\rm{opt}}} \right)}  _{{\rm{BS}}}} \in \left[ {\lambda _{{\rm{BS}}}^{\left( {{\rm{min}}} \right)},\lambda _{{\rm{BS}}}^{\left( {{\rm{max}}} \right)}} \right]$. (a) Load Model 1 and (b) Load Model 2. Setup: ${\rm{R}}_{{\rm{cell}}}^{\left( {{\rm{min}}} \right)} = 10$ m, ${\rm{R}}_{{\rm{cell}}}^{\left( {{\rm{max}}} \right)} = 2000$ m, ${\rm{P}}_{{\rm{tx}}}^{\left( {{\rm{min}}} \right)} = -20$ dBm, ${\rm{P}}_{{\rm{tx}}}^{\left( {{\rm{max}}} \right)} = 60$ dBm.} \label{Fig_Convergence}
\end{figure}
\paragraph{Validation of \textit{Theorem \ref{Theorem__OptimalPower}} and \textit{Theorem \ref{Theorem__OptimalLambda}}} In Figs. \ref{Fig_Theo1} and \ref{Fig_Theo2}, we compare the optimal transmit power and BSs' density obtained from \textit{Theorem \ref{Theorem__OptimalPower}} and \textit{Theorem \ref{Theorem__OptimalLambda}}, i.e., by computing the unique zero of \eqref{Eq_15} and \eqref{Eq_16}, respectively, against a brute-force search of the optimum of \eqref{Eq_15pre} and \eqref{Eq_16pre}, respectively. We observe the correctness of \textit{Theorem \ref{Theorem__OptimalPower}} and \textit{Theorem \ref{Theorem__OptimalLambda}} for the load models analyzed in the present paper. Figures \ref{Fig_Theo1} and \ref{Fig_Theo2}, in addition, confirm two important remarks that we have made throughout this paper. The first is that a joint pair of transmit power and BSs' density exists. This is highlighted by the fact that the EE evaluated at the optimal transmit power, given the BSs' density, and at the optimal BSs' density, given the transmit power, is still a unimodal and pseudo-concave function. This motivates one to use the alternating optimization algorithm proposed in Section \ref{JointOptimization}. The second is related to the difficulty of obtaining an explicit closed-form expression of the optimal transmit power as a function of the BSs' density and of the BSs' density as a function of the transmit power. Figure \ref{Fig_Theo2}(a), for example, clearly shows that the behavior of the optimal transmit power is not monotonic as a function of the BSs' density. This is in contrast with heuristic optimization criteria based on the coverage probability metric \cite{CellBreathing2010}. Figure \ref{Fig_Theo1}(a), on the other hand, provides more intuitive trends according to which the optimal transmit power increases as the density of the BSs decreases. This is, however, just a special case that is parameter-dependent. A counter-example is, in fact, illustrated in Fig. \ref{Fig_Theo1bis}, where, for a different set of parameters, it is shown that the optimal transmit power may increase, decrease and then increase again as a function of the average inter-site distance of the BSs ($\rm{R_{cell}}$). In this case, the density of the MTs coincides with the average density of inhabitants in Sweden and a large path-loss exponent is assumed to highlight the peculiar performance trend. These numerical examples clearly substantiate the importance of \textit{Theorem \ref{Theorem__OptimalPower}} and \textit{Theorem \ref{Theorem__OptimalLambda}}, and highlight the complexity of the optimization problem that is analyzed and successfully solved in the present paper.
\paragraph{Validation of the Alternating Optimization Algorithm in Table \ref{Table_AlternatingOpt}} In Figs. \ref{Fig_DensityMT} and \ref{Fig_GammaIA}, we provide numerical evidence of the convergence of the alternating optimization algorithm introduced in Section \ref{JointOptimization} towards the global optimum of the optimization problem formulated in \eqref{Eq_19}. The study is performed by computing the joint optimal transmit power and BSs' density as a function of the density of the MTs (Fig. \ref{Fig_DensityMT}) and of the reliability thresholds (Fig. \ref{Fig_GammaIA}). We observe a very good agreement between the algorithm in Table \ref{Table_AlternatingOpt} and a brute-force search of the optimum of \eqref{Eq_19}. Similar studies have been conducted as a function of other system parameters, but they are not reported in the present paper due to space limitations.
\paragraph{Comparison Between Load Model 1 and 2} With the exception of Figs. \ref{Fig_MC1} and \ref{Fig_MC2}, all the figures reported in this section illustrate the achievable EE of the two load models analyzed in the present manuscript when they operate at their respective optima. Based on the obtained results, we conclude that, for the considered system setup, the first load model outperforms the second one in terms of EE. Figures \ref{Fig_DensityMT} and \ref{Fig_GammaIA} show, for example, that this may be obtained by transmitting a higher power but, at the same time, by reducing the deployment density of the BSs. It is worth mentioning that, even though both load models provide the same PSE and serve, in the long time-horizon, all the MTs of the network, they have one main difference: the MTs under the first load model experience a higher latency (i.e., the MTs experience a longer delay before being served, since they are randomly chosen among all the available MTs in the cell), since a single MT is served at any time instance. We evince, as a result, that the higher EE provided by the first load model is obtained at the price of increasing the MTs' latency. The analysis and optimization of energy-efficient cellular networks with latency constraints is, therefore, an important generalization of the study conducted in the present paper.
\paragraph{Analysis of the EE vs. PSE Trade-Off} In Fig. \ref{Fig_EEvsPSE}, we illustrate the trade-off between EE and PSE, which is obtained by setting the transmit power and density of the BSs at the optimal values that are obtained by solving the optimization problem in \eqref{Eq_19} with the aid of the algorithm in Table \ref{Table_AlternatingOpt}. Figure \ref{Fig_EEvsPSE} provides a different view of the comparison between Load Model 1 and 2 introduced in Section \ref{LoadModeling}. The Load Model 1 is a suitable choice to obtain a high EE at low-medium PSEs, while the Load Model 2 is a more convenient option torequired to converge within  obtain a good EE at medium-high PSEs. Based on these results, the optimization of the EE vs. PSE trade-off constitutes an interesting generalization of the study carried out in the present paper.
\paragraph{Convergence Analysis of the Maximization Algorithm in Table \ref{Table_AlternatingOpt}} Motivated by \textit{Remark \ref{Remark__ConvergeRate}}, Fig. \ref{Fig_Convergence} shows the average number of iterations of the alternating optimization algorithm in Table \ref{Table_AlternatingOpt} as a function of the convergence accuracy $\epsilon$. We observe that the algorithm necessitates more iterations for Load Model 1. In general, however, we observe that the number of iterations that are required to converge within the defined convergence accuracy is relatively small.
\section{Conclusion} \label{Conclusion}
In the present paper, we have introduced a new closed-form analytical expression of the potential spectral efficiency of cellular networks. Unlike currently available analytical frameworks, we have shown that the proposed approach allows us to account for the tight interplay between transmit power and density of the base stations in cellular networks. Therefore, the proposed approach is conveniently formulated for the optimization of the network planning of cellular networks, by taking into account important system parameters. We have applied the new approach to the analysis and optimization of the energy efficiency of cellular networks. We have mathematically proved that the proposed closed-form expression of the energy efficiency is a unimodal and strictly pseudo-concave function in the transmit power, given the density, and in the density, given the transmit power of the base stations. Under these assumptions, as a result, a unique transmit power and density of the base stations exist, which can be obtained by finding the unique zero of a simple non-linear function that is provided in a closed-form expression. All mathematical derivations and findings have been substantiated with the aid of numerical simulations. We argue that the applications of the proposed approach to the system-level modeling and optimization of cellular networks are countless and go beyond the formulation of energy efficiency problems.

Extensions and generalizations of the analytical and optimization frameworks proposed in the present paper, include, but are not limited to, the system-level analysis and optimization of i) the energy efficiency versus spectral efficiency trade-off, ii) uplink cellular networks, iii) three-dimensional network topologies with elevated base stations and spatial blockages, iv) cache-enabled cellular networks, v) cellular networks with network slicing, vi) cellular networks with renewable energy sources and energy harvesting, and vii) multi-tier (heterogeneous) cellular networks.
\appendices
\section{Proof of Proposition \ref{Proposition__PSE}} \label{Appendix_PropPSE}
Under the assumption that ${\rm{M}}{{\rm{T}}_{\rm{0}}}$ is selected, from \eqref{Eq_3} and \eqref{Eq_4}, we have:
\setcounter{equation}{21}
\begin{equation}
\label{Eq_AppA_1}
\begin{split}
& {{\rm{P}}_{{\rm{cov}}}}\left( {{\gamma _{\rm{D}}},{\gamma _{\rm{A}}}} \right) = \Pr \left\{ {\frac{{{{{g_0}} \mathord{\left/
 {\vphantom {{{g_0}} {{L_0}}}} \right.
 \kern-\nulldelimiterspace} {{L_0}}}{\mathbbm{1}}\left( {{L_0} \le {{{{\rm{P}}_{{\rm{tx}}}}} \mathord{\left/
 {\vphantom {{{{\rm{P}}_{{\rm{tx}}}}} {\left( {{\gamma _{\rm{A}}}\sigma _{\rm{N}}^2} \right)}}} \right.
 \kern-\nulldelimiterspace} {\left( {{\gamma _{\rm{A}}}\sigma _{\rm{N}}^2} \right)}}} \right)}}{{\sum\nolimits_{{\rm{B}}{{\rm{S}}_i} \in \Psi _{{\rm{BS}}}^{\left( {\rm{I}} \right)}} {{{{g_i}} \mathord{\left/
 {\vphantom {{{g_i}} {{L_i}}}} \right.
 \kern-\nulldelimiterspace} {{L_i}}}{\mathbbm{1}}\left( {{L_i} > {L_0}} \right)} }} \ge {\gamma _{\rm{D}}}} \right\} \\
 & = \hspace{-0.6cm} \int\limits_0^{{{{{\rm{P}}_{{\rm{tx}}}}} \mathord{\left/
 {\vphantom {{{{\rm{P}}_{{\rm{tx}}}}} {\left( {{\gamma _{\rm{A}}}\sigma _{\rm{N}}^2} \right)}}} \right.
 \kern-\nulldelimiterspace} {\left( {{\gamma _{\rm{A}}}\sigma _{\rm{N}}^2} \right)}}} \hspace{-0.25cm} \underbrace{\Pr \left\{ {\frac{{{{{g_0}} \mathord{\left/
 {\vphantom {{{g_0}} x}} \right.
 \kern-\nulldelimiterspace} x}}}{{\sum\nolimits_{{\rm{B}}{{\rm{S}}_i} \in \Psi _{{\rm{BS}}}^{\left( {\rm{I}} \right)}} {{{{g_i}} \mathord{\left/
 {\vphantom {{{g_i}} {{L_i}}}} \right.
 \kern-\nulldelimiterspace} {{L_i}}}{\mathbbm{1}}\left( {{L_i} > x} \right)} }} \ge {\gamma _{\rm{D}}}} \right\}}_{{\mathcal{G}}\left( {{\gamma _{\rm{D}}};x} \right)} {f_{{L_0}}}\left( x \right)dx
\end{split}
\end{equation}
\noindent where ${f_{{L_0}}}\left( x \right) = 2\pi {\lambda _{{\rm{BS}}}}{\left( {{\kappa ^{{2 \mathord{\left/ {\vphantom {2 \beta }} \right. \kern-\nulldelimiterspace} \beta }}}\beta } \right)^{ - 1}}{x^{{2 \mathord{\left/ {\vphantom {2 \beta }} \right. \kern-\nulldelimiterspace} \beta } - 1}} {\rm{e}}^{ { - \pi {\lambda _{{\rm{BS}}}}{{\left( {{x \mathord{\left/ {\vphantom {x \kappa }} \right. \kern-\nulldelimiterspace} \kappa }} \right)}^{{2 \mathord{\left/ {\vphantom {2 \beta }} \right. \kern-\nulldelimiterspace} \beta }}}} }$ is the probability density function of $L_0$ that is obtained by applying the displacement theorem of PPPs \cite[Eq. (21)]{MDR_IM}. It is worth mentioning that \eqref{Eq_AppA_1} is exact if the Crofton cell is considered, while it is an approximation if the typical cell is considered (see \textit{Remark \ref{Remark__CroftonCell}} for further details).

The probability term, ${\mathcal{G}}\left( {\cdot;\cdot} \right)$, in the integrand function of \eqref{Eq_AppA_1} can be computed as follows:
\setcounter{equation}{22}
\begin{equation}
\label{Eq_AppA_2}
\begin{split}
{\mathcal{G}}\left( {{\gamma _{\rm{D}}};x} \right) & \mathop  = \limits^{\left( a \right)} \exp \left( { - \int\nolimits_x^{ + \infty } {{{\left( {1 + \frac{y}{{x{\gamma _{\rm{D}}}}}} \right)}^{ - 1}}2\pi \lambda _{{\rm{BS}}}^{\left( {{\rm{tx}}} \right)}\frac{{{y^{{2 \mathord{\left/
 {\vphantom {2 \beta }} \right.
 \kern-\nulldelimiterspace} \beta } - 1}}}}{{{\kappa ^{{2 \mathord{\left/
 {\vphantom {2 \beta }} \right.
 \kern-\nulldelimiterspace} \beta }}}\beta }}dy} } \right) \\
& \mathop  = \limits^{\left( b \right)} \exp \left( { - \pi \lambda _{{\rm{BS}}}^{\left( {{\rm{tx}}} \right)}{{\left( {{x \mathord{\left/
 {\vphantom {x \kappa }} \right.
 \kern-\nulldelimiterspace} \kappa }} \right)}^{{2 \mathord{\left/
 {\vphantom {2 \beta }} \right.
 \kern-\nulldelimiterspace} \beta }}}\Upsilon } \right)
 \end{split}
\end{equation}
\noindent where (a) follows from the probability generating functional theorem of PPPs \cite{AndrewsNov2011} by taking into account that, based on \eqref{Eq_8}, the interfering BSs constitute a PPP of intensity equal to $\lambda _{{\rm{BS}}}^{\left( {{\rm{tx}}} \right)} = {\lambda _{{\rm{BS}}}}{\mathop{\mathbb P}\nolimits} _{{\rm{BS}}}^{\left( {{\rm{tx}}} \right)} = {\lambda _{{\rm{BS}}}}{\mathcal{L}}\left( {{{{\lambda _{{\rm{MT}}}}} \mathord{\left/ {\vphantom {{{\lambda _{{\rm{MT}}}}} {{\lambda _{{\rm{BS}}}}}}} \right. \kern-\nulldelimiterspace} {{\lambda _{{\rm{BS}}}}}}} \right)$, and (b) follows by solving the integral. The intensity of the interfering PPP, $\lambda _{{\rm{BS}}}^{\left( {{\rm{tx}}} \right)}$, is obtained by taking into account that only that BSs that are in transmission mode contribute to the inter-cell interference. The analytical expression of $\lambda _{{\rm{BS}}}^{\left( {{\rm{tx}}} \right)}$ is, in particular, obtained with the aid of the independent thinning theorem of PPPs, similar to \cite{MDR_IM} and \cite{LoadAwareIndependent}. The impact of the spatial correlation that exists among the BSs that operate in transmission mode \cite{LoadAwareCorrelated}, is, on the other hand, postponed to future research.

By inserting \eqref{Eq_AppA_2} in \eqref{Eq_AppA_1} and by applying some changes of variable, we obtain:
\setcounter{equation}{23}
\begin{equation}
\label{Eq_AppA_3}
\begin{split}
& {{\rm{P}}_{{\rm{cov}}}}\left( {{\gamma _{\rm{D}}},{\gamma _{\rm{A}}}} \right) = \pi {\lambda _{{\rm{BS}}}}{\kappa ^{ - {2 \mathord{\left/
 {\vphantom {2 \beta }} \right.
 \kern-\nulldelimiterspace} \beta }}} \\ & \times \hspace{-0.5cm} \int\limits_0^{{{\left( {{{{{\rm{P}}_{{\rm{tx}}}}} \mathord{\left/
 {\vphantom {{{{\rm{P}}_{{\rm{tx}}}}} {\left( {{\gamma _{\rm{A}}}\sigma _{\rm{N}}^2} \right)}}} \right.
 \kern-\nulldelimiterspace} {\left( {{\gamma _{\rm{A}}}\sigma _{\rm{N}}^2} \right)}}} \right)}^{{2 \mathord{\left/
 {\vphantom {2 \beta }} \right.
 \kern-\nulldelimiterspace} \beta }}}} \hspace{-0.5cm} {\exp \left( { - \pi {\lambda _{{\rm{BS}}}}{\kappa ^{ - {2 \mathord{\left/
 {\vphantom {2 \beta }} \right.
 \kern-\nulldelimiterspace} \beta }}}\left( {1 + \Upsilon {\mathcal{L}}\left( {{{{\lambda _{{\rm{MT}}}}} \mathord{\left/
 {\vphantom {{{\lambda _{{\rm{MT}}}}} {{\lambda _{{\rm{BS}}}}}}} \right.
 \kern-\nulldelimiterspace} {{\lambda _{{\rm{BS}}}}}}} \right)} \right)z} \right)dz}.
 \end{split}
\end{equation}

The proof follows from \eqref{Eq_6} and \eqref{Eq_7} with the aid of some simplifications and by using the identity $\sum\nolimits_{u = 0}^{ + \infty } {{{\left( {u + 1} \right)}^{ - 1}}\Pr \left\{ {{{{\rm{\bar N}}}_{{\rm{MT}}}} = u} \right\}} = {\left( {{{{\lambda _{{\rm{MT}}}}} \mathord{\left/ {\vphantom {{{\lambda _{{\rm{MT}}}}} {{\lambda _{{\rm{BS}}}}}}} \right. \kern-\nulldelimiterspace} {{\lambda _{{\rm{BS}}}}}}} \right)^{ - 1}}{\mathcal{L}}\left( {{{{\lambda _{{\rm{MT}}}}} \mathord{\left/ {\vphantom {{{\lambda _{{\rm{MT}}}}} {{\lambda _{{\rm{BS}}}}}}} \right. \kern-\nulldelimiterspace} {{\lambda _{{\rm{BS}}}}}}} \right)$ \cite[Proposition 2]{WiOPT_2013}.
\section{Proof of Theorem \ref{Theorem__OptimalPower}} \label{Appendix_TheoPtx}
In this section, we are interested in the functions that depend on ${{\rm{P}}_{{\rm{tx}}}}$. For ease of writing, we adopt the simplified notation: ${{\rm{P}}_{{\rm{tx}}}} \to {\rm{P}}$, ${\mathcal{L}}\left(  \cdot  \right) \to {\mathcal{L}}$, ${\mathcal{M}}\left(  \cdot  \right) \to {\mathcal{M}}$, ${\mathcal{Q}}\left( { \cdot ,{{\rm{P}}_{{\rm{tx}}}}, \cdot } \right) \to {\mathcal{Q}}\left( {\rm{P}}\right)$, ${{{\mathcal{\dt Q}}}_{{{\rm{P}}_{{\rm{tx}}}}}}\left( { \cdot ,{{\rm{P}}_{{\rm{tx}}}}, \cdot } \right) \to {\mathcal{\dt Q}}\left( {\rm{P}} \right)$, ${{\rm{P}}_{{\rm{circ}}}} = {{\rm{P}}_{\rm{c}}}$, ${{\rm{P}}_{\rm{idle}}} = {{\rm{P}}_{\rm{i}}}$, ${\rm{EE}}\left( {{{\rm{P}}_{{\rm{tx}}}}, \cdot } \right) \to {\rm{EE}}\left( {\rm{P}} \right)$, and ${\dt{\rm{EE}}_{{{\rm{P}}_{{\rm{tx}}}}}}\left( {{{\rm{P}}_{{\rm{tx}}}}, \cdot } \right) \to {\dt {\rm{EE}}}\left( {\rm{P}} \right)$. A similar notation is adopted for higher-order derivatives with respect to $\rm P$.

The stationary points of \eqref{Eq_14} are the zeros of the first-order derivative of ${\rm{EE}}\left( \cdot \right)$ with respect to $\rm{P}$. From \eqref{Eq_14}, we obtain ${{\dt{\rm{EE}}}}\left( {\rm{P}} \right) = 0 \Leftrightarrow {{\rm{P}}_{\rm{i}}} - {{\mathcal{S}}_{\mathcal{P}}}\left( {\rm{P}} \right) = 0$, which can be re-written as follows:
\setcounter{equation}{24}
\begin{equation}
\label{Eq_AppB_1}
\underbrace {{{{\mathcal{Q}}\left( {\rm{P}} \right)} \mathord{\left/
 {\vphantom {{{\mathcal{Q}}\left( {\rm{P}} \right)} {{\mathcal{\dt Q}}\left( {\rm{P}} \right)}}} \right.
 \kern-\nulldelimiterspace} {{\mathcal{\dt Q}}\left( {\rm{P}} \right)}} - {\rm{P}}}_{{W_{{\rm{left}}}}\left( {\rm{P}} \right)} = \underbrace {\Delta {\rm{P}} + {{{{\rm{P}}_{\rm{i}}}} \mathord{\left/
 {\vphantom {{{{\rm{P}}_{\rm{i}}}} {\mathcal{L}}}} \right.
 \kern-\nulldelimiterspace} {\mathcal{L}}} + {{{{\rm{P}}_{\rm{c}}}{\mathcal{M}}} \mathord{\left/
 {\vphantom {{{{\rm{P}}_{\rm{c}}}{\mathcal{M}}} {\rm{L}}}} \right.
 \kern-\nulldelimiterspace} {\mathcal{L}}}}_{{W_{{\rm{right}}}}}.
\end{equation}

With the aid of some algebraic manipulations and by exploiting \textit{Lemmas \ref{Lemma__LoadFunction}-\ref{Lemma__PilotFunctionDensity}}, the following holds: i) ${{W_{{\rm{right}}}}} \ge 0$ is a non-negative function that is independent of $\rm P$, ii) ${{W_{{\rm{left}}}}\left( {\rm{P}} \right)} \ge 0$ is a non-negative and increasing function of $\rm P$, i.e., ${{\dt W}_{{\rm{left}}}}\left( {\rm{P}} \right) \ge 0$, since ${\mathcal{Q}}\left( {\rm{P}} \right) \ge 0$ and ${\mathcal{\ddt Q}}\left( {\rm{P}} \right) \le 0$ from \textit{Lemma \ref{Lemma__PilotFunctionPower}}, iii) ${W_{{\rm{left}}}}\left( {{\rm{P}} \to 0} \right) = 0$ and ${W_{{\rm{left}}}}\left( {{\rm{P}} \to \infty } \right) = \infty$. This implies that ${{W_{{\rm{left}}}}\left( \cdot \right)}$ and ${{W_{{\rm{right}}}}}$ intersect each other in just one point. Therefore, a unique stationary point, ${{\rm{P}}^{\rm{*}}}$, exists. Also, ${\dt{\rm{EE}}}\left( {\rm{P}} \right) > 0$ for $\rm P < {{\rm{P}}^{\rm{*}}}$ and ${\dt{\rm{EE}}}\left( {\rm{P}} \right) < 0$ for $\rm P > {{\rm{P}}^{\rm{*}}}$. Finally, by taking into account the constraints on the transmit power, it follows that the unique optimal maximizer of the EE is ${{\rm{P}}^{\left( {{\rm{opt}}} \right)}} = \max \left\{ {{{\rm{P}}^{\left( {{\rm{min}}} \right)}},\min \left\{ {{{\rm{P}}^*},{{\rm{P}}^{\left( {{\rm{max}}} \right)}}} \right\}} \right\}$, since ${\rm{P}} \in \left[ {{{\rm{P}}^{\left( {{\rm{min}}} \right)}},{{\rm{P}}^{\left( {{\rm{max}}} \right)}}} \right]$. This concludes the proof.
\section{Proof of Theorem \ref{Theorem__OptimalLambda}} \label{Appendix_TheoLambda}
In this section, we are interested in the functions that depend on ${\lambda _{{\rm{BS}}}}$. For ease of writing, we adopt the simplified notation: ${\lambda _{{\rm{BS}}}} \to \lambda$, ${\mathcal{L}}\left(  \cdot/{\lambda _{{\rm{BS}}}}  \right) \to {\mathcal{L}}\left( \lambda \right)$, ${\mathcal{M}}\left(  \cdot/{\lambda _{{\rm{BS}}}}  \right) \to {\mathcal{M}}\left( \lambda \right)$, ${\mathcal{Q}}\left( { {\lambda _{{\rm{BS}}}} ,\cdot, \cdot/{\lambda _{{\rm{BS}}}} } \right) \to {\mathcal{Q}}\left( \lambda \right)$, ${{{\mathcal{\dt Q}}}_{{\lambda _{{\rm{BS}}}}}}\left( { {\lambda _{{\rm{BS}}}} ,\cdot, \cdot/{\lambda _{{\rm{BS}}}} } \right) \to {\mathcal{\dt Q}}\left( \lambda \right)$, ${{\rm{P}}_{{\rm{circ}}}} = {{\rm{P}}_{\rm{c}}}$, ${{\rm{P}}_{\rm{idle}}} = {{\rm{P}}_{\rm{i}}}$, ${\rm{EE}}\left( {\cdot, {\lambda _{{\rm{BS}}}} } \right) \to {\rm{EE}}\left( \lambda \right)$, ${\dt{\rm{EE}}_{{\lambda _{{\rm{BS}}}}}}\left( {\cdot, {\lambda _{{\rm{BS}}}} } \right) \to {\dt {\rm{EE}}}\left( \lambda \right)$, ${{\rm{P}}_{{\rm{tx}}}} \to {\rm{P}}$. Similar notation applies to higher-order derivatives.

The proof is split in two parts: i) ${{{\lambda _{{\rm{MT}}}}} \mathord{\left/ {\vphantom {{{\lambda _{{\rm{MT}}}}} \lambda }} \right. \kern-\nulldelimiterspace} \lambda } \ge 2.8$ and ii) ${{{\lambda _{{\rm{MT}}}}} \mathord{\left/ {\vphantom {{{\lambda _{{\rm{MT}}}}} \lambda }} \right. \kern-\nulldelimiterspace} \lambda } \le 2.8$. This is necessary because, from \textit{Lemma \ref{Lemma__LoadFunction}}, ${\mathcal{L}}\left(  \cdot  \right)$ is concave in $\lambda$ if ${\lambda _{{\rm{MT}}}}/\lambda  \ge 2.8$ and convex in $\lambda$ if ${\lambda _{{\rm{MT}}}}/\lambda  \le 2.8$.
\paragraph{Case Study ${\lambda _{{\rm{MT}}}}/\lambda  \ge 2.8$} The stationary points of \eqref{Eq_14} are the zeros of the first-order derivative of ${\rm{EE}}\left( \cdot \right)$ with respect to $\lambda$. From \eqref{Eq_14}, we obtain ${{\dt{\rm{EE}}}}\left( \lambda \right) = 0 \Leftrightarrow {{\mathcal{S}}_{\mathcal{D}}}\left( \lambda \right) - {{\rm{P}}_{\rm{i}}} = 0$. This stationary equation can be re-written as follows (${W_{{\rm{right}}}}\left( \lambda  \right) = \sum\nolimits_{\ell  = 1}^5 {{W_\ell }\left( \lambda  \right)}$):
\setcounter{equation}{25}
\begin{equation}
\label{Eq_AppC_1}
\begin{split}
\underbrace {{{\rm{P}}_{\rm{i}}}}_{{W_{{\rm{left}}}}} & = \underbrace {- \left( {{{{\mathcal{L}}\left( \lambda  \right)} \mathord{\left/
 {\vphantom {{{\mathcal{L}}\left( \lambda  \right)} {{\mathcal{\dt L}}\left( \lambda  \right)}}} \right.
 \kern-\nulldelimiterspace} {{\mathcal{\dt L}}\left( \lambda  \right)}}} \right)\left( {{{{\mathcal{\dt Q}}\left( \lambda  \right)} \mathord{\left/
 {\vphantom {{{\mathcal{\dt Q}}\left( \lambda  \right)} {{\mathcal{Q}}\left( \lambda  \right)}}} \right.
 \kern-\nulldelimiterspace} {{\mathcal{Q}}\left( \lambda  \right)}}} \right)}_{{W_1}\left( \lambda  \right)} \\ & \times \underbrace {\left[ {1 + \Upsilon {\mathcal{L}}\left( \lambda  \right)} \right]\left[ {{\mathcal{L}}\left( \lambda  \right)\left( {{\rm{P}} + \Delta {\rm{P}}} \right) + {{\rm{P}}_{\rm{i}}} + {{\rm{P}}_{\rm{c}}}{\mathcal{M}}\left( \lambda  \right)} \right]}_{{W_2}\left( \lambda  \right)} \\
 & + \underbrace {{{\rm{P}}_{{\rm{circ}}}}\left( {{{{\mathcal{\dt M}}\left( \lambda  \right){\mathcal{L}}\left( \lambda  \right)} \mathord{\left/
 {\vphantom {{{\mathcal{\dt M}}\left( \lambda  \right){\mathcal{L}}\left( \lambda  \right)} {{\mathcal{\dt L}}\left( \lambda  \right)}}} \right.
 \kern-\nulldelimiterspace} {{\mathcal{\dt L}}\left( \lambda  \right)}} - {\mathcal{M}}\left( \lambda  \right)} \right)}_{{W_3}\left( \lambda  \right)} \\ & + \underbrace {\Upsilon {{\mathcal{L}}^2}\left( \lambda  \right)\left( {{\rm{P}} + \Delta {\rm{P}}} \right)}_{{W_4}\left( \lambda  \right)} + \underbrace {\Upsilon {{\rm{P}}_{\rm{c}}}{{\mathcal{L}}^2}\left( \lambda  \right){{{\mathcal{\dt M}}\left( \lambda  \right)} \mathord{\left/
 {\vphantom {{{\mathcal{\dt M}}\left( \lambda  \right)} {{\mathcal{\dt L}}\left( \lambda  \right)}}} \right.
 \kern-\nulldelimiterspace} {{\mathcal{\dt L}}\left( \lambda  \right)}}}_{{W_5}\left( \lambda  \right)}.
\end{split}
\end{equation}

With the aid of some algebraic manipulations and by exploiting \textit{Lemmas \ref{Lemma__LoadFunction}-\ref{Lemma__PilotFunctionDensity}}, the following holds: i) ${{W_{{\rm{left}}}}} \ge 0$ is a non-negative function that is independent of $\lambda$, ii) ${{W_{{\rm{right}}}}\left( \lambda \right)} \ge 0$ is a non-negative function of $\lambda$, since ${{W_\ell }\left( \lambda  \right)} \ge 0$ for $\ell=1,\ldots,5$ if ${\lambda _{{\rm{MT}}}}/\lambda  \ge 2.8$. In particular, ${W_3}\left( \lambda  \right) \ge 0$ if ${\lambda _{{\rm{MT}}}}/\lambda  \ge 1.4$ and ${{W_\ell }\left( \lambda  \right)} \ge 0$ for $\lambda \ge 0$ if $\ell=1,2,4,5$, iii) ${W_{{\rm{right}}}}\left( {{\lambda} \to 0} \right) = \infty$ and ${W_{{\rm{right}}}}\left( {{\lambda} \to \infty } \right) = 0$. This implies that ${{W_{{\rm{left}}}}}$ and ${{W_{{\rm{right}}}}\left( \cdot \right)}$ would intersect each other in just a single point if ${{W_{{\rm{right}}}}}$ is a decreasing function in $\lambda$, i.e., ${{\dt W}_{{\rm{right}}}}\left( \lambda  \right) \le 0$ for ${\lambda _{{\rm{MT}}}}/\lambda  \ge 2.8$. A sufficient condition for this to hold is that ${{W_\ell }\left( \cdot  \right)}$ for $\ell=1,\ldots,5$ are decreasing functions in $\lambda$, i.e., ${{\dt{W_\ell }}\left( \lambda  \right)} \le 0$ for ${\lambda _{{\rm{MT}}}}/\lambda  \ge 2.8$. This holds to be true and can be proved as follows. ${\dt{{W_2}}\left( \lambda  \right)} \le 0$ for $\lambda \ge 0$ and ${\dt{{W_4}}\left( \lambda  \right)} \le 0$ for $\lambda \ge 0$ because ${{\mathcal{L}}\left( \cdot  \right)}$ and ${{\mathcal{M}}\left( \cdot  \right)}$ are decreasing functions in $\lambda$ (see \textit{Lemma \ref{Lemma__LoadFunction}} and \textit{Lemma \ref{Lemma__LoadFunctionModified}}). ${\dt{{W_3}}\left( \lambda  \right)} \le 0$ for $\lambda \ge 0$ and ${\dt{{W_5}}\left( \lambda  \right)} \le 0$ for $\lambda \ge 0$ immediately follow by inserting into them the first-order derivatives of ${{\mathcal{L}}\left( \cdot  \right)}$ and ${{\mathcal{M}}\left( \cdot  \right)}$ with respect to $\lambda$ and with the aid of simple algebraic manipulations. Less evident is the behavior of ${{W_1}\left( \cdot  \right)}$ as a function of $\lambda$. Using some algebra, the first-order derivative satisfies the following:
\setcounter{equation}{26}
\begin{equation}
\label{Eq_AppC_2}
\begin{split}
{{\dt W}_1}\left( \lambda  \right){\left( {{\mathcal{Q}}\left( \lambda  \right){\mathcal{\dt L}}\left( \lambda  \right)} \right)^2} & = \underbrace { - {\mathcal{L}}\left( \lambda  \right){\mathcal{\dt L}}\left( \lambda  \right){\mathcal{Q}}\left( \lambda  \right){\mathcal{\ddt Q}}\left( \lambda  \right)}_{{A_1}\left( \lambda  \right)} \\ &+ \underbrace {\left( { - {{{\mathcal{\dt L}}}^2}\left( \lambda  \right){\mathcal{Q}}\left( \lambda  \right){\mathcal{\dt Q}}\left( \lambda  \right)} \right)}_{{A_2}\left( \lambda  \right)}\\
& + \underbrace {{\mathcal{L}}\left( \lambda  \right){\mathcal{\dt L}}\left( \lambda  \right){{{\mathcal{\dt Q}}}^2}\left( \lambda  \right)}_{{A_3}\left( \lambda  \right)} \\ & + \underbrace {{\mathcal{L}}\left( \lambda  \right){\mathcal{\ddt L}}\left( \lambda  \right){\mathcal{Q}}\left( \lambda  \right){\mathcal{\dt Q}}\left( \lambda  \right)}_{{A_4}\left( \lambda  \right)}.
\end{split}
\end{equation}

A sufficient condition for ${{W_1}\left( \cdot  \right)}$ to be a decreasing function in $\lambda$ is that ${{A_\ell }\left( \lambda  \right)} \le 0$ for $\ell=1,\ldots,4$. From \textit{Lemmas \ref{Lemma__LoadFunction}-\ref{Lemma__PilotFunctionDensity}}, this can be readily proved. In particular, ${{A_\ell }\left( \lambda  \right)} \le 0$ for $\lambda \ge 0$ if $\ell=1,2,3$ and ${{A_4 }\left( \lambda  \right)} \le 0$ for ${\lambda _{{\rm{MT}}}}/\lambda  \ge 2.8$. Therefore, a unique stationary point, ${{\lambda}^{\rm{*}}}$, exists. Also, ${\dt{\rm{EE}}}\left( {\lambda} \right) > 0$ for $\lambda < {{\lambda}^{\rm{*}}}$ and ${\dt{\rm{EE}}}\left( {\lambda} \right) < 0$ for $\lambda > {{\lambda}^{\rm{*}}}$. Finally, by taking into account the constraints on the density of BSs, it follows that the unique optimal maximizer of the EE is ${\lambda ^{\left( {{\rm{opt}}} \right)}} = \max \left\{ {{\lambda ^{\left( {{\rm{min}}} \right)}},\min \left\{ {{\lambda ^*},{\lambda ^{\left( {{\rm{max}}} \right)}}} \right\}} \right\}$, since $\lambda  \in \left[ {{\lambda ^{\left( {{\rm{min}}} \right)}},{\lambda ^{\left( {{\rm{max}}} \right)}}} \right]$.
\paragraph{Case Study ${\lambda _{{\rm{MT}}}}/\lambda  \le 2.8$}
As for this case study, we leverage a notable result in fractional optimization \cite{AlessioMonograph2015}: the ratio between a i) non-negative, differentiable and concave function, and a ii) positive, differentiable and convex function is a pseudo-concave function. It is, in addition, a unimodal function with a finite maximizer if the ratio vanishes when the variable of interest (i.e., the BSs' density) tends to zero and to infinity. As for the case study under analysis, the EE in \eqref{Eq_14} can be re-written, by neglecting unnecessary constants that are independent of $\lambda$ and do not affect the properties of the function, as follows:
\setcounter{equation}{27}
\begin{equation}
\label{Eq_AppC_3}
\begin{split}
& {\rm{EE}}\left( \lambda  \right) = \\ & \frac{{{\mathcal{Q}}\left( \lambda  \right)}}{{\left[ {1 + \Upsilon {\mathcal{L}}\left( \lambda  \right)} \right]\left[ {\left( {{\rm{P}} + \Delta {\rm{P}}} \right) + {{{{\rm{P}}_{\rm{i}}}} \mathord{\left/
 {\vphantom {{{{\rm{P}}_{\rm{i}}}} {{\mathcal{L}}\left( \lambda  \right)}}} \right.
 \kern-\nulldelimiterspace} {{\mathcal{L}}\left( \lambda  \right)}} + {{\rm{P}}_{\rm{c}}}{{{\mathcal{M}}\left( \lambda  \right)} \mathord{\left/
 {\vphantom {{{\mathcal{M}}\left( \lambda  \right)} {{\mathcal{L}}\left( \lambda  \right)}}} \right.
 \kern-\nulldelimiterspace} {{\mathcal{L}}\left( \lambda  \right)}}} \right]}}.
 \end{split}
\end{equation}

From \textit{Lemma \ref{Lemma__PilotFunctionDensity}}, the numerator of \eqref{Eq_AppC_3} is a non-negative, differentiable, increasing and concave function for $\lambda \ge 0$. From \textit{Lemma \ref{Lemma__EE}}, the EE in \eqref{Eq_AppC_3} tends to zero if $\lambda \to 0$ and $\lambda \to \infty$. Therefore, a sufficient condition to prove the unimodality and pseudo-concavity of the EE is to show that the denominator of \eqref{Eq_AppC_3} is a positive, differentiable and convex function in $\lambda$ for ${\lambda _{{\rm{MT}}}}/\lambda  \le 2.8$. From \textit{Lemma \ref{Lemma__LoadFunction}} and \textit{Lemma \ref{Lemma__LoadFunctionModified}}, the first two properties are immediately verified. To complete the proof, the convexity of the denominator of \eqref{Eq_AppC_3} needs to be analyzed.

Let ${\rm{Den}}\left(  \cdot  \right)$ be the denominator of \eqref{Eq_AppC_3}. Let us introduce the function ${\mathcal{K}}\left( \lambda  \right) = 2{{{{{\mathcal{\dt L}}}^2}\left( \lambda  \right)} \mathord{\left/
{\vphantom {{{{{\mathcal{\dt L}}}^2}\left( \lambda  \right)} {{\mathcal{L}}\left( \lambda  \right)}}} \right. \kern-\nulldelimiterspace} {{\mathcal{L}}\left( \lambda  \right)}} - {\mathcal{\ddt L}}\left( \lambda  \right)$. The second-order derivative of ${\rm{Den}}\left(  \cdot  \right)$, as a function of $\lambda$, is as follows:
\setcounter{equation}{28}
\begin{equation}
\label{Eq_AppC_4}
\begin{split}
{\rm{\ddt Den}}\left( \lambda  \right) &= \underbrace {\Upsilon ({\rm {P}} + \Delta {\rm{P}}) {\mathcal{\ddt L}}\left( \lambda  \right)}_{{D_1}\left( \lambda  \right)} + \underbrace {\Upsilon {{\rm{P}}_{\rm{c}}}{\mathcal{\ddt M}}\left( \lambda  \right)}_{{D_2}\left( \lambda  \right)} \\ & + \underbrace {\left( {{{{{\rm{P}}_{\rm{c}}}} \mathord{\left/
 {\vphantom {{{{\rm{P}}_{\rm{c}}}} {{{\mathcal{L}}^2}\left( \lambda  \right)}}} \right.
 \kern-\nulldelimiterspace} {{{\mathcal{L}}^2}\left( \lambda  \right)}}} \right)\left( {{{2{\lambda _{{\rm{MT}}}}} \mathord{\left/
 {\vphantom {{2{\lambda _{{\rm{MT}}}}} {{\lambda ^3}}}} \right.
 \kern-\nulldelimiterspace} {{\lambda ^3}}}} \right)}_{{D_3}\left( \lambda  \right)}\underbrace {\left( {{\mathcal{L}}\left( \lambda  \right) + \lambda {\mathcal{\dt L}}\left( \lambda  \right)} \right)}_{{D_4}\left( \lambda  \right)}\\
& + \underbrace {{{\rm{P}}_{\rm{c}}}\left( {{{{\mathcal{M}}\left( \lambda  \right)} \mathord{\left/
 {\vphantom {{{\mathcal{M}}\left( \lambda  \right)} {{{\mathcal{L}}^2}\left( \lambda  \right)}}} \right.
 \kern-\nulldelimiterspace} {{{\mathcal{L}}^2}\left( \lambda  \right)}}} \right)}_{{D_5}\left( \lambda  \right)}{\mathcal{K}}\left( \lambda  \right) + \underbrace {\left( {{{{{\rm{P}}_{\rm{c}}}} \mathord{\left/
 {\vphantom {{{{\rm{P}}_{\rm{c}}}} {{{\mathcal{L}}^3}\left( \lambda  \right)}}} \right.
 \kern-\nulldelimiterspace} {{{\mathcal{L}}}\left( \lambda  \right)}}} \right)}_{{D_6}\left( \lambda  \right)}{\mathcal{K}}\left( \lambda  \right) \\ &+ \underbrace {\left( {{{{{\rm{P}}_{\rm{i}}}} \mathord{\left/
 {\vphantom {{{{\rm{P}}_{\rm{i}}}} {{{\mathcal{L}}^2}\left( \lambda  \right)}}} \right.
 \kern-\nulldelimiterspace} {{{\mathcal{L}}^2}\left( \lambda  \right)}}} \right)}_{{D_7}\left( \lambda  \right)}{\mathcal{K}}\left( \lambda  \right).
\end{split}
\end{equation}

A sufficient condition for proving that ${\rm{Den}}\left(  \cdot  \right)$ is a convex function in $\lambda$ is to show that ${{D_{\ell}}\left( \lambda  \right)} \ge 0$ for $\ell =1,2, \ldots, 7$ and $\mathcal{K}\left( \lambda  \right) \ge 0$ if ${\lambda _{{\rm{MT}}}}/\lambda  \le 2.8$. This can be proved as follows. ${D_1}\left( \lambda  \right) \ge 0$ for ${\lambda _{{\rm{MT}}}}/\lambda  \le 2.8$ follows from \textit{Lemma \ref{Lemma__LoadFunction}}. ${D_{\ell}}\left( \lambda  \right) \ge 0$ for $\ell=2,5$ if $\lambda  \ge 0$  follows from \textit{Lemma \ref{Lemma__LoadFunctionModified}}. ${D_{\ell}}\left( \lambda  \right) \ge 0$ for $\ell = 3, 6, 7$ if $\lambda  \ge 0$ follows from \textit{Lemma \ref{Lemma__LoadFunction}}. ${D_4}\left( \cdot  \right)$ and $\mathcal{K}\left( \cdot  \right)$ require deeper analysis. Define $\xi  = {{{\lambda _{{\rm{MT}}}}} \mathord{\left/ {\vphantom {{{\lambda _{{\rm{MT}}}}} \lambda }} \right. \kern-\nulldelimiterspace} \lambda }$. ${D_4}\left( \cdot  \right)$ and $\mathcal{K}\left( \cdot  \right)$ are positive functions in $\xi$ if:
\setcounter{equation}{29}
\begin{equation}
\label{Eq_AppC_5}
\begin{split}
& {D_4}\left( \xi  \right) \ge 0 \Leftrightarrow \\ & {{\underline D}_4}\left( \xi  \right) = 1 - {\left( {1 + {\xi  \mathord{\left/
 {\vphantom {\xi  \alpha }} \right.
 \kern-\nulldelimiterspace} \alpha }} \right)^{ - \alpha }} - x{\left( {1 + {\xi  \mathord{\left/
 {\vphantom {\xi  \alpha }} \right.
 \kern-\nulldelimiterspace} \alpha }} \right)^{ - \left( {\alpha  + 1} \right)}} \ge 0
\end{split}
\end{equation}
\setcounter{equation}{30}
\begin{equation}
\label{Eq_AppC_6}
\begin{split}
& {\mathcal{K}}\left( \xi  \right) \ge 0 \Leftrightarrow \\ & {\mathcal{\underline K}}\left( \xi  \right) = {\left( {1 + {\xi  \mathord{\left/
 {\vphantom {\xi  \alpha }} \right.
 \kern-\nulldelimiterspace} \alpha }} \right)^{ - \alpha }} \\ & \hspace{0.75cm} + \left[ {2 + \left( {1 + {1 \mathord{\left/
 {\vphantom {1 \alpha }} \right.
 \kern-\nulldelimiterspace} \alpha }} \right)x} \right]{\left[ {2 - \left( {1 - {1 \mathord{\left/
 {\vphantom {1 \alpha }} \right.
 \kern-\nulldelimiterspace} \alpha }} \right)x} \right]^{ - 1}} \ge 1.
\end{split}
\end{equation}

By direct inspection of \eqref{Eq_AppC_5} and \eqref{Eq_AppC_6}, it is not difficult to prove the following: i) ${{{ \underline D}}_4}\left( {\xi  \to 0} \right) = 0$ and ${{{{\underline {\dt D}}}}_4}\left( \xi  \right) \ge 0$ for $\xi \ge 0$, and ii) ${\mathcal{ \underline K}}\left( {\xi  \to 0} \right) = 1$ and ${\mathcal{\underline {\dt K}}}\left( \xi  \right) \ge 0$ for $\xi \le 2.8$. These two conditions imply ${{D_4}\left( \lambda  \right)} \ge 0$ for $\lambda \ge 0$ and ${\mathcal{K}}\left( \lambda  \right) \ge 0$ for ${\lambda _{{\rm{MT}}}}/\lambda  \le 2.8$. This concludes the proof.
\end{document}